\documentclass[onecolumn]{emulateapj}
\usepackage{graphicx}

\newcommand{\lsim }{{\lower0.8ex\hbox{$\buildrel <\over\sim$}}}
\newcommand{\gsim }{{\lower0.8ex\hbox{$\buildrel >\over\sim$}}}
\newcommand{\up}{\ifmmode{u^{\prime}} \else $u^{\prime}$\fi}
\newcommand{\gp}{\ifmmode{g^{\prime}} \else $g^{\prime}$\fi}
\newcommand{\rp}{\ifmmode{r^{\prime}} \else $r^{\prime}$\fi}
\newcommand{\ip}{\ifmmode{i^{\prime}} \else $i^{\prime}$\fi}
\newcommand{\zp}{\ifmmode{z^{\prime}} \else $z^{\prime}$\fi}
\newcommand{\us}{\ifmmode{u^{\ast}} \else $u^{\ast}$\fi}
\newcommand{\gs}{\ifmmode{g^{\ast}} \else $g^{\ast}$\fi}
\newcommand{\rs}{\ifmmode{r^{\ast}} \else $r^{\ast}$\fi}
\newcommand{\is}{\ifmmode{i^{\ast}} \else $i^{\ast}$\fi}
\newcommand{\zs}{\ifmmode{z^{\ast}} \else $z^{\ast}$\fi}
\newcommand{\gmr}{\ifmmode{(g^{\prime}-r^{\prime})} \else $(g^{\prime}-r^{\prime})$\fi}
\newcommand{\rmi}{\ifmmode{(r^{\prime}-i^{\prime})} \else $(r^{\prime}-i^{\prime})$\fi}
\newcommand{\gmrs}{\ifmmode{(g^{\ast}-r^{\ast})} \else $(g^{\ast}-r^{\ast})$\fi}
\newcommand{\rmis}{\ifmmode{(r^{\ast}-i^{\ast})} \else $(r^{\ast}-i^{\ast})$\fi}
\def\Chandra     {{\em Chandra}}
\newcommand{\fcgs}{\ifmmode {\rm \,erg~cm}^{-2}~{\rm s}^{-1}\else 
\,erg~cm$^{-2}$~s$^{-1}$\fi}
\newcommand{\lcgs}{\ifmmode {\rm \,erg~s}^{-1}\else \,erg~s$^{-1}$\fi}
\newcommand{\flamcgs}{\ifmmode {\rm erg~cm}^{-2}~{\rm s}^{-1}~Hz^{-1}\else 
erg~cm$^{-2}$~s$^{-1}~$\AA$^{-1}$\fi}
\newcommand{\fnucgs}{\ifmmode {\rm erg~cm}^{-2}~{\rm s}^{-1}~{\rm
Hz}^{-1}\else  erg~cm$^{-2}$~s$^{-1}$~Hz$^{-1}$\fi}
\newcommand{\lnucgs}{\ifmmode {\rm erg~s}^{-1}~{\rm Hz}^{-1}\else 
erg~s$^{-1}$~Hz$^{-1}$\fi}
\newcommand{\kms}{\ifmmode~{\rm km~s}^{-1}\else ~km~s$^{-1}~$\fi}
\newcommand{\mone}{\ifmmode ^{-1}\else$^{-1}$\fi}
\newcommand{\mtwo}{\ifmmode ^{-2}\else$^{-2}$\fi}
\newcommand{\mv}{\ifmmode {m_{V}}\else${m_{V}}$\fi}
\newcommand{\Mv}{\ifmmode {M_{V}}\else${M_{V}}$\fi}
\newcommand{\msun}{\ifmmode {M_{\odot}}\else${M_{\odot}}$\fi}
\newcommand{\rsun}{\ifmmode {R_{\odot}}\else${R_{\odot}}$\fi}
\newcommand{\lsun}{\ifmmode {L_{\odot}}\else${L_{\odot}}$\fi}
\newcommand{\lapprox }{{\lower0.8ex\hbox{$\buildrel <\over\sim$}}}
\newcommand{\gapprox }{{\lower0.8ex\hbox{$\buildrel >\over\sim$}}}
\newcommand{\lognlogs}{{log{\em N}-log{\em S}}}
\newcommand{\cmsq}{\ifmmode{\rm ~cm^{-2}} \else cm$^{-2}$\fi}
\newcommand{\ebmv}{\ifmmode{\rm E}_{B-V} \else E$_{B-V}$\fi}
\newcommand{\nh}{\ifmmode{\rm N_{H}} \else N$_{H}$\fi}
\newcommand{\lognh}{\ifmmode{\rm log\,N_{H}} \else log\,N$_{H}$\fi}
\newcommand{\nhgal}{\ifmmode{ N_{H}^{Gal}} \else N$_{H}^{Gal}$\fi}
\newcommand{\nhintr}{\ifmmode{ N_{H}^{intr}} \else N$_{H}^{intr}$\fi}
\newcommand{\nhtot}{\ifmmode{ N_{H}^{tot}} \else N$_{H}^{tot}$\fi}
\newcommand{\meangamma}{\ifmmode{\langle\Gamma\rangle} \else 
$\langle\Gamma\rangle$\fi}
\newcommand{\fx}{\ifmmode f_X \else $~f_X$\fi}
\newcommand{\fxfo}{\ifmmode \frac{f_X}{f_{opt}} \else
 $\frac{f_X}{f_{opt}}$\fi}
\newcommand{\fxfr}{\ifmmode \frac{f_X}{f_{r}} \else
 $\frac{f_X}{f_r}$\fi}
\newcommand{\logfx}{\ifmmode{\rm log}~f_X \else log$~f_X$\fi}
\newcommand{\logfxfo}{\ifmmode{\rm log}\,(\frac{f_X}{f_{opt}}) \else
 ${\rm log}\,(\frac{f_X}{f_{opt}})$ \fi}
\newcommand{\logxr}{\ifmmode{\rm log}\,\frac{f_X}{f_{r}} \else
 ${\rm log}\,\frac{f_X}{f_{r}}$ \fi}
\newcommand{\lopt}{\ifmmode L_{opt} \else $~L_{opt}$\fi}
\newcommand{\loglopt}{\ifmmode{\rm log}~L_{opt} \else log$~L_{opt}$\fi}
\newcommand{\lx}{\ifmmode L_X \else $~L_X$\fi}
\newcommand{\loglx}{\ifmmode{\rm log}~L_X \else log$~L_X$\fi}
\newcommand{\aox}{\ifmmode{\alpha_{ox}} \else $\alpha_{ox}$\fi} 
\newcommand{\snr}{\ifmmode{\frac{S}{N}} \else $\frac{S}{N}$\fi} 

\shorttitle{The ChaMP: Optical Followup}
\shortauthors{Green et al.}

\begin{document}

\title{The Chandra Multiwavelength Project: \\  
Optical Followup of Serendipitous Chandra Sources}


\author{P. J. Green,$^{1,14}$
J. D. Silverman,$^{1,14,15}$
R. A. Cameron,$^1$
D.-W. Kim,$^1$
B. J. Wilkes,$^1$   
W. A. Barkhouse,$^1$
A. LaCluyz\'{e},$^{4,14}$
D. Morris,$^2$
A. Mossman,$^1$
H. Ghosh,$^1$
J. P. Grimes,$^{16}$
B. T. Jannuzi,$^3$
H. Tananbaum,$^1$
T. L. Aldcroft,$^1$
J. A. Baldwin,$^4$ 
F. H. Chaffee,$^5$ 
A. Dey,$^3$
A. Dosaj,$^{6,14}$
N. R. Evans,$^1$
X. Fan,$^7$
C. Foltz,$^8$
T. Gaetz,$^1$ 
E. J. Hooper,$^9$
V. L. Kashyap,$^1$
S. Mathur,$^{10}$
M. B. McGarry,$^1$
E. Romero-Colmenero,$^{17}$
M. G. Smith,$^{12}$
P. S. Smith$^{11,14}$
R. C. Smith,$^{12}$
G. Torres,$^1$
A. Vikhlinin$^1$ 
and
D. R. Wik$^{13}$}
  
\email{pgreen@cfa.harvard.edu}

\begin{abstract}
We present followup optical \gp, \rp, and \ip\, imaging and
spectroscopy of serendipitous X-ray sources detected in 6 archival
\Chandra\, images included in the \Chandra\, Multiwavelength Project
(ChaMP).  Of the 486 X-ray sources detected between $3\times10^{-16}$
and $2\times10^{-13}$ (with a median flux of $3\times10^{-15}$) \fcgs,
we find optical counterparts for 377 (78\%), or 335 (68\%) counting
only unique counterparts.  We present spectroscopic classifications
for 125 objects, representing 75\% of sources with $\rs<21$ optical
counterparts (63\% to $\rs=22$).  Of all classified objects, 63 (50\%)
are broad line AGN, which tend to be blue in \gmrs\ colors.  X-ray
information efficiently segregates these quasars from stars, which
otherwise strongly overlap in these SDSS colors until $z>3.5$.  We
identify 28 sources (22\%)  as galaxies that show narrow emission
lines, while 22 (18\%) are absorption line galaxies.  Eight galaxies
lacking broad line emission have X-ray luminosities that require
they host an AGN (log$L_X>43$).  Half of these have hard X-ray
emission suggesting that high gas columns obscure both the X-ray
continuum and the broad emission line regions.   We find objects
in our sample that show signs of X-ray or optical absorption, or both, 
but with no strong evidence that these properties are coupled.
ChaMP's deep X-ray and optical imaging enable multiband selection of
small and/or high-redshift groups and clusters.  In these 6 fields we
have discovered 3 new clusters of galaxies, two with $z>0.4$, and one 
with photometric evidence for a similar redshift. 
\end{abstract}

\section{Introduction}

\subsection{X-ray Surveys and the Cosmic X-ray Background}

X-ray surveys provide fundamental advances in our knowledge of
the X-ray universe and indeed the universe as a whole (e.g., the
Einstein Medium Sensitivity Survey \citep{SJT91}; the
Cambridge-Cambridge ROSAT Serendipitous Survey \citep{BB97}; ROSAT
International X-ray/Optical Survey \citep{PM96}, the ASCA Large Sky
Survey \citep{ASCALSS}. ROSAT (0.1-2.4\,keV) and more recently
\Chandra\, (0.3-8\,keV) have resolved 80-90\% of the Cosmic X-ray
background (CXRB) into discrete sources (Hasinger et al. 1998; Rosati
et al. 2002; Moretti et al. 2003) most of which are the unobscured AGN
familiar from optical and soft X-ray surveys.  However, the high
energy spectrum of the CXRB is much harder ($\Gamma\sim
1.4$\footnote{$\Gamma$ is the 
photon number index  of an assumed power-law continuum such that
$N_E(E)=N_{E_0}\,E^{\Gamma}$. In terms of a spectral index $\alpha$ from
$f_{\nu}=f_{\nu_0}\nu^{\alpha}$, we define $\Gamma= (1-\alpha)$.})
than that of known AGN ($\Gamma\sim 1.9$). Population synthesis models
that satisfy both the CXRB spectrum and the observed X-ray number
counts vs. flux relation (\lognlogs; Comastri et al. 1995;
Hasinger et al. 1998; Tozzi et al. 2001) favor
absorbed AGN as a dominant component of the CXRB.  In these
models, {\em most} of the accretion luminosity in the universe is from
obscured sources \citep{HG00,FAIK99}, which appear to have hard
X-ray spectra because circumnuclear gas absorbs low energy X-rays.
X-ray spectral analyses and optical follow-up of faint hard X-ray
sources detected by  \Chandra\ and XMM have confirmed this
interpretation generally  \citep{BW1MSEC01,ADOPTFNT01} but also
presented some surprises. Type~2 active galactic nuclei are mostly
found at $z<1$, and the required space density of such objects must be
much greater \citep{RPTP02,ADOPTFNT01,HA01} than those of standard
unabsorbed broad-line AGN (BLAGN). For the standard AGN unification
model to survive, wherein optical Type~1 and Type~2 classifications
represent different viewing angles on identical objects \citep{AR93},
it must begin to encompass the population of X-ray absorbed AGN and its
evolution in number density, covering fraction and/or optical depth of
absorbers.  

Absorption may be increasing with luminosity or redshift 
\citep{EM98,RJ97}, and could be associated with early circumnuclear
starbursts \citep{GMETAL00}.  The intriguing suggestions of a $z\sim
1$ peak in X-ray-selected (X-S) galaxies \citep{BAJ02,TP01} has been
suggested as  evidence for an epoch of enhanced activity related to
the assembly of massive galaxies \citep{FXRB02}.  
At high redshift ($z>4$), a significant dropoff in the co-moving space
density of quasars seen in optical (e.g., \citealt{SSG95, wa94, os82})
and radio surveys \citep{sh96} hints at 
either the detection of the onset of accretion onto supermassive black
holes, or a missed high-redshift population, possibly due to
intrinsic absorption.  Based on preliminary evidence for constant
space densities of X-ray selected quasars beyond a redshift of 2,
deep (ROSAT) soft X-ray surveys \citep{MIY00} have been used to
support the latter interpretation. Unfortunately, the sample
size is small with only 8 quasars beyond a redshift of 3.  
At these early epochs, higher rates of galaxy interactions and 
mergers are expected to have triggered nuclear activity in galaxies
(e.g. Blain et al. 1999, Osterbrock 1993). As gas-rich protogalaxies
grow by merging, they may induce growth in the central black holes.
Preliminary models tie together AGN evolution with hierarchical growth
of clustering (e.g., Cole \& Kaiser 1989) and galaxy formation models.
\citet{WFN00} use the assumption that seed blocks for clustering each
contain black holes of mass $1.6\times 10^6\msun$.  During these early
high accretion phases, quasars may be self-cloaking, naturally copious
producers of metals \citep{KG02} and dust \citep{EMK02}.  A truncation
of growth in baryonic mass is perhaps achieved by the wind-driven gas
from the AGN, yielding today's observed black hole/bulge mass
correlation \citep{MDFL01}.   

At more recent epochs, the bright optically-selected (O-S) quasar
population appears to fall dramatically between $z\sim2.5$ and the
present, on a timescale of about 2\,Gyr, quite rapid compared to the
fading of galaxy formation in hierarchical clustering models. Strong
luminosity evolution dominates the luminosity function (LF) during
this epoch across radio, optical, and X-rays, possibly associated with
depletion of gas reservoirs by accretion of smaller group companions
\citep{CV00}.  

\subsection{The need for the Chandra Multiwavelength Project (ChaMP)}

To understand the formation and evolution of accretion onto
supermassive black holes, and its links to galaxy formation, an
accurate census of accreting objects is required.  Effectively, 
we must account for the accretion radiation energy density across
cosmic time consistent with the mass density of supermassive black
holes in the Universe (e.g., Yu \& Tremaine 2002; Elvis,
Risaliti \& Zamorani 2002; Mainieri et al. 2002).  Galaxies without a
dominant active nucleus also produce X-rays from 
by-products of star formation such as low- and high-mass X-ray
binaries (LMXBs and HMXBs), supernova remnants, and hot diffuse gas. 
The distinction between star-formation and AGN-induced emission
is a subject of long debate that relates intimately to the
above discussion.  Progress on the composition of the CXRB, 
and the accretion and star formation history of the universe
all require a wide area, multiwavelength survey with greater
sensitivity and completeness than previously achieved, and with
reduced bias against absorbed  objects.  

Both soft X-ray and optical surveys suffer strong selection effects
due to intrinsic obscuration and the intervening Ly$\alpha$ forest. 
Consistent with the hypothesis that the CXRB is dominated by absorbed
AGN, a substantial population of reddened AGN is being found in IR
\citep{GMLM02,CR01,MFDM99,WBSG02} and radio \citep{BR97,MD99,VJETAL01}
surveys.  While radio surveys are least affected by intrinsic
absorption, radio-loud objects constitute a minority of AGN.  
Optical classification rests on secondary properties (e.g. emission
lines from ionized plasma).  By contrast, the {\em primary signature
of accretion} is emission from close (10-100 gravitational radii) to
the supermassive black hole, which is why {\em only sensitive hard
X-ray selection 
samples all known varieties of AGN}. Furthermore, for \Chandra\, and {\em
XMM,} sensitivity to emission up to 10~keV (observed frame) can reveal
hidden populations of active galactic nuclei (AGN) including heavily
obscured quasars \citep{NCTYP2X02, ST01}. Absorption ($\sim 10^{22-23}$
cm$^{-2}$) has $10-3000\times$ less impact on detection by \Chandra\,
(0.3-8~keV) than on ROSAT (0.3-2.4~keV). High-$z$ objects can be
detected through an even larger intrinsic absorbing column of gas and
dust because the observed-frame X-ray bandpass corresponds to higher
energy, and hence more penetrating X-rays at the source.\footnote{The
observed-frame, effective absorbing column is $N_{\rm H}^{\rm eff}\sim
N_{\rm H}/(1+z)^{2.6}$ (Wilman \& Fabian 1999).}  Therefore, optical
and X-ray surveys will complement each other, providing a fair census
of mass accretion onto black holes at high redshift.

With its low background and tight PSF (\lapprox$1\arcsec$ FWHM on
axis) \Chandra\, is probing the high energy universe with unparalleled
sensitivity and resolution.  The \Chandra\ Multiwavelength
Project (ChaMP) is a wide-area study of a large sample of
serendipitous X-ray sources detected in \Chandra\, archival images.
The ChaMP has selected more than 100 \Chandra\, images, in
which we will detect several thousand serendipitous (non-target)
extragalactic sources.  A first X-ray catalog of these sources
and early results are published in accompanying papers
(\citealt{kim03a, kim03b}).  The ChaMP AGN 
sample includes faint nearby AGN, absorbed quasars to $z\sim3$ and
unobscured AGN to $z\sim5$, populating new regions of $L-z$ space, and
providing a broad census of the population that will reap
statistically significant samples for rare source types.  The two
0.1deg$^2$ \Chandra\, 1Msec Deep Fields probe far too little
volume at low redshift where many of the absorbed AGN probably reside
\citep{BAJ01,FF00} and contain few bright examples to easily
characterize these sources.  The ChaMP will effectively bridge the gap
between flux limits achieved with the \Chandra\, deep field (CDF)
observations and those of past surveys.  The flux regime where the
ChaMP has greatest sky area ($10^{-14}\fcgs$) coincides with the break 
between power-law slopes fitting the hard (2-10~keV) \lognlogs\
\citep{RPTP02}.  Furthermore, as shown in Figure~1 of
\citealt{kim03b}, a gap in coverage remains between deep ASCA and
\Chandra\ surveys near the break.  The wide-area and medium depth of
\Chandra\ fields included in the ChaMP will span these fluxes with
good statistics.    

A number of new high galactic latitude serendipitous surveys are
underway both with the XMM-Newton X-ray observatory and \Chandra.  
The XMM-Newton Survey Science Center \citep{WM01} targets several
flux ranges, including a medium sensitivity survey
($f$90.5-4.5keV)$>2\times 10^{-14}$\fcgs) with first results described
by \citet{barcons02}.  The XMM EPIC instrument covers a large ($\sim
15\arcmin$ radius) field of view and has sensitivity out to higher
energies than \Chandra. However, given the higher background and
larger PSF of XMM (5-7\arcsec\ FWHM), the minimum detectable flux for
an on-axis  point source with a typical AGN spectrum is
similar.  For cross-correlation to faint sources in other wavebends,
the tight \Chandra\ PSF is a great advantage.  

While the \Chandra\, Multiwavelength Project (ChaMP) emphasizes study
of the nature and evolution of active galaxies, as sketched above, it
will also provide to the scientific community well-defined wide-area
survey products suitable for detailed multiwavelength studies of
quasars, stars, star formation, galaxies, clusters of galaxies, and
clustering.  

We have chosen to analyze a subsample of 6 ChaMP fields for this paper
as an example of what can be expected from the full ChaMP sample.
After describing  in \S\ref{fields} the overall ChaMP field
selection process, we briefly describe the 
processing and analysis of the X-ray images in \S\ref{xanalysis},
which is detailed in our companion X-ray paper \citep{kim03a}.
We then outline our ongoing optical imaging campaign (\S\ref{optim}),
X-ray/optical source matching (\S\ref{matching}), and optical
spectroscopy (\S\ref{ospec}).  We describe overall sample results
in \S\ref{results}, with separate emphasis on AGN, galaxies,
and clusters of galaxies.  Prospects and plans for ChaMP science
are sketched in \S\ref{prospects}.  We discuss selected individual
objects in each of the 6 fields in Appendix~1.  Throughout this
paper, we assume H$_{\circ}$=70 km s$^{-1}$ Mpc$^{-1}$, 
$\Omega_{\Lambda}= 0.7$, and $\Omega_{M}=0.3$.

\section{Field Selection and X-ray Image Processing}
\label{fields}

\begin{figure*}[ht!]
\includegraphics[width=14cm,angle=-90]{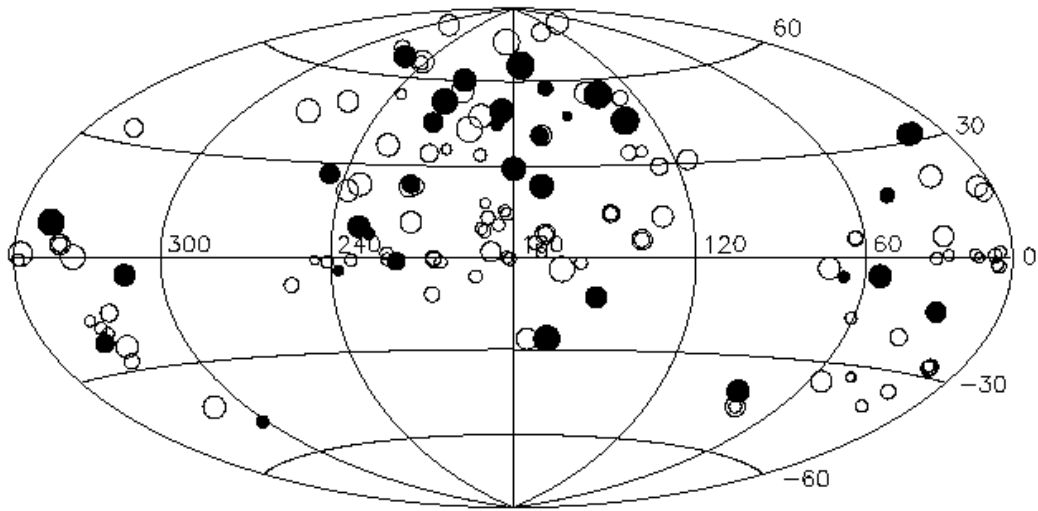}
\caption{Location of all 137 ChaMP high latitude ($|b|>20\deg$) fields
is shown in an Aitoff equatorial projection, with RA and Dec in decimal
degrees.  Filled circles represent ACIS-I at the aimpoint, and open
circles ACIS-S.  The average area for ChaMP-selected ObsIDs 
corresponds to 5.8 CCDs turned on, each of which is 8.3\arcmin\ on
a side.  Point size crudely indicates \Chandra\, exposure times,
varying from 2 to 190 ksec.   
}
\label{aitoff}
\end{figure*}

Our field selection, and X-ray image processing strategies are detailed
in the companion X-ray paper of \citet{kim03a},  Briefly, fields are
restricted to high 
Galactic latitudes ($|b|>20^{\circ}$).  We omit a handful
of pointings from other serendipitous surveys like the \Chandra\,
Deep Fields, and exclude fields with large bright optical or X-ray
sources.  This selection yields 146 ACIS exposures in Cycles 1 \& 2,
with \Chandra\, exposure times ranging from 2 to 190~ksec, and
covering about 14\,deg$^2$.   The sky placement of these fields is
shown in an equatorial  Aitoff projection in Figure~\ref{aitoff}. The
full field list is public, available at \\ 
\hbox{\url http://hea-www.harvard.edu/CHAMP/}.

The fields studied in this pilot paper include 6 early (Cycle~1)
fields listed in Table~\ref{tfields} for which we have already reduced
and calibrated both archival X-ray images and photometric NOAO
4\,meter/Mosaic imaging.  These 6 fields represent about 12\% of
ChaMP fields of similar exposure time or greater.  Four of the 6 fields
have clusters as their intended \Chandra\, targets.  We do not
include the intended (PI) targets in our analysis, their presence may
bias the sample of source types analyzed here.  The large number of
fields in the full ChaMP and the variety of PI target types allow
important tests for possible sample biases.  While an excess of point
sources in cluster fields has been claimed anecdotally
\citep{CMMP01,MPKD02}, no significant difference is found using the 
larger sample of \Chandra\ images (29 with clusters and 33 without)
in \citet{kim03b}.   

\begin{deluxetable}{clrrccccl}
\tabletypesize{\scriptsize}
\tablecaption{\Chandra\, Fields \label{tfields}}
\tablewidth{0pt}
\tablehead{
\colhead{ObsID} & 
\colhead{PI Target}  &
\colhead{~~ Exposure\tablenotemark{a}} & 
\colhead{ACIS CCDs\tablenotemark{b}} & 
\colhead{RA}  & 
\colhead{DEC} & 
\colhead{UT Date} &
\colhead{Galactic $N_H$} \\
\colhead{}  &
\colhead{}  &
\colhead{(ksec)}  &
\colhead{}  &
\multicolumn{2}{c}{ J2000\tablenotemark{c} } & 
\colhead{}  &
\colhead{($10^{20}$cm$^{-2}$)} \\
}
\startdata
536  &   MS1137.5+6625  & 114.6 & 012{\em 3}67 & 11:40:23.3 &
+66:08:42.0 & 1999 Sep 30 & 1.18  \\
541  &   V1416$+$4446   & 29.8  & 012{\em 3}7  & 14:16:28.8 &
+44:46:40.8 & 1999 Dec 02 & 1.24  \\
624  &   LP944$-$20     & 40.9  & 236{\em 7}8  & 03:39:34.7 &
$-$35:25:50.0 & 1999 Dec 15 & 1.44  \\
861  &   Q2345$+$007    & 65.0  & 236{\em 7}8  & 23:48:19.6 &
+00:57:21.1 & 2000 Jun 27 & 3.81  \\
914  &   CLJ0542$-$4100 & 48.7  & 012{\em 3}7  & 05:42:50.2 &
$-$41:00:06.9 & 2000 Jul 26 & 3.59  \\
928  &   MS2137$-$2340  & 29.1 &  236{\em 7}8  & 21:40:12.7 &
$-$23:39:27.0 & 1999 Nov 18 & 3.57  \\
\enddata
\tablenotetext{a}{Effective screened exposure time for the
on-axis chip. BI chips 5 and 7 generally have lower screened exposures 
since they are more susceptible to solar flares.}
\tablenotetext{b}{The ACIS CCD chips used in the observation, with
the aimpoint chip in italics. The ACIS has 10 chips total, of which
CCDs 5 and 7 are back-illuminated.  Due to telemetry limitations, at
most 6 chips can be read out.} 
\tablenotetext{c}{Nominal target position, not including any
\Chandra\, pointing offsets.}
\end{deluxetable}

\section{X-ray Image Processing and Analysis}
\label{xanalysis}

We have used \Chandra\ archival data reprocessed (in April
2001) at CXC\footnote{CXCDS versions R4CU5UPD14.1 or later, along with
ACIS calibration data from the Chandra CALDB~2.0b.}. 
The foundation of the ChaMP rests on our own \Chandra\, image analysis
pipeline XPIPE \citep{kim03a}, built mainly with CIAO tools
({\url http://asc.harvard.edu/ciao}), and designed to uniformly and carefully
screen and analyze the X-ray data.  Details of XPIPE can be found in our
X-ray companion paper, \citet{kim03a}.  Briefly, XPIPE screens high particle
background periods, cosmic rays and hot pixels from each ACIS CCD
exposure.  Wavelet transform detection is performed (using
{\tt wavdetect}; Freeman et al. 2002) at 7 scales from 0.5 to
32\arcsec\, thus accommodating the \Chandra\, PSF variation as a
function of off-axis angle from the mirror axis. We select a
significance threshold parameter of $10^{-6}$ per pixel, which
corresponds to about one spurious pixel detection per CCD.  We
have confirmed this rate by an extensive simulation showing that the
number of spurious sources detected is always less than this rate.  An
additional flux criterion of signal to noise (\snr) greater than 2
from our X-ray photometry further decreases the number of spurious
sources in the sample.  We avoid spurious sources near detector edges
by generating an exposure map for each CCD with the appropriate aspect
histogram, and then applying a minimum exposure threshold of 10\% to
all pixels included for analysis. 

The absolute celestial positions of sources detected with {\tt wavdetect}
are accurate to about 1\arcsec, which is the specified absolute
position accuracy for the Chandra observatory.\footnote{
See {\url http://asc.harvard.edu/cal/ASPECT/} for a detailed
discussion of \Chandra\, aspect solutions.}  To check for and
minimize any systematic offset in the absolute source positions
returned by {\tt wavdetect}, the positions of X-ray sources are
correlated to optical sources in the field. The average position
offset from X-ray sources to matched optical sources is then applied
to each X-ray source position. This procedure is described in more
detail in \S\ref{match}.

We visually inspect all the X-ray images to validate sources and
flag them for contamination (by hot pixels, readout streaks, bad
bias values) or excessive uncertainty due to contamination
of source or background regions.  Throughout this paper and for ChaMP
statistics, the target sources are flagged and excluded from analysis
in all fields.  Source catalogs and FITS images are
being made publicly available, with details at the ChaMP web site,
{\url http://hea-www.harvard.edu/CHAMP/}. 

At the X-ray source positions derived from {\tt wavdetect}, we perform
aperture photometry. We use a circular source aperture whose radius
encompasses 95\% of deposited energy, determined at 1.5~keV, but restrict
this radius to the range $3\arcsec < r_{ap} < 40\arcsec$.
We use a background annulus extending from 2 to 5 times $r_{ap}$,
excluding flux within $r_{ap}$ of any other detected sources
in the annulus.  We perform photometry in 3 energy bands: soft ($S$;
0.3-2.5~keV), hard ($H$; 2.5-8~keV), and broad ($B$; 0.3-8~keV).  We
calculate net count rates using the effective exposure and vignetting
factors both at the source and background regions. 

Only sources with $\snr>2$ in at least one of the bands $B$, $S$, or
$H$ are included for analysis here.\footnote{\snr\ is calculated from
	$\sigma = \sqrt{\sigma_{tot}^2 + R^2\sigma_{bkg}^2}$
where $\sigma_{tot}$ is the error in the total number of counts
in the source aperture, and $\sigma_{bkg}$ is the error on counts
in the source-excised background aperture. $R$ is the ratio of areas
of source and background regions. The \snr\ criterion is a flux, {\em
not} a detection criterion, since detection is based on probability in
{\tt wavdetect}.  Sources with $\snr=2$ can be easily detected in many
cases, since the background rates are typically 1 broadband
(0.5-10\,keV) count per 40 pixels in 100 ksec \citep{CXC4}.} 
Flux is calculated from the net count rate in the $B$, or if necessary from
another ($H$ or $S$) band to satisfy our $\snr$ criterion. 
We assume a power-law of photon index $\Gamma=1.4$ absorbed by a
neutral Galactic column density $N_H$ taken from \citet{DJLF90} for
the \Chandra\, aimpoint position on the sky.  The effect of varying
the source model between reasonable powerlaw slopes from $\Gamma=1.0$
to 2.5 causes (for $\nhgal = 3\times 10^{20}$cm$^{-2}$)
an increase in the derived $f_X$(0.5-2~keV) of $\sim
12\%$ for FI chips, and 6\% for BI chips.

The input bandpass energy range is passed to PIMMS (Mukai 1993; v3.2d,
which includes \Chandra\, Cycle~4 effective area curves) along with
the adjusted count rate and spectral model parameters to calculate a
de-absorbed flux $f_X$ in the 0.5-2~keV band, since this bandpass is
most useful for comparison to numerous published results from ROSAT,
XMM, and \Chandra.  The error in $f_X$ is computed by applying the
same counts-to-flux conversion factor to the error in counts,
increased by a further 15\% to qualitatively account for errors
induced by the spectral model assumptions.  These results are listed
on-line along with other quantities described below.

For every source, we wish to calculate an effective on-axis ACIS-I
hardness ratio 
	$$HR=\frac{H-S}{H+S}$$  
that may be fairly compared among sources to study their X-ray
properties even in the low counts regime.  Many (especially weak) 
sources have derived $H$ or $S$ counts that are formally $<2\sigma$,
or even negative.  Naive inclusion of these yields nonsensical $HR$
values outside the physical range from -1 to 1.  Therefore, we first
replace such counts by a $2\sigma$ upper limit.  This yields for $HR$
either a `detection' (when both $H$ and $S$ counts are $>2\sigma$), an
upper limit (when $H<2\sigma_H$), a lower limit (when
$S<2\sigma_S$), or very rarely an indeterminate value (both).  

Mirror reflectance changes as a function of energy and 
of the angle $\theta$ between the source and \Chandra's aimpoint, 
causing  energy-dependent vignetting.  A different energy dependence
arises from the detecting CCD - the quantum efficiency of backside- (BI) vs.
frontside-illuminated (FI) chips.  Therefore, for an identical
source spectrum, different $HR$ values result, depending on
$\theta$ and chip.  To correct for this, we perform a second adjustment
of $H$ and $S$ counts, normalizing them to refer uniformly to an
on-axis source falling on an FI chip.  To do this, we have modified
the PIMMS code to include vignetting.  Where the integration of source
model is convolved with the instrument effective area (EA), 
we incorporate an extra multiplicative factor - the percentage of the
on-axis EA as a function of photon energy and $\theta$.  We
interpolate between analytic fits (courtesy Diab Jerius) to EA as a
function of $\theta$ from 0 to 30\arcmin, performed for 4 energies (0.49,
1, 1.48, 2.02, 2.99, and 6.4~keV).  Similar curves are shown in the
CXC Proposer's Guide Rev4.0 \citep{CXC4}.  

To derive the corrected hardness ratio $HR_0$, we first convert 
counts to de-absorbed flux in each band using our standard source
model, the source $\theta$ and chip.  Using the same source model, we
then convert the derived flux back counts at $\theta=0$ on an
FI chip.   The resulting corrections in $HR_0$ do not depend strongly
on the assumed $\Gamma$; the difference in the mean
$HR_0$ derived assuming $\Gamma=0.5$ and 3.0, is only
0.02 (3\%) out to 15\arcmin\, off-axis.   

\begin{figure*}[ht!]
\epsscale{1.5}
\plottwo{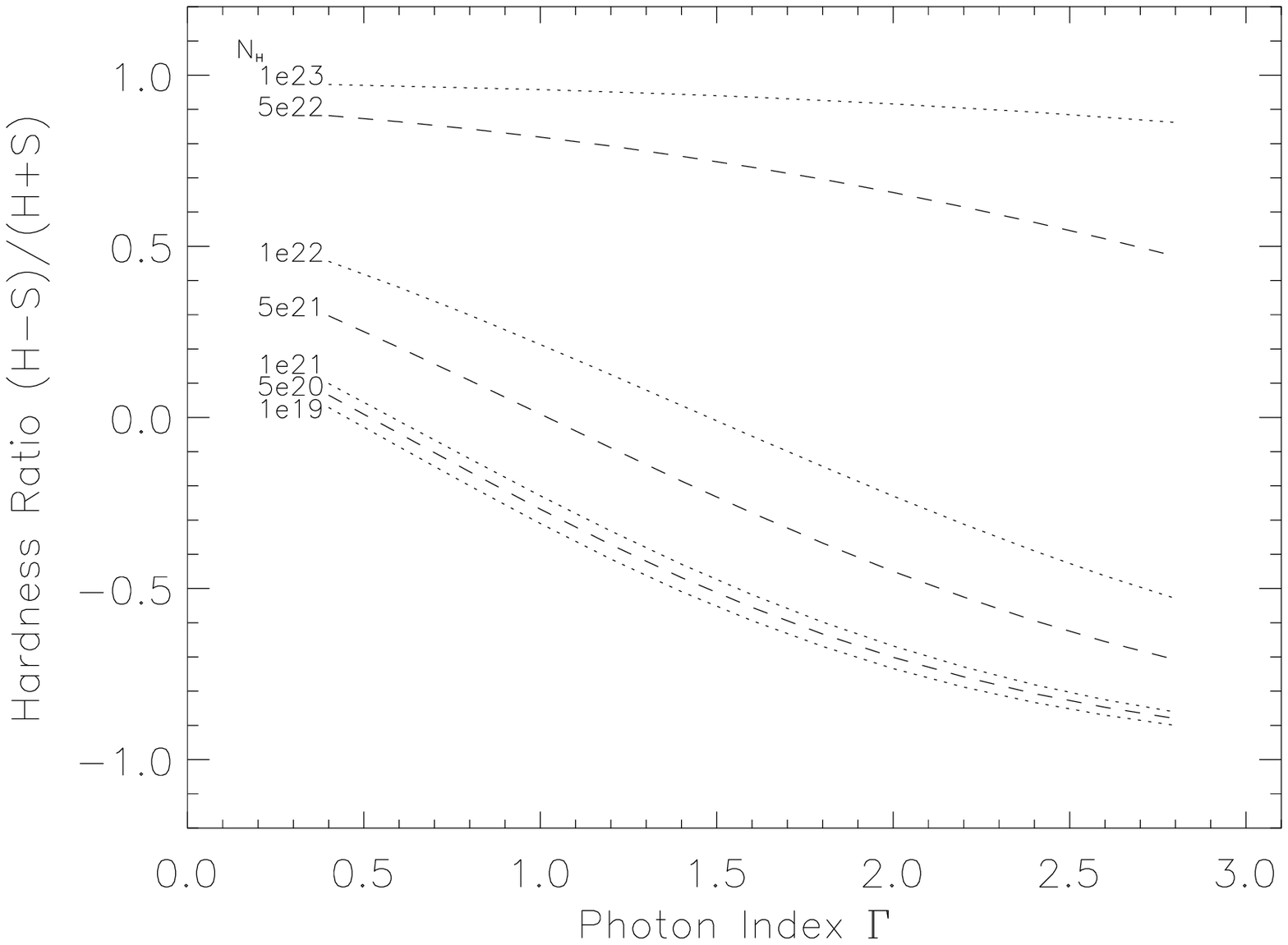}{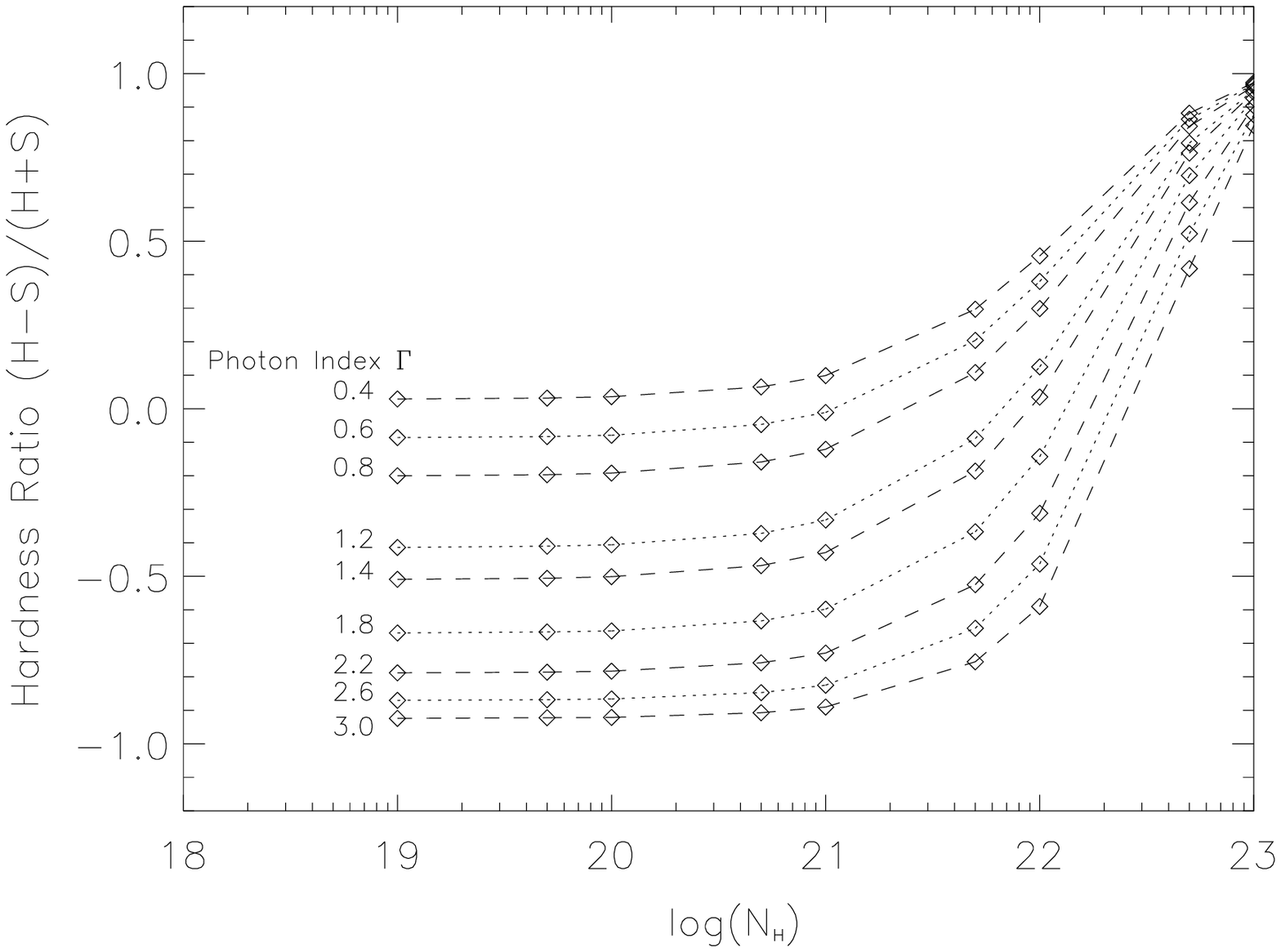}
\caption{\small
Simulated on-axis FI chip Hardness Ratios for high $\snr$ sources. 
{\em Top.} The family of curves show the resulting $HR$ for different
($z=0$) absorbing columns as a function of power-law Photon Index
$\Gamma$.  {\em Bottom.} The family of curves show the resulting $HR$ for
different values of power-law Photon Index $\Gamma$ as a function of  
absorbing column.
\label{hr_grid}}
\end{figure*}

We have performed a series of simulations to derive the expected $HR_0$
values for high $\snr$ sources across a grid of $\Gamma$ and $N_H$
values, as shown in Figure~\ref{hr_grid}. A comparison to the median
corrected value of $\overline{HR_0}=-0.5$ for all sources detected in both $H$
and $S$, and assuming no absorption implies $\Gamma\sim 1.4$.

Finally, if at least one of $H$ and $S$ are significant, then we
calculate errors for $HR_0$ from the $H$ and $S$ counts (or limits) as 
	$$\sigma_{HR} = 
	\frac{2}{(H+S)^2}\sqrt{S^2\sigma^2_H + H^2\sigma^2_S}.$$

\section{Optical Imaging }
\label{optim}

Optical imaging is crucial to provide optical fluxes, preliminary
source classification, and accurate centroiding for spectroscopic
followup.  Optical centroids supersede X-ray centroids for 
extended or multiple objects, for cluster galaxies, or for \Chandra\,
sources with large off-axis angles and hence less accurate X-ray positions.
Multiband colors can provide source classifications and photometric 
redshifts for the majority of fainter ChaMP sources ($\rs>22$)
for which high quality spectra are more difficult to obtain.

\subsection{Exposure Times}

For every ChaMP field, we scale our optical exposure times to the
\Chandra\, X-ray exposure times, to provide a uniform sensitivity to
X-ray/optical flux ratios.  Our {\em color-limited} survey thus
minimizes optical telescope usage, while accessing for every field a
similar fraction of every object type.  To convert X-ray counts to
flux, we first assume 80\% of the exposure time remains after cleaning
high particle background periods.\footnote{Since X-ray photons are
tagged with arrival times, high background periods can be excised.}
The low background and small PSF of Chandra allows high confidence
detection of point sources with as few as 4-5 photons, but to reduce
uncertainties on \fx\, and thereby improve object class
discrimination, we adopt 10 counts as the ChaMP minimum detection
limit.  We convert the detectable number of counts for each \Chandra\,
field to an X-ray flux limit, and then scale the corresponding optical
magnitude limit to include $>90\%$ of ROSAT AGN even at the X-ray flux
limit, based on 1448 ROSAT-detected quasars from \citet{YWBW98}.  The
typical resulting magnitude limit as a function of \Chandra\, exposure
time $T$ can be expressed as $V= 14.17 + 2.5\,{\rm log}T$.  This
criterion (${\rm log}\frac{f_X}{f_V}<0.9$) should include {\em larger}
fractions of other source types with brighter relative optical
emission like stars and most galaxies.  Sources that are relatively
weak optically compared to ROSAT quasars will have a lower
completeness and a brighter effective limiting magnitude in our survey
(e.g., sources suffering from substantial optical extinction).
Details of our optical imaging for these 6 fields are presented in
Table~\ref{timages}.  The resulting desired limits range from $V$ mags
(note that $V\sim (g^{\prime} + r^{\prime})/2$) of 21 to 26 with a
median of 23.5\,mag for all ChaMP fields.  However, to achieve
reasonable limits on the amount of required optical imaging time, we
limit our maximum desired optical magnitude limit to $V\sim25$ in any
field.

\begin{deluxetable}{rccccccccccc}
\tabletypesize{\scriptsize}
\tablecaption{Optical Imaging \label{timages}}
\tablewidth{0pt}
\tablehead{
\colhead{ObsID} & 
\colhead{E(B-V)} & 
\colhead{Telescope} & 
\colhead{UT Date}  &
\colhead{Filter}  &
\colhead{Dithers} &
\colhead{~ ~ Exposure} & 
\colhead{Airmass}  & 
\colhead{FWHM\tablenotemark{a}}  & 
\colhead{m$_{\rm TO}$\tablenotemark{b}} &
\colhead{m$_{5\sigma}$\tablenotemark{c}} \\
\colhead{}  &
\colhead{}  &
\colhead{}  &
\colhead{}  &
\colhead{}  &
\colhead{}  &
\colhead{(total sec)}  &
\colhead{(Mean)}  &
\colhead{(\arcsec)}  &
\colhead{Limit}  &
\colhead{Limit}  &
\colhead{}  \\
}
\startdata
536 & 0.0131 & KPNO 4m & 12 Jun 2000 & \gp & 2 & 1200 & 1.39 & 2.2  & 22.875 & 23.9 \\
    &        &         &             & \rp & 2 & 1200 & 1.43 & 2.0  & 22.875 & 24.2 \\
    &        &         &             & \ip & 2 & 1200 & 1.48 & 1.9  & 22.625 & 23.8 \\
541 & 0.008  & KPNO 4m & 12 Jun 2000 & \gp & 2 & 1000 & 1.33 & 1.9  & 22.875 & 23.9 \\
    &        &         &             & \rp & 1 &  500 & 1.40 & 2.0  & 22.125 & 23.7 \\
    &        &         &             & \ip & 1 &  500 & 1.50 & 1.7  & 22.125 & 23.5 \\
624 & 0.014  & CTIO 4m & 28 Sep 2000 & \gp & 3 &  810 & 1.05 & 1.2  & 24.875 & 25.9 \\
    &        &         &             & \rp & 3 &  630 & 1.01 & 1.1  & 24.125 & 25.3 \\
    &        &         &             & \ip & 3 &  720 & 1.00 & 1.1  & 23.375 & 24.4 \\
861 & 0.0246 & CTIO 4m & 28 Sep 2000 & \gp & 3 & 1260 & 1.23 & 1.4  & 24.625 & 25.8 \\
    &        &         &             & \rp & 3 & 1080 & 1.17 & 1.4  & 24.375 & 25.3 \\
    &        &         &             & \ip & 3 & 1170 & 1.17 & 1.1  & 23.375 & 24.3 \\
914 & 0.0383 & CTIO 4m & 28 Sep 2000 & \gp & 3 &  990 & 1.04 & 1.2  & 25.125 & 26.1 \\
    &        &         &             & \rp & 3 &  810 & 1.06 & 1.2  & 24.375 & 25.4 \\
    &        &         &             & \ip & 3 &  900 & 1.09 & 1.2  & 23.375 & 24.2 \\
928 & 0.051  & CTIO 4m & 28 Sep 2000 & \gp & 3 &  900 & 1.01 & 1.6  & 24.375 & 25.6 \\
    &        &         &             & \rp & 3 &  720 & 1.02 & 1.3  & 24.125 & 25.2 \\
    &        &         &             & \ip & 3 &  810 & 1.05 & 1.0  & 23.375 & 24.3 \\
%
\enddata
\tablenotetext{a}{FWHM of point sources in final stacked images.}
\tablenotetext{b}{Turnover magnitude limit at $\sim90\%$ completeness,
using 0.25\,mag bins before extinction correction, as described in the text.} 
\tablenotetext{c}{Magnitude limit for a $\sim5\,\sigma$ detection.} 
\end{deluxetable}


NOAO 4-meter imaging with the Mosaic CCD cameras \citep{mosaic98}
is key to the ChaMP,
since it provides adequate depth, spatial resolution 
($\sim0.26\arcsec$/pixel), and a large field of view ($36\arcmin\times
36\arcmin$) over the full \Chandra\, FoV.  For shallower northern fields, 
for secondary calibration of deep imaging from non-photometric
4\,meter (4m) nights, and for imaging bright objects within deep fields,
the ChaMP also uses the FLWO~1.2m with the 4shooter CCD on Mt Hopkins
($0.32\arcsec$/pixel, 22\arcmin\, field).  In this paper, we
report only deep 4m fields observed in photometric conditions. 

We dither the 4m/Mosaic pointings to allow for better sky flat
construction, averaging over defects, cosmic ray removal, and gap
coverage.  While the optical data reduction is conceptually
straightforward, it is in practice complicated by the size of the data
set (285 MB per image), and by the need to interactively manage
bad pixel flagging.  After photometric analysis, our final reduced
images are stored in (short) integer format, which reduces their size
by a factor of two, with $<1\%$ effect on measured magnitudes.

\subsection{Filter Choice}

The ChaMP uses Sloan Digital Sky Survey (SDSS) $g^{\prime}$,
$r^{\prime}$, and $i^{\prime}$ filters \citep{FMIT96}, whose steep
transmission cutoffs improve object classification and photometric redshift
determination \citep{GAXPHOTOZ02,RGT01b}. While \fxfo\, is clearly
our primary discriminant of AGN from stars, at least 2 optical filters
are required to also constrain stellar spectral class.  Still, with
just 2 filters, high-z, reddened or obscured AGN may share the stellar 
locus with M dwarfs \citep{LDNH98}, so we chose 3 filters to separate
such quasars from stars because these types of AGN represent a prime
science goal of the ChaMP.  Furthermore, it has already 
been demonstrated that AGN of redshift $3.5<z<5$ can be constrained to
$\Delta z\sim0.2$ with SDSS $g^{\prime}, r^{\prime},$ and $i^{\prime}$
colors \citep{RGT01b}.  Since $g^{\prime}$ and $r^{\prime}$ also bracket the
4000\AA\, break in $z>0.8$ cluster ellipticals, this aids distance
estimation and cluster membership evaluation. The bright $\lambda
5577$ night sky line falls 
between the $g^{\prime}$ and $r^{\prime}$ filters, allowing fainter
limits to be achieved in shorter integration times.  Finally, 
someday the need for NOAO northern imaging followup will be relieved
by the SDSS itself for the {\em shallower}\footnote{ The SDSS achieves
$\snr\sim10$ to $g^{\prime}\lapprox22$mag at best \citep{RGT01b}.}  ChaMP
fields.  The huge database of SDSS magnitudes and colors help
with deeper fields and core science as well by offering (1) bootstrap
calibration of deep ChaMP imaging observed in non-photometric
conditions (2) calibration and testing of accurate photometric
redshifts \citep{RGT01b} and source classification \citep{NHJ99} (3)
detailed comparisons between SDSS optical and ChaMP X-ray sample
selection. 

\subsection{Optical Image Reduction}
\label{reduce}

The reduction of the raw optical CCD images is primarily done using
standard techniques in IRAF.\footnote{IRAF is distributed by the
National Optical Astronomy Observatory, which is operated by the
Association of Universities for Research in 
Astronomy, Inc., under cooperative agreement with the National Science
Foundation. }  The Mosaic cameras utilize 
the multi-extension FITS format, allowing the eight individual CCD
images to be treated as a single image for most reduction purposes.
The IRAF package {\tt mscred} \citep{valdes02} is used to properly handle
the multi-extension FITS images and to simplify the data reduction.
Although an overview is presented here, a detailed reduction
walkthrough is available \footnote{
http://www.noao.edu/noao/noaodeep/ReductionOpt/frames.html} for the
NOAO Deep Wide Field Survey (NDWFS) data \citep{JD99}
using the same instruments.

Standard calibration images are taken at the telescope (bias, dome
flats, etc.) excluding twilight sky flats.  We construct a master
super-sky flat by combining multiple object 
frames, thereby rejecting real objects in the frame and leaving us
with a high $\snr$ image of ``blank" sky.  These improve the flat-field
correction provided by conventional dome flats, mostly by
accounting for differences in each filter between the night-sky
and our dome lamp color.  Before the super-sky flat can be made,
images from the KPNO 4m also require subtraction of a pupil ghost
caused by light back-scattered from the telescope optics, which affects
primarily the inner four CCDs.  Because the pupil ghost changes
with the amount of light entering the telescope, its strength varies
with the filter used, and with ambient light from the moon or 
bright stars near the field of view.  For this reason we
create a template pupil ghost seperately for each filter, taking care
to exclude images with extremely bright saturated objects near the
center of the field of view.  Once a template pupil has been generated,
it must be scaled and subtracted from individual object frames in each
filter, a process that is now largely automated.

\subsection{Astrometry and Photometry}
\label{phot}

We first derive ($\sim0.3\arcsec$ RMS) astrometry from our optical
imaging to enable accurate X-ray source matches, and optimal
positioning of spectroscopic fibers or slits.  The (J2000) position of
the optical sources are referenced to the Guide Star Catalog
II.\footnote{The Guide Star Catalogue-II is a joint project of the
Space Telescope Science Institute and the Osservatorio Astronomico di
Torino.}  Final positions used by the ChaMP for all matched sources
are the average of positions measured from all optical filters in which an
object is detected.

We use SExtractor \citep{BE96} to detect sources, and measure their
positions and magnitudes.  A first run of SExtractor is used to
measure the mode $\Gamma$ of the FWHM of stellar objects in each
image.  For this step, our filters first  remove objects that are
unreliable (any with processing error flag $>3$ -- cosmic rays,
spurious blended, or contaminated sources, or sources near a chip
edge), sources that are too faint (typically $<2\times10^4$ net
counts) or too bright (typically $>2\times10^5$~counts), and extended
sources (SExtractor {\tt  stellarity} $<$ median).   
For program fields, we typically use a minimum detection threshold 
(SExtractor {\tt detect\_thresh}) of $0.8\sigma$ above background per
pixel, and require a minimum grouping ({\tt detect\_minarea})
of $\Gamma^2/4$~pixels at this threshold for a detection.  
Since we are cross-correlating to X-ray sources - rare on the sky
compared to optical sources - we are somewhat more concerned with
completeness than with spurious source rejection.  From detailed
comparisons of dithered and deep stacked images of the same fields, we
find that these parameter settings forge a good compromise between
detecting most real sources, but not too many spurious sources.

A second run of SExtractor measures 3 instrumental magnitudes: one
from a circular aperture with diameter FWHM (for highest $\snr$; see
Howell et al. 1989), another with 4 times the FWHM (which we find
encompasses 95\% flux for point sources), and finally SExtractor's
{\tt mag\_auto}.  The latter is a elliptical aperture magnitude (al\`a
\citealt{kron80}) that is robust to seeing variations, as described in
\citet{nonino99}.  Following the convention of the early data release
of the SDSS quasar catalog \citep{SDP02}, we present the optical
photometry here as $g^{\ast}$, $r^{\ast}$ and $i^{\ast}$ since the
SDSS photometry system is not yet finalized and the NOAO filters are
not a perfect match to the SDSS filters.  Standard stars calibrated
for the SDSS are available from \citet{SJA02} but are generally too
few per field to practically allow for a complete photometric
calibration, especially given the long readout times for the Mosaic.
To facilitate the use of the many standard stars listed by
\citet{LAU92}, we transform them to the SDSS photometric system using
\citet{FMIT96} to derive our nightly photometric solutions.  In the
standard star fields, we find standard stars automatically by
positional matching of accurate coordinates for Landolt standards
\citep{HA02} to SExtractor source positions, using a high
($\sim7\sigma$) SExtractor detection threshold to provide unambiguous
detections.

We merge instrumental photometry results from each filter into a
multicolor file by positional matching using a search radius of
1\arcsec.  If multiple matches are found within this radius, the
closest match is retained.  If no matching detection is found in some
filter, the RMS of the background for that filter at that object's
position is recorded for later conversion to a limiting magnitude.
In these 6 fields, our merged catalog of Mosaic photometry contains
about 343,000 objects with flags $<4$ indicating good photometry in at
least one band.

We then perform photometric calibration of the standard stars for each
night via a 2-step multilinear method \citep{HRH62}, adapted from an
implementation written in IDL by James Higdon. For a given night, for
the instrumental magnitude $v$ in each filter, we solve the linear
equation 
        $$v_0-v = (k_0 + k_1 X) + e (c_0) + z$$
where $v_0$ and $c_0$ are the standard star's cataloged magnitude
and color, respectively, $e$ the color coefficient, $z$ a zeropoint
correction, and $k_0$ and $k_1$ are extinction coefficients.  We
iteratively solve for these 4 coefficients, first freezing $k_0$ and
$k_1$ and solving for $e$ and 
$z$, then freezing $e$ and $z$ and solving for $k_0$ and $k_1$, until
the solution converges (typically 3 or 4 iterations). We then delete
measurements with residuals exceeding $3\sigma$, and repeat the
solution until no such residuals remain.

Similarly, for each color $c$ we iteratively solve the linear equation
       $$c_0 = e (c + k_0 + k_1 X) + z$$
for the four coefficients $e, z, k_0$ and $k_1$, deleting measurements
with residuals exceeding $3\sigma$ as before. We tested the final
results from our procedure against that of \citet{HFR81} as
implemented in the IRAF {\tt photcal} procedure, and find results identical
to within the errors.  Our calibrations produce $e, z, k0, k1$
coefficients for each of \gs, \rs, \is, \gmrs, and \rmis.  Since these
calibrations for these NOAO filters have not been published
previously, we list them in Table~\ref{tphotcal} as a reference.
The final transformations are applied to the standard stars to derive
RMS residuals for each night for \gs, \rs, \is, \gmrs, and \rmis.
These are listed in Table~\ref{tphotcal} for each night.  The nightly
transformations applied to our program objects are
	$$c^{\ast} = e (c + k_0 + k_1 X) + z$$
	$$v^{\ast} = v + (k_0 + k_1 X) + e (c\prime) + z$$
\noindent where $c^{\ast}$ and $v^{\ast}$ are the final calibrated
color and magnitude, respectively.

\begin{deluxetable}{crrrrcc}
\tabletypesize{\scriptsize}
\tablecaption{Nightly Photometric Calibration Coefficients \label{tphotcal}}
\tablewidth{0pt}
\tablehead{
\colhead{Fit Variable}  &
\colhead{$e$}  &
\colhead{$z$}  &
\colhead{$k_0$}  &
\colhead{$k_1$}  &
\colhead{} N Stars &
\colhead{Fit RMS}  \\
\colhead{}  &
\colhead{(mag)}  &
\colhead{}  &
\colhead{}  &
\colhead{}  &
\colhead{}  &
\colhead{(mag)}  \\
}
\startdata
\multicolumn{7}{c}{KPNO 4m 12 June 2000} \\
\gs\tablenotemark{a} &  0.060 & 25.314 &  0.284 & -0.217 &  52 &  0.057 \\ 
\rs\tablenotemark{b} & -0.055 & 25.424 &  0.207 & -0.155 &  49 &  0.068 \\ 
\is\tablenotemark{b} & -0.185 & 25.194 &  0.191 & -0.117 &  54 &  0.250 \\ 
\gs--\rs & 1.021 & -0.087 &  0.044 & -0.036 &  54 &  0.111 \\ 
\rs--\is & 0.978 &  0.362 &  0.064 & -0.050 &  47 &  0.031 \\ 
  \multicolumn{7}{c}{} \\
  \multicolumn{7}{c}{CTIO 4m 28 September 2000} \\
\gs\tablenotemark{a} &  0.038 & 25.695 &  0.257 & -0.210 &  43 &  0.027 \\ 
\rs\tablenotemark{b} & -0.020 & 25.794 &  0.145 & -0.118 &  43 &  0.033 \\ 
\is\tablenotemark{b} &  0.083 & 25.358 &  0.089 & -0.072 &  41 &  0.041 \\ 
\gs--\rs & 1.011 & -0.087 &  0.110 & -0.090 &  41 &  0.037 \\ 
\rs--\is & 0.903 &  0.410 &  0.078 & -0.064 &  43 &  0.027 \\ 
\enddata
\tablenotetext{a}{Uses (\gs--\rs) colors in fit.}
\tablenotetext{b}{Uses (\rs--\is) colors in fit.}
\end{deluxetable}

For objects not detected in any optical filter, we assign a magnitude
limit $m_{5\sigma}$ corresponding to a flux of $5\sigma$ detection
where $\sigma$ is the background RMS at that position.  Starting from
the standard CCD equation, we find
  $$m_{D\sigma} = -2.5\,{\rm log}\Bigl(\frac{D~\Gamma~\sigma}{T}\Bigr) + z $$
Here, $D$ is the desired \snr.  The RMS noise at the position of an
object on the CCD is denoted as $\sigma$;  $\Gamma$ is again the FWHM
in pixels, $T$ is the total exposure time, and $z$ indicates that the
photometric calibration is applied, for which we assume the mean color
for objects in the field.  This limiting point source magnitude is
listed for each field in Table~\ref{timages} for the median background
RMS in the field. Objects at these magnitudes are not detected with high
completeness.  From our comparisons of individual dithered and deep
stacked images, we find that the magnitude where the number counts
peak in a differential (0.25~mag bin) number counts histogram 
corresponds approximately to 90\% completeness in the magnitude
range 20 -- 25.  This turnover magnitude $m_{\rm TO}$ is
typically about 1~mag brighter than the $5\sigma$ limiting magnitude. 
Both magnitude limits are listed in Table~\ref{timages}.

We perform several sanity checks on the final photometry. First, we
plot color-color diagrams against a crude mean stellar locus from the
SDSS Early Data Release (EDR).  The locus we derive by eye (from
objects with $<0.1$mag error among 30,000 high latitude objects) in
the SDSS EDR corresponds to one line primarily following the SDSS
halo/thick disk locus \citep{CB01,YN01} 
        $$\rmis \sim 0.4029\gmrs - 0.007$$ 
and another following the thin disk locus
        $$\rmis \sim 14.286\gmrs - 18.5$$
An example of these color loci is shown in Figure~\ref{gmr_rmi},
to be discussed in more detail in \S~\ref{colors}. 
By studying a variety of SDSS fields at 7 different galactic
latitudes ($b=$ 30, 35, 45, 60, --45, --55, --65) we find 
variations of about 0.2mag in \rmi\ occur in the thin disk locus. 

We also compare our magnitudes in \gs\, and \rs\, to magnitudes
derived from the GSC2.2.  We have derived the transformation between
SDSS EDR and GSC2.2 magnitudes by direct comparison of the two surveys,
using only stars between 17th and 19th magnitude.  After $3\sigma$
clipping, the typical residuals are 0.2~mag, with zeropoints differing
by $\sim1\%$. 
        $$\gs = 1.032\,J + 0.1255 (J-F) - 0.851 $$
        $$\rs = 0.982\,F + 0.1518 (J-F) + 0.451 $$
        $$\is = 0.994\,F - 0.1235 (J-F) + 0.281 $$

We note that most optical point sources brighter than 18th mag suffer
from saturation effects in our NOAO 4m imaging.  While these objects
constitute a small fraction ($\sim 6\%$) of our optically identified
X-ray sources, they are a larger fraction ($\sim 10\%$) of our
spectroscopically-classified objects.  We expect 
a full release of the SDSS will substantially alleviate this problem, 
as will our ongoing imaging study of these fields with the SAO
FLWO1.2m and 4-shooter CCD on Mt Hopkins.

Since most of our X-ray sources are extragalactic, before calculations
of optical luminosities, we correct 
the optical magnitude measurements for Galactic extinction using the
maps of \citet{SFD98}.  We assume an $R_V=3.1$ absorbing medium, and
absorption in the SDSS  bands as given by \citet{SDP02}. 

\section{X-ray to Optical Source Matching}
\label{matching}
For the ChaMP, \citet{kim03a} have carried out extensive simulations of
point sources generated using the SAOSAC raytrace program
({\url http://hea-www.harvard.edu/MST/}) and detected using
CIAO/{\tt{wavdetect}} (Freeman et al.\ 2002).  For weak sources of
$\sim 20$ counts between 8-10\arcmin\, off-axis from the aim point, the
reported X-ray centroid position is correct within $1.8\arcsec$,
corresponding to a $1\sigma$ confidence contour.

The source naming convention of the ChaMP as
officially registered with the IAU is given with a prefix CXOMP
(Chandra X-ray Observatory Multiwavelength 
Project) and affixed with the truncated J2000 position of the X-ray
source (CXOMP\,Jhhmmss.s$\pm$ddmmss) after a mean field offset 
correction is applied, derived from the positional matching of optical
and X-ray sources in each field.  

\subsection{Matching criteria}
\label{match}

\subsubsection{Automated Matching}
\label{automatch}

\begin{figure*}[ht!]
\epsscale{0.9}
\plotone{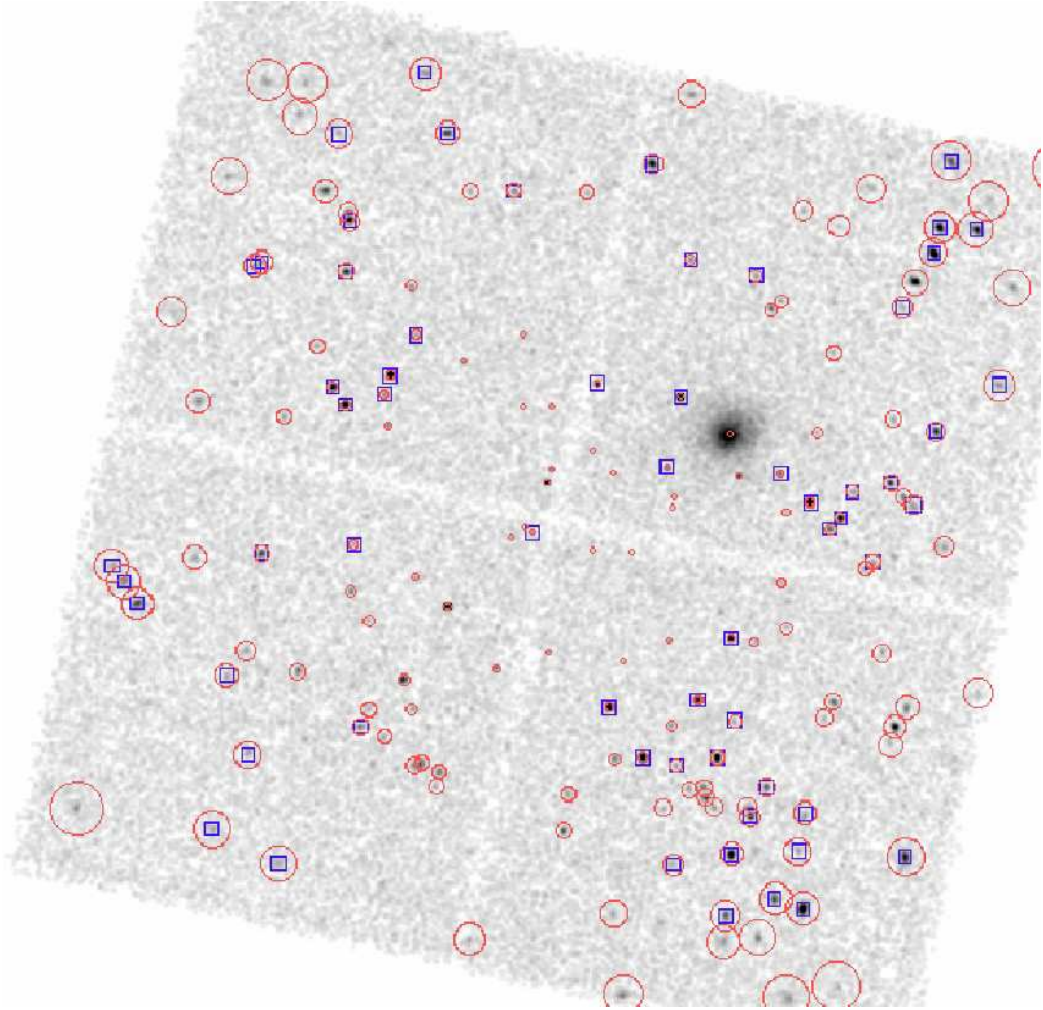}
\caption{\small \Chandra\, image of the field of ObsID 536 (target
MS1137.5+6625).  The  $16\arcmin \times 16\arcmin$ \hbox{ACIS-I} (0.3-5~keV)
image (N up, E to the left as in all images) was smoothed with a 3
pixel (1.5$\arcsec$) gaussian for clarity. 156 X-ray sources detected 
to $2\times10^{-16}~{\rm erg~cm}^{-2}~{\rm s}^{-1}$ using the ChaMP's
X-ray pipeline are marked with circles sized to the \Chandra\, PSF
(90\% EE). To the 125 $\snr>2$ X-ray sources detected in these 4 chips, 
we have matched 80 optical sources detected in our combined 4m/Mosaic
images to $\rs\sim 25$, indicated by 15\arcsec\ boxes. The target
cluster is easily visible as the extended source NW of center, and is
detected as a single source by XPIPE, but like all target objects is
excluded from ChaMP analysis.  
}  
\label{ms1137ox}
\end{figure*}

We obtain candidate optical identifications of X-ray sources in each
field using a 2-stage automated position matching process between
optical and X-ray source positions. In the first stage, the subset of
X-ray sources within 5\arcmin\ of the Chandra aimpoint and with X-ray
$B$ (0.3--8~keV) detections 
$>20$~counts are matched to optical sources from our CCD imaging,
using position matching 
with a radial offset matching tolerance, in arcsec, $R_{\rm match} =
\theta/\log_{10}(B)$, where $\theta$ is the off-axis angle in arcminutes,
and $B$ is the X-ray source counts, but $R_{\rm match}$ is 
limited to a minimum of 3\arcsec\ for all sources.  This matching tolerance
encompasses the {\tt wavdetect} centroid positional uncertainty
variation with off-axis angle and source counts, using the refined
source position estimator described in \citet{kim03a}. The optical
position errors are neglegible in comparision.  In those cases
where there are multiple optical matches, the closest match is
used. The median X-ray to optical position translational offset of
these matched sources is then applied to all the X-ray sources for the
field, to bring the optical and X-ray source lists onto the same
coordinate frame. The X-ray position corrections are typically less
than 1\arcsec, which is consistent with the specified absolute
position accuracy\footnote{ See {\url
http://asc.harvard.edu/cal/ASPECT/} for a detailed discussion of
\Chandra\, aspect solutions.}  for the Chandra observatory and {\tt
wavdetect} source positions.  Figure~\ref{ms1137ox} shows optical
matches overlaid directly on the ACIS-I image of ObsID 536.

In the second stage of position matching, optical matching is
performed for all X-ray sources with X-ray $B$ detections $>5$~counts,
using the same matching tolerance $R_{\rm match}$ as above. Multiple
optical matches for an X-ray source are allowed in this stage, and are
passed to the visual inspection process described next.  Extended
X-ray source centroids may have larger position errors than are
predicted by our point source formula.  Furthermore, in some cases
no corresponding optical cluster galaxy may exist close to the cluster
X-ray centroid.  If no optical counterpart is found in the automated
matching routine for these or other cases, visual inspection provides 
a second chance to identify counterparts.

\subsubsection{Visual Inspection}

\begin{figure}[ht]
\centering
\epsscale{0.85}
\plotone{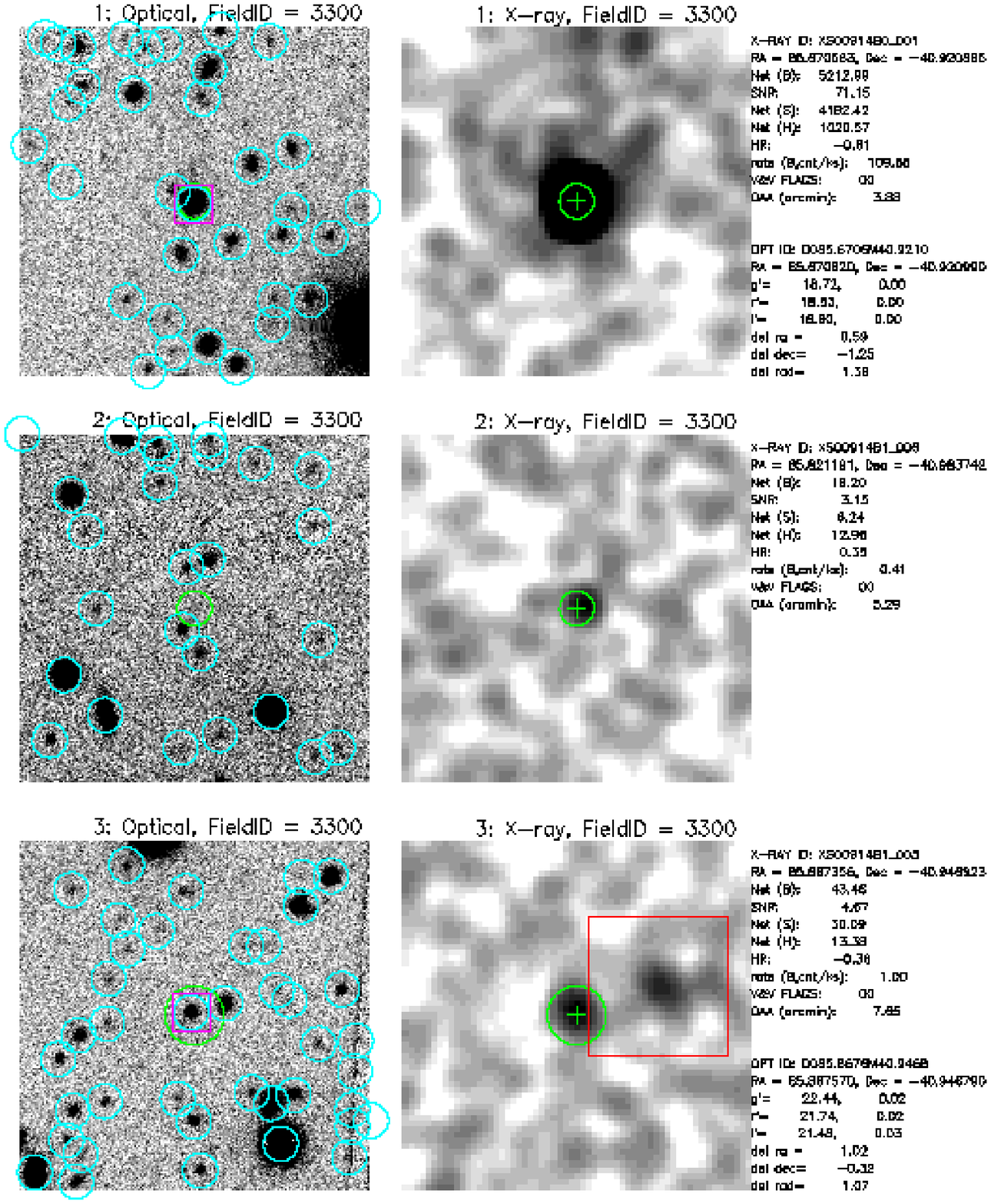}
\caption{\small 
An example from ObsID 914 (target CL\,J0542.8-4100) of the pages used
for initial visual inspection of X-ray source matches.   
RIGHT PANELS: A 1\arcmin\, field centered on a detected X-ray source
is shown, with N up and E to the left.  
The 0.3-8\,~keV X-ray image is smoothed with a 2 pixel (1$\arcsec$)
gaussian for clarity, which facilitates source inspection and
examination of background and confusion effects. 
Green circles around the X-ray source have size equal to the
PSF (50\% EE) at the source position.  Crosses show the
{\tt wavdetect} X-ray centroid.  If other X-ray sources are detected
within the field, they are marked by a 20\arcsec\, box.  LEFT PANELS:
The corresponding optical \rs\, images shows all optical sources detected
by SExtractor, marked with $3\arcsec$ circles.  The \Chandra\ 50\% EE
circle is reproduced at the X-ray source position.  When our automated
matching procedure finds an optical counterpart, the optical source is
marked with a 3\arcsec\, box.  X-ray and optical photometric data are
displayed to the far right.  
\label{VIexamp}}
\end{figure}

After our automated O/X matching between detected sources from XPIPE
and SExtractor source positions has been performed, we examine the
images by eye.  An example of the plots used for visual inspection is
shown in Fig~\ref{VIexamp}.  In many cases, no obvious optical
counterpart can be detected, while in other cases there are several
(source confusion).  
Even with typical astrometric accuracy of an arcsec, at these faint
optical magnitudes and X-ray fluxes we frequently find several optical
counterpart candidates.  Beyond $\rs\sim20$, galaxies dominate the
optical number counts, and at by 24th magnitude, cumulative source
counts are near $10^5\,$deg$^{-2}$ \citep{KW01}.  At these faint
magnitudes, the probability of finding an unrelated optical source by
chance within any 3\arcsec\, circle becomes $\sim20\%$.  The small PSF
and accurate centroiding of \Chandra\, become crucial at these faint
magnitudes, as does careful source inspection.

\subsection{Fraction with Optical Counterparts}

During visual inspection of the optical field of each X-ray source,
we assign a match confidence to each plausible counterpart.
To consider a sample as uncontaminated as possible by incorrect matches,
we have removed from further optical analysis below 45 X-ray sources
with poor or ambiguous counterpart assignments.

The resulting fraction of X-ray sources with confidently matched optical
counterparts is shown in Table~\ref{tmatch}.  We have not yet achieved
the desired optical magnitude limits (set to match optical
counterparts for 90\% of ROSAT AGN) in all these deep \Chandra\, fields,
but the imaging has served to facilitate spectroscopy to $\rs\sim21$,
and to provide fluxes and colors to still deeper limits.  Lower
achieved match fractions also reflect the inclusion by \Chandra\ of
new \fxfo\ populations atypical for ROSAT such as heavily absorbed AGN.


\begin{deluxetable}{crrccc}
\tabletypesize{\scriptsize}
\tablecaption{Source Matching \label{tmatch}}
\tablewidth{0pt}
\tablehead{
\colhead{ObsID}  &
\colhead{$N_X$\tablenotemark{a}} & 
\colhead{$N_{XO}$\tablenotemark{b}} & 
\colhead{$f_{XO}$\tablenotemark{b}} & 
\colhead{$N_{XOS}$\tablenotemark{c}} &
\colhead{$f_{XOS}$\tablenotemark{d}} \\
\colhead{}  &
\colhead{}  &
\colhead{}  &
\colhead{\%}  &
\colhead{}  &
\colhead{\%}  \\
}
\startdata
 536  &  140    &   77  & 55   & 33 	 &  70 	\\
 541  &  65     &   45  & 69   & 24 	 &  72 	\\			   
 624  &  58     &   45  & 78   & 19 	 &  74 	\\			   
 861  &  80     &   57  & 71   & 14 	 &  59 	\\			   
 914  &  88     &   65  & 74   & 20 	 &  50 	\\			   
 928  &  55     &   42  & 76   & 15 	 &  50 	\\
Total &  486    &   331 &  71  & 125 	 &  63 	\\
\enddata
\tablenotetext{a}{Number of X-ray sources detected in each \Chandra\
image and also having $\snr>2$ in counts as described in
\S~\ref{xanalysis}.} 
\tablenotetext{b}{Number $N$ or percent $f$ of X-ray
 sources with high confidence optical counterparts as described in
 \S~\ref{matching}.}  
\tablenotetext{c}{Number $N$ of optical counterparts with
spectroscopic classification.}  
\tablenotetext{d}{Percent $f$ of $\rs\leq 22$ optical counterparts with
spectroscopic classification.}  
\end{deluxetable}


Excluding the $\sim10\%$ of sources with more than 200 X-ray counts,
the mean/median X-ray flux for matched sources (3.5/2.4 in units of
10$^{-15}$ \fcgs) exceeds the flux for unmatched sources (2.6/1.6)
as expected, but not strongly.  The mean off-axis angle of
unmatched sources ($\theta=6.7\pm3.9$\arcmin) is indistinguishable
from that of matched sources ($6.2\pm3.2$\arcmin; both quoted errors
are 1$\sigma$ RMS). 

For a typical ChaMP source in these fields, the median X-ray (0.5 -
2~keV) flux is $2.4\times 10^{-15}$ \fcgs, where according to the
\lognlogs, the cumulative (0.5-2~keV) X-ray source density attains
about 30deg$^{-2}$ (e.g., \citet{RPTP02}).  The probability
of two X-ray sources falling within 10\arcsec\, of each other
by chance is $<0.001$.  On the other hand, the number of true multiple
sources is of great interest; they may either be lensed images
or perhaps binary AGN, possibly interaction-triggered (e.g., Green et
al. 2002).  Since these latter types are expected to be rare, a
large-area survey like the ChaMP is needed to find them.  
New close pairs of X-ray sources with optical counterparts are
described in Appendix~A.

\section{Optical Spectroscopy}
\label{ospec}

The nature of many individual sources, especially those with unusual 
properties, cannot be reliably determined until spectroscopy is
obtained.  Our spectroscopy targets all objects in a representative
subset of 40 ChaMP Cycle1+2 fields, yielding $\sim1000$ objects for
which we will obtain the most detailed information. With our
spectroscopic subsample, we will classify sources by type,
determine redshifts and hence luminosities and look-back times for
evolutionary studies, and test photometric determinations of redshift 
and classification.

The mean magnitude of optical counterparts is $r^{\prime}\sim22$ for a
23 ksec high galactic latitude \Chandra\, exposure, and so fields of this
\Chandra\, exposure or larger typically have at least 15 sources
suitable for fiber spectroscopy. Multi-fiber spectroscopy, even on 4 to 6m
telescopes, allows us to consistently obtain source classification and
redshifts to $r^{\prime}\sim20$, and up to 1.5mag fainter for objects
with strong emission lines.  Our overall spectroscopic observing
strategy has been to use single-slit (FLWO1.5m w/ FAST) spectroscopy
for isolated $r^{\prime}<17$ sources in shallow \Chandra\, fields, and
in our deeper fields, which might otherwise introduce scattered light
problems for multi-object spectroscopy.  Multi-fiber wide-field
spectroscopy with the HYDRA spectrographs \citep{BSC94} on the WIYN
3.5m and CTIO~4m telescopes enables us to acquire spectra for sources
with $17<r^{\prime}<$21.  A subset of 
faint ChaMP sources ($21<r^{\prime}<$23) in fields designated for
spectroscopic follow-up are being observed with 6m class telescopes
such as Magellan and the MMT. Table~\ref{tspectels} lists the
technical specifications and configurations for each spectroscopic
facility used for results 
presented herein.

\begin{deluxetable}{lllllcc}
\tabletypesize{\scriptsize}
\tablecaption{Spectroscopic Facilities \label{tspectels}}
\tablewidth{0pt}
\tablehead{
\colhead{Telescope} &\colhead{Instrument} &\colhead{Mode}
&\colhead{Grating/Grism}
&\colhead{$\lambda$ range} &\colhead{R} &\colhead{Spectral} \\
\colhead{} &\colhead{} &\colhead{} &\colhead{} &\colhead{(\AA)}
&\colhead{($\lambda/\Delta\lambda$)} &\colhead{Resolution (\AA)}\\
}
\startdata
WIYN\tablenotemark{1}&HYDRA/RBS&multi-fiber&316@7.0&4500-9000&950&7.8\\
CTIO Blanco 4m&HYDRA&multi-fiber&KPGL3&4600-7400&1300&4.6\\
MMT&Blue Channnel&single slit&300~l/mm&3500-8300&800&8.8\\
Magellan&Boller \& Chivens&single slit&1200~l/mm&4200-5800&2083&2.4\\
Magellan&LDSS-2&single slit&med/red&4000-8500\tablenotemark{2}&520&13.5\\
Keck\tablenotemark{3}&LRIS\tablenotemark{4}&multi slit&300~l/mm&4000-9000\tablenotemark{2}\\
FLWO 1.5m&FAST&single slit&300~l/mm&3600-7500&850&5.9\\
\enddata
\tablenotetext{1}{The WIYN Observatory is a joint facility of the
University of Wisconsin Madison, Indiana University, Yale University,
and the National Optical Astronomy Observatory.}
\tablenotetext{2}{Spectral coverage can vary as a function of slit
position in the mask.}
\tablenotetext{3}{The W. M. Keck Observatory is operated as a scientific
partnership among the California Institute of Technology, the
University of California, and the National Aeronautics and Space
Administration, and was made possible by the generous financial
support of the W. M. Keck Foundation.}
\tablenotetext{4}{LRIS; Oke et al. 1995}
\end{deluxetable}

Of the 6 fields studied here, each has been observed with a
multi-fiber spectrograph on either the WIYN or CTIO 4m
through 2\arcsec\ fibers (Table~\ref{tspecruns}).  An average of 15
spectra per field are acquired with this  configuration.  
We obtained multi-slit observations for 17 sources with Keck-I/LRIS
or Magellan/LDSS2.  Single slit observations of 19 sources were
acquired with the FLWO 1.5m when optically bright ($\rs<17$) and MMT
or Magellan when faint.  We perform standard optical spectroscopic
reductions within the  IRAF environment.  For multi-fiber reductions,
we use the IRAF task {\tt dohydra}. To optimize sky subtraction, we
cross-correlated all dispersion-corrected (wavelength calibrated) sky spectra from random
sky locations in the field against the object+sky spectrum from each
object fiber using the task {\tt xcsao} within the external IRAF package
{\tt rvsao}.  We then median-combine the 9 sky fibers whose spectra
correlate most closely with the total object+sky spectrum.  We found
this method to be superior to removal of a simple field-averaged sky
spectrum for faint objects.  Slit spectra were reduced using the IRAF
task {\tt apall} with a background signal measured locally for each object
within the same slit.  The multi-fiber spectra are only nominally
flux-calibrated, since we use just 1 or 2 standards per night down a
single central fiber. One or two standards are used for the
single slit spectra. 

\begin{deluxetable}{llllc}
\tabletypesize{\scriptsize}
\tablecaption{Optical Spectroscopy \label{tspecruns}}
\tablewidth{0pt}
\tablehead{
\colhead{ObsID} &
\colhead{Telescope} &
\colhead{Instrument} &
\colhead{UT Date} &
\colhead{\# of spectra}\\
}
\startdata
536
 & Keck & LRIS  & 15 May 2000 & 5 \\
 & WIYN & HYDRA & 07 Apr 2001 & 7 \\
 & WIYN & HYDRA & 30 Jan 2003 & 13 \\
 & WIYN & HYDRA & 31 Jan 2003 & 4 \\
541 
 & WIYN & HYDRA & 02 Apr 2001 & 12 \\
 & WIYN & HYDRA & 07 Apr 2001 & 9 \\
 & MMT & Blue Channel & 26 May 2001 & 1 \\
 & MMT & Blue channel & 12 July 2002 & 2 \\
624 
 & CTIO 4m & HYDRA & 16 Oct 2001 & 11 \\
 & Magellan & LDSS-2 & 03 Dec 2002 & 5 \\
861 
 & WIYN & HYDRA & 19 Sep 2001 & 10 \\
 & WIYN & HYDRA & 20 Sep 2001 & 2 \\
 & MMT & Blue channel & 11 July 2002 & 1 \\
 & FLWO 1.5m & FAST & 15 Oct 2001 & 2 \\
914 
 & CTIO 4m & HYDRA & 17 Oct 2001 & 15 \\
 & Magellan & LDSS-2 & 01 Dec 2002 & 5 \\
928 
 & CTIO 4m & HYDRA & 15 Oct 2001 & 7 \\
 & Magellan & LDSS-2 & 21 May 2001 & 1 \\
 & Magellan & BSC & 15 July 2002 & 5 \\
 & Magellan & BSC & 15 July 2002 & 1 \\
\enddata
\end{deluxetable}

\subsection{Object types \& Redshifts}

We classify objects spectroscopically simply as broad
line AGN (BLAGN; FWHM$>1000\kms$) narrow emission line galaxy (NELG;
FWHM$<1000\kms$; some of which may be AGN), absorption line galaxy
(ALG), or Star.  Objects with high \snr\ ($>20$) and no identifying features
are denoted as BL~Lac candidates, as are those whose identified CaII
4000\AA\, region shows a break contrast of $<0.25$ (see \S~\ref{bllacs}).
Our classification confidence level is denoted as well (0 for not 
classifiable, 1 insecure, 2 secure). These are all {\em strictly optical} 
classifications.  

Redshifts are measured for extragalactic objects using the radial
velocity package {\tt rvsao} under the IRAF environment.  This method
depends on the optical classification.  Quasar spectra are quite
similar over a broad range of redshifts and luminosities
\citep{FKGPAT01}, so that - absent the effects of broad absorption
lines (BALs) or a strong Ly$\alpha$ forest absorption - a single high
$\snr$ composite serves well as a redshift template.  For BLAGN, we
use the quasar template created from over 2200 quasar spectra collected
by the SDSS \citep{BV01}.
We run {\tt xcsao} over the full redshift range in 0.1 redshift
intervals, choosing the strongest correlation as the true 
redshift.  Since the correlation strength depends strongly on
$\snr$ and artifacts such as poorly subtracted sky lines,
we inspect and visually verify all redshifts and classifications 
by comparison with overplotted templates. For ALG we use {\tt xcsao}
with a synthetic absorption line template. Whenever narrow emission
lines are present, we elect to measure the redshift from these
features to reduce the associated error.  For NELG, the task {\tt
emsao} is used to identify emission lines using a line list.  For
stars, we assume an effective radial velocity of zero and allow {\tt
xcsao} to correlate against stellar template spectra from
\citet{JHC84}.  Our comparisons between by-eye classifications and
these automated results suggest an uncertainty of about one full
spectral type.   


For X-ray sources with $\rs<22$ optical counterparts, 
we have already achieved 54\% completeness.  All sources with $r>21$
that we have identified spectroscopically to date are BLAGN or NELG,
because their strong emission lines facilitate classification and
redshift determination.  We are currently extending to 90\%
completeness at $\rs=21$ in at least 20 ChaMP fields specifically to
map the space density of X-S AGN.  

\section{Results}
\label{results}

\subsection{Fluxes, Redshifts, and Object Types}
\label{rxz}

\begin{figure}[ht]
\epsscale{1}
\plotone{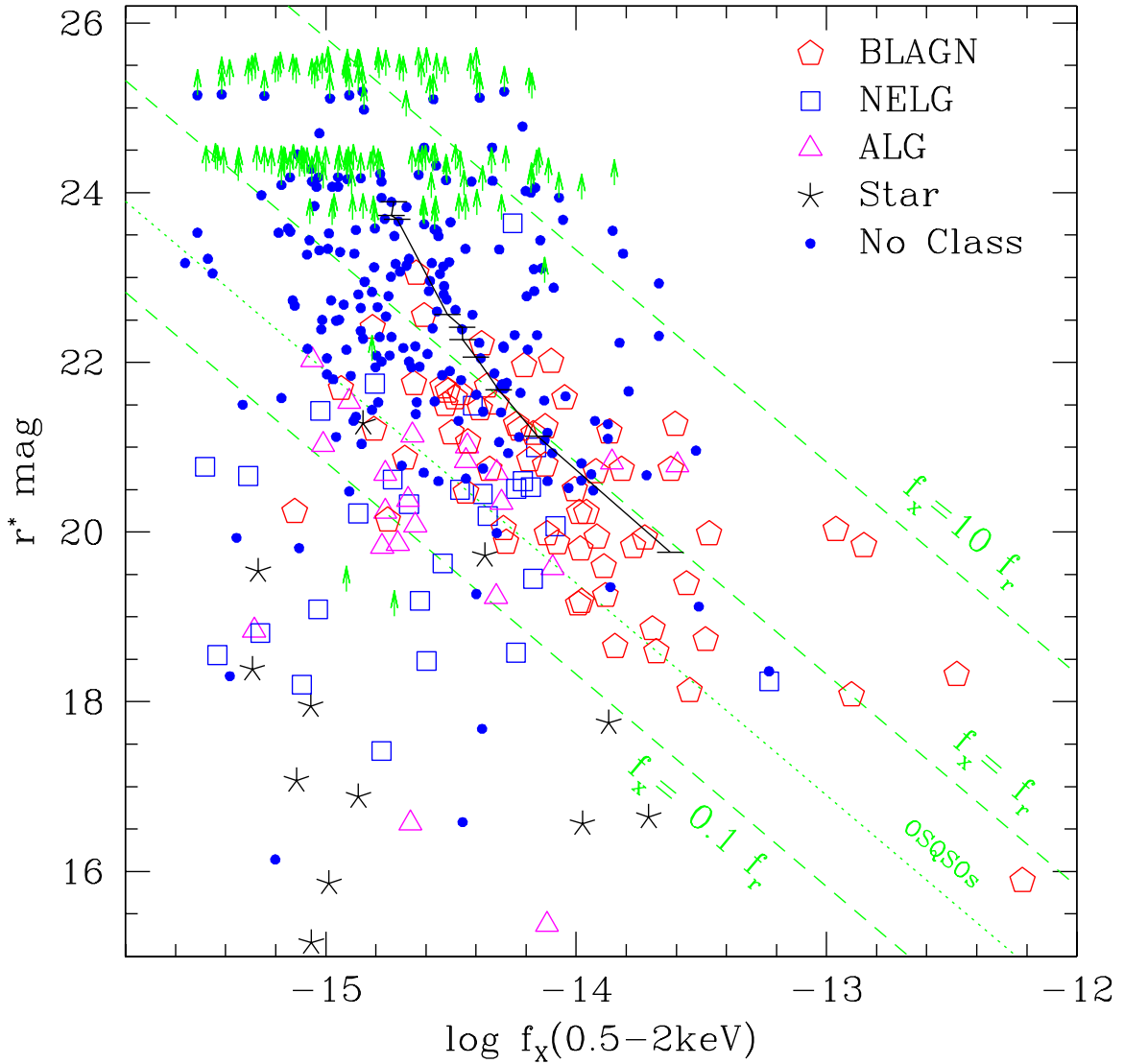}
\vspace*{0.5in}
\caption{\small 
SDSS \rs\, band magnitudes vs. \Chandra\, \logfx\, (calculated 
for the 0.5--2\,~keV band from the broadband 0.3 - 8\,keV counts)
for all X-ray sources in the 6 fields studied here.
Overplotted are 3 dashed lines of constant \fxfr\, at 0.1, 1, and 10 
calculated as described in \S~\ref{rxz}.  A fourth (dotted) line
marked OSQSOs shows the typical mean values for optically-selected
quasars ($\logxr=-0.57$).  Large symbols for different optical source
types are shown as determined from our spectroscopy: broad line AGN
(BLAGN), narrow 
emission line  galaxies (NELG), absorption line galaxies (ALG) and stars.
Any X-ray source not detected in our \rs-band images is marked with an
arrow at the position of its \rs\ magnitude limit.  Small filled
circles indicate sources with an optical counterpart in at least
one filter.  The curve traces the mean expected quasar track for $z=0.5$
redshift bins for the simulation described in \S~\ref{results},
imposed flux limits of $\logfx= -15$ and $\rs=25$.  The
bright end point has $\overline{z}=0.37$, while the faint end
has $\overline{z}=4.6.$  Many of the optically faint objects are likely
to be high$-z$ AGN.  Magnitudes of point sources brighter than 18 
suffer from saturation effects.
\label{rfx}}
\end{figure}

Figure~\ref{rfx} shows a plot of the \rs\, band magnitude
against \logfx\, calculated in the 0.5--2\,~keV band.  Overplotted are 
3 lines of constant \fxfr\, at 10, 1, and 0.1.  We calculate this X-ray
to optical flux ratio similarly to \citet{HA01} or \citet{MJC02},
using our SDSS \rs\, magnitudes: $${\rm log}\frac{f_X}{f_r} = {\rm
log}f_X + 5.67 + 0.4\,\rs $$ 
\noindent and use this as our measure of \fxfo\ throughout.
For comparison, we also plot a line showing the typical 
relation expected for O-S quasars \citep{GPJETAL95},  assuming a
constant $\aox=1.5$, which corresponds to $\logxr = -0.57$.  ChaMP
quasars clearly differ from O-S quasars in the expected sense - X-S
objects have typically stronger X-ray emission relative to optical
($\overline{\logxr}=-0.15$).  Alternatively, even optically 'dull'
quasars are still detected by the ChaMP, since luminous X-ray emission
is the {\em primary} signature of AGN activity. 
BLAGN are X-ray bright, being largely limited to $\fxfr>0.1$.
Galaxies tend to appear towards fainter X-ray fluxes but span a wide
range of optical brightness.  Stars are essentially limited to
$\fxfr<0.1$ and $\rs<20$, which corresponds to a typical M dwarf
at distances of $400 - 1000$\,pc.  

Apparent trends between fluxes or colors in any survey can be due to a
combination of survey sensitivity limits. To investigate selection
effects in a first approximation, we have generated a very deep
simulated SDSS quasar catalog as described in \citep{FXSIMS99}, based
on their observed optical luminosity function and its evolution, and
including the effects of emission lines and intervening absorption.
We have extended the simulation to include AGN with $M_B<-21$ out to a
redshift of 6, corresponding to $\zp$ magnitudes to 28.  We then
assume that the mean observed ratio \logxr\ has a constant value of
--0.14 with a dispersion $\sigma_{xr} = 0.4$.  (These are the actual
values we find for all 41 quasars in our sample in the flux regime
$\fx >5\times 10^{-15}$\fcgs\ where our optical matching rate is high.)
From the $\rp$ mag for each simulated quasar, we then generate X-ray
fluxes, luminosities, etc.  Finally, we can cut the sample at any
desired mag or X-ray flux.  Primarily, we use a cut 
of $\rp<25$ and $\logfx=-15$, similar to our actual final sample
limits.  For most figures which require redshifts, we use a cut of
$\rp<22$ more appropriate to the spectroscopic sample.  The simulated
sample is thus based on a quasar OLF with an assumed constant 
observed \fxfr\ for an X-S sample, rather than on the
actual quasar XLF.  Indeed, a prime goal of the ChaMP is to
measure the XLF of X-S quasars with improved
statistics at fainter fluxes where it is not yet well-characterized,
and to compare the XLF and OLF particularly at redshifts beyond two.

In Figure~\ref{rfx}, we plot the mean $\rp$ vs. \logfx\ values of
quasars in this simulation subsample, derived in redshift bins of
$\Delta z=0.5$.  The track matches well with the overall trend of BLAGN.
There is large dispersion in both optical and X-ray fluxes about this
mean track.  Substantial incompleteness in optical IDs sets in fainter
than $\rs\sim23.5$, where a decreasing fraction of X-ray sources have
optical counterparts.  However, as discussed below in \S~\ref{lums},
the characteristics of X-S quasars clearly differ from the
O-S quasars on which the OLF is based.

\begin{figure}[ht]
\centering
\plotone{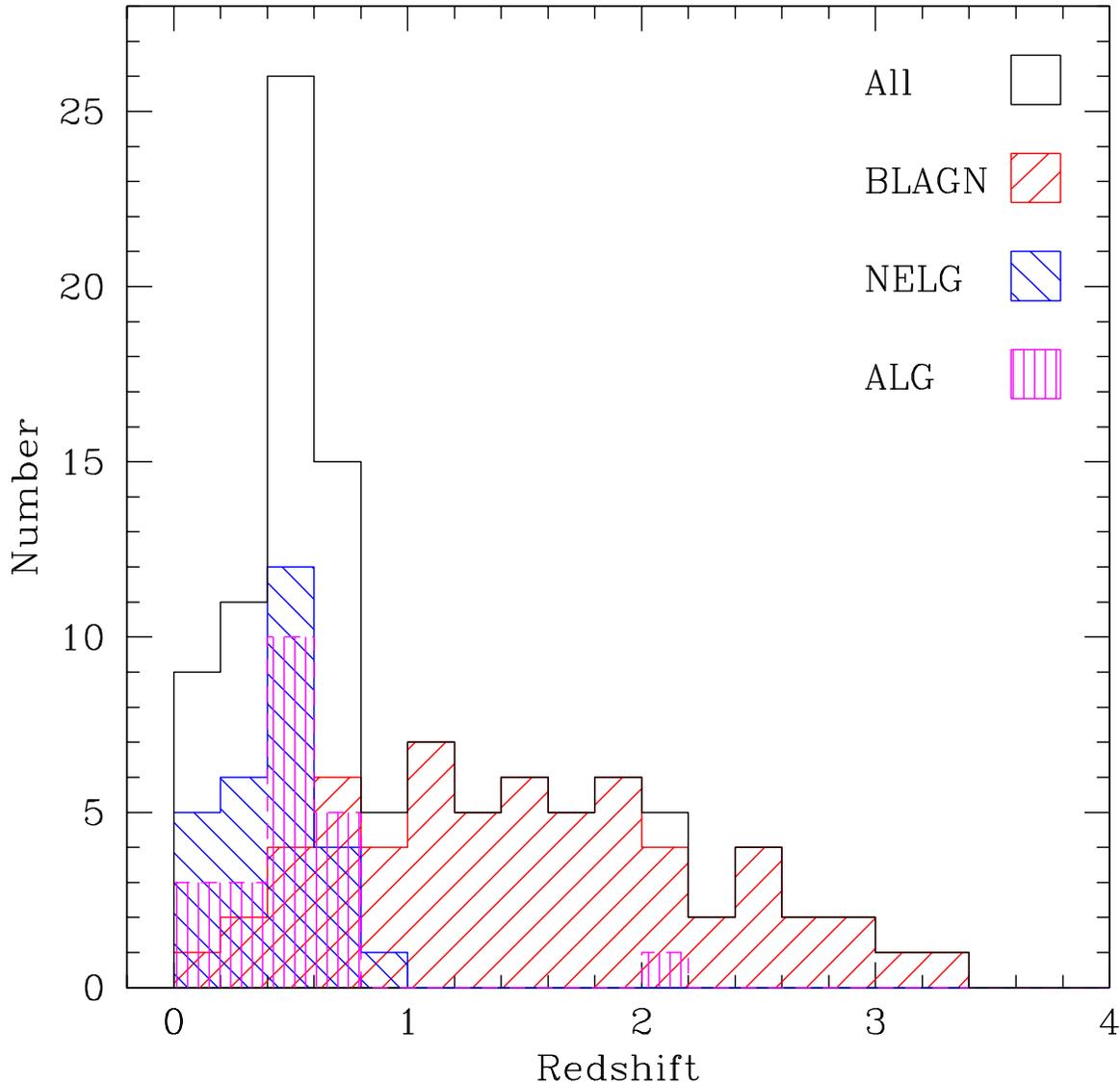}
\caption{\small 
Redshift histogram of classified extragalactic sources in these 6
fields. Most objects showing no broad line component are below
$z=0.8$, likely caused by our typical spectroscopic magnitude limit
near 21.  
\label{zhisto}}
\end{figure}

\begin{figure}[ht]
\plotone{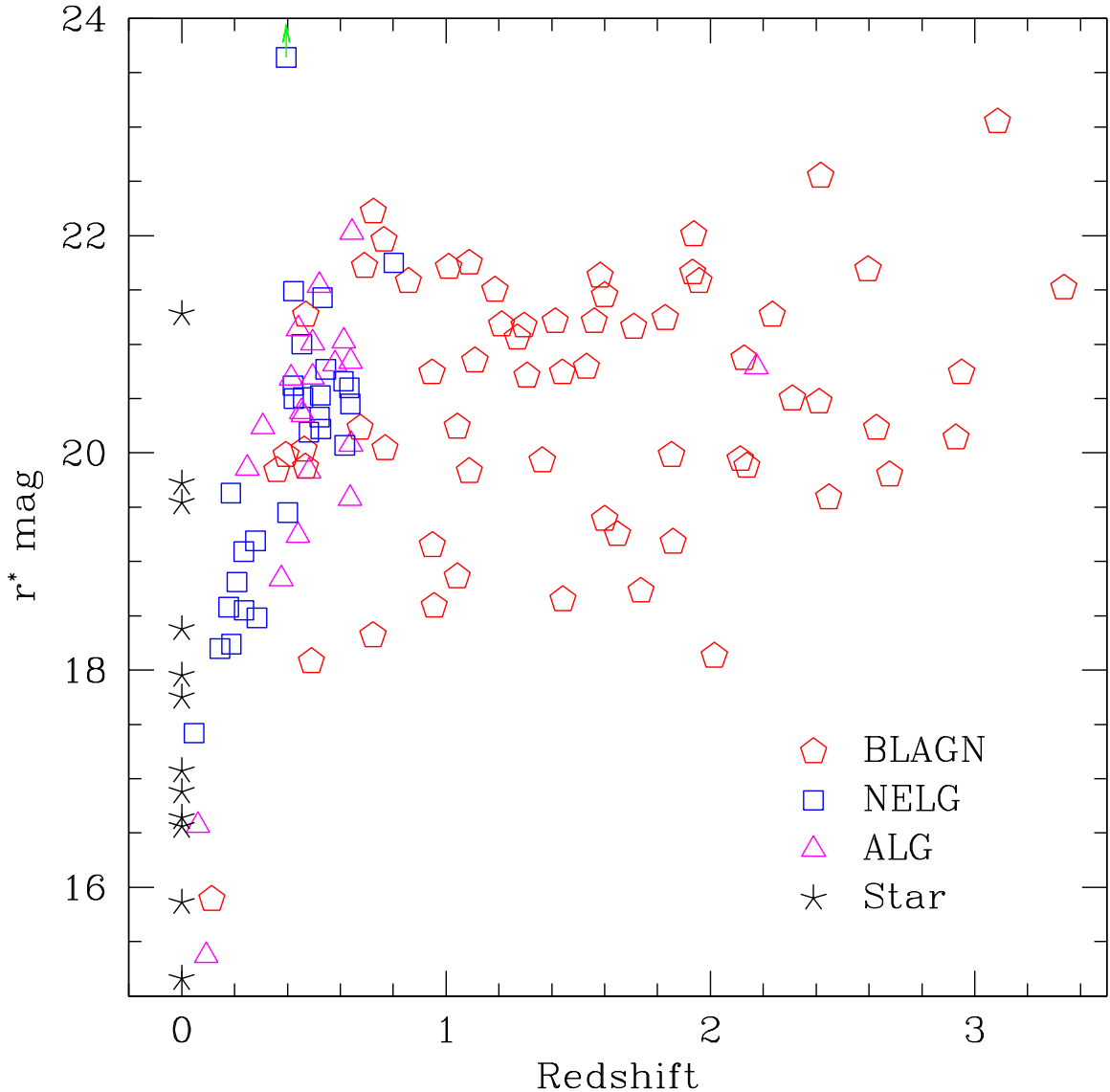}
\vspace*{1in}
\caption{\small 
SDSS \rs\, band magnitudes vs. redshift.  Galaxies of all types are
seen out to $z\sim0.8$, beyond which they are typically too faint for
spectroscopic classification.  BLAGN can be detected at all redshifts 
and extend well beyond our nominal $\rs=21$ spectroscopic limit
due to their strong broad emission lines.  The wide range of \rs\ mags
at each redshift reflects both the breadth of object types detected,
and the range of exposure time across and between fields.
\label{rz}}
\end{figure}

The proportional representation of different object types changes
strongly with redshift.  Figure~\ref{zhisto} shows the overall
redshift distribution of objects classified in our spectroscopy.  
BLAGN are from $z\sim0.2$ to beyond $z\sim 3$.  We can detect and
classify quasars to \rs\ magnitudes considerably fainter than galaxies
due to their strong broad emission lines.  We only see galaxies (both
NELGs and ALGs) to $z\sim 0.8.$  This effect is due primarily to
magnitude limits; our spectroscopic limit so far is $\rs\sim21$ for
objects with no strong  emission lines within the covered optical
wavelength range. Figure~\ref{rz} highlights this by showing that the
faintest classified ALGs are at $\rs=21$. 

From galaxies in the Hubble Deep Field with good completeness  to
$R=23$ \citep{CohenHDF00}, the {\em median} redshift of galaxies at
$R=21$ is about 0.5.  By contrast, our median $\rs$ mag for $z=0.5$
galaxies is significantly brighter at that redshift.  This may not be
due to incompleteness of optical followup; given both the possibility of an
AGN contribution and the known correlation between bulge luminosity
and black hole mass, we note that X-S galaxies are likely to be
atypically luminous in the optical band.   

\subsection{Luminosities and X-ray/Optical Ratio}
\label{lums}

\begin{figure}[ht]
\centering
\plotone{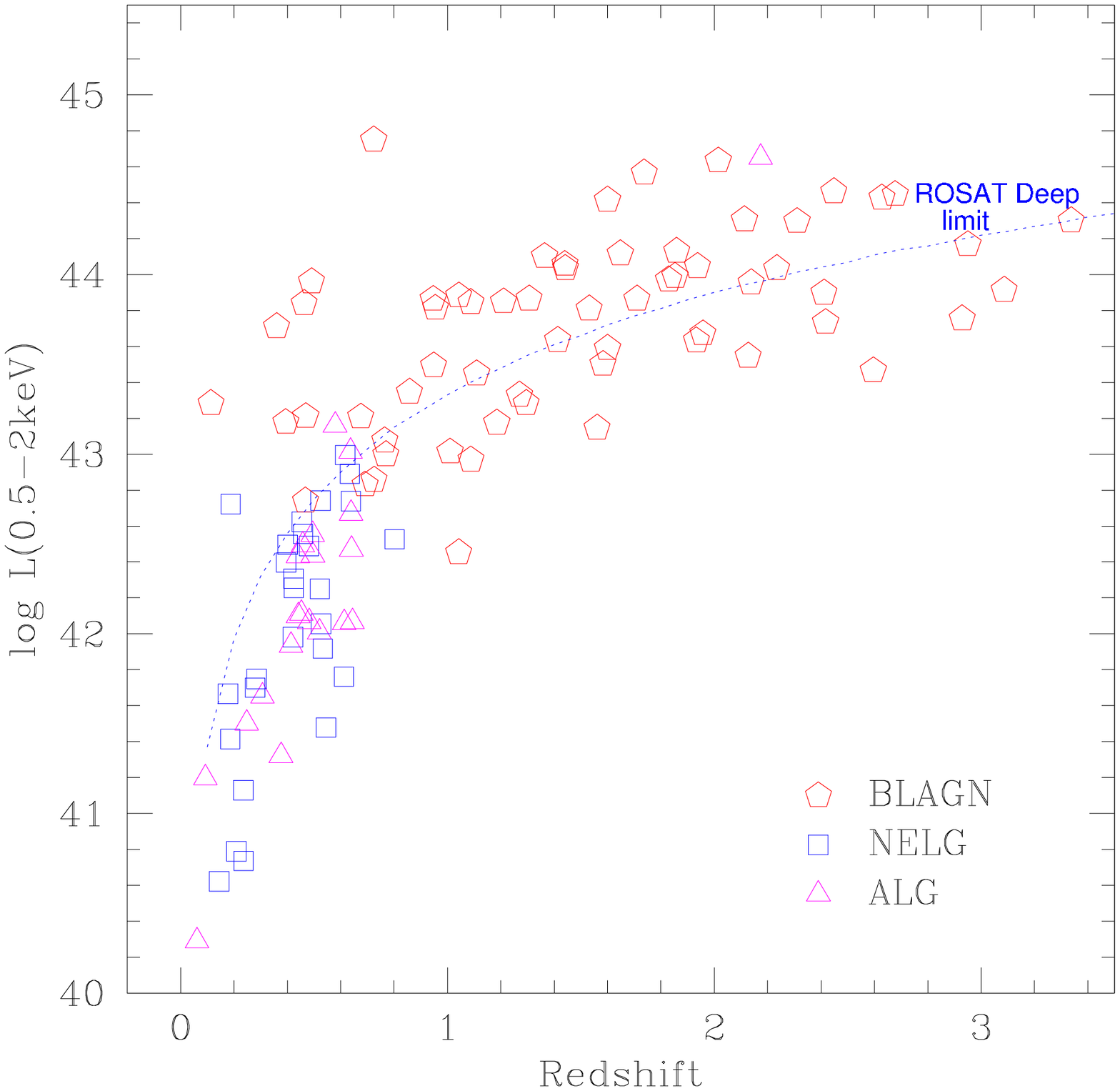}
\caption{\small 
Wide coverage of the \loglx\, vs. redshift plane highlights the
advantages of varying depths and wide sky area of the ChaMP.
BLAGN span a higher range of \lx\, than galaxies but with some
overlap. The dashed line shows the minimum luminosity corresponding
to which the ROSAT Deep Survey was sensitive (see \S~\ref{lums}).
\label{z_xebl}}
\end{figure}

Figure~\ref{z_xebl} shows the 0.5-2~keV X-ray luminosity of
spectroscopically classified galaxies and AGN as a function of
redshift.  BLAGN span a higher range of luminosities than do galaxies,
but with some overlap. Overplotted is the sensitivity of the ROSAT
Deep Survey  assuming their quoted limit $f_x=5.5\times10^{-15}$;
\citep{LROSDEEP01}.  The ChaMP samples luminosities 5--10 times
fainter at a given redshift even assuming a more optimistic
limit (from their Figure~3) of $2\times10^{-15}$.

The slope of a hypothetical power law connecting 2500\AA\, and 2~keV
is defined as $\aox\, = 0.384~{\rm log} \frac{\lopt}{\lx}$ (where
these are monochromatic luminosities at 2500\AA\, and 2~keV), so that
\aox\, is larger for objects with stronger optical emission relative
to X-ray.  We note that our definition of \aox\ here includes no mean
$k$-corrections for emission lines, intervening or intrinsic
absorbers, so \aox\ is directly related to \logxr, which we
prefer.\footnote{For convenience, the exact conversion between the two
is $ \aox\ = -0.379\logxr\, + 1.286 $.}

O-S AGN samples show typical \aox\ values of about 1.5
\citep{GPJETAL95} corresponding to $\logxr= -0.57$. We find for the
current sample of ChaMP BLAGN a somewhat lower value of
$\overline{\aox}=1.36\pm0.15$ ($\logxr= -0.19$).  \citet{LFFA95} found
a similar mean \aox\ of 1.3 for X-S AGN ($\logxr= 0)$.
X-ray surveys select X-ray bright objects by definition, and so are
complementary to O-S samples.   

A significant correlation between optical luminosity and \aox\ has
been noted in many studies based on O-S quasars (Vignali et al. 2001,
2003; Yuan et al. 1998; Green et al. 1995; Wilkes et al. 1994).   One
cause might be low-energy X-ray absorption as reported in quasars at
higher redshifts (Vignali et al. 2001, Yuan et al. 2000; Fiore et
al. 1998; Elvis et al. 1994).  Hypothesized to be due to the presence
of large gas quantities in primeval galaxies, such absorption 
might also be related to strong dense gas outflows 
akin to those seen in broad absorption line (BAL) QSOs.   However, the
trend in these  samples is primarily with $L_{opt}$ and not redshift.
Another cause of the apparent trend may be a combination of wider
intrinsic population dispersions in $L_{opt}$ than in $L_X$ along with
superposed selection effects (Yuan et al. 1998, La Franca et al. 1995).

\begin{figure}[ht]
\centering
\vspace*{0.0in}
\plottwo{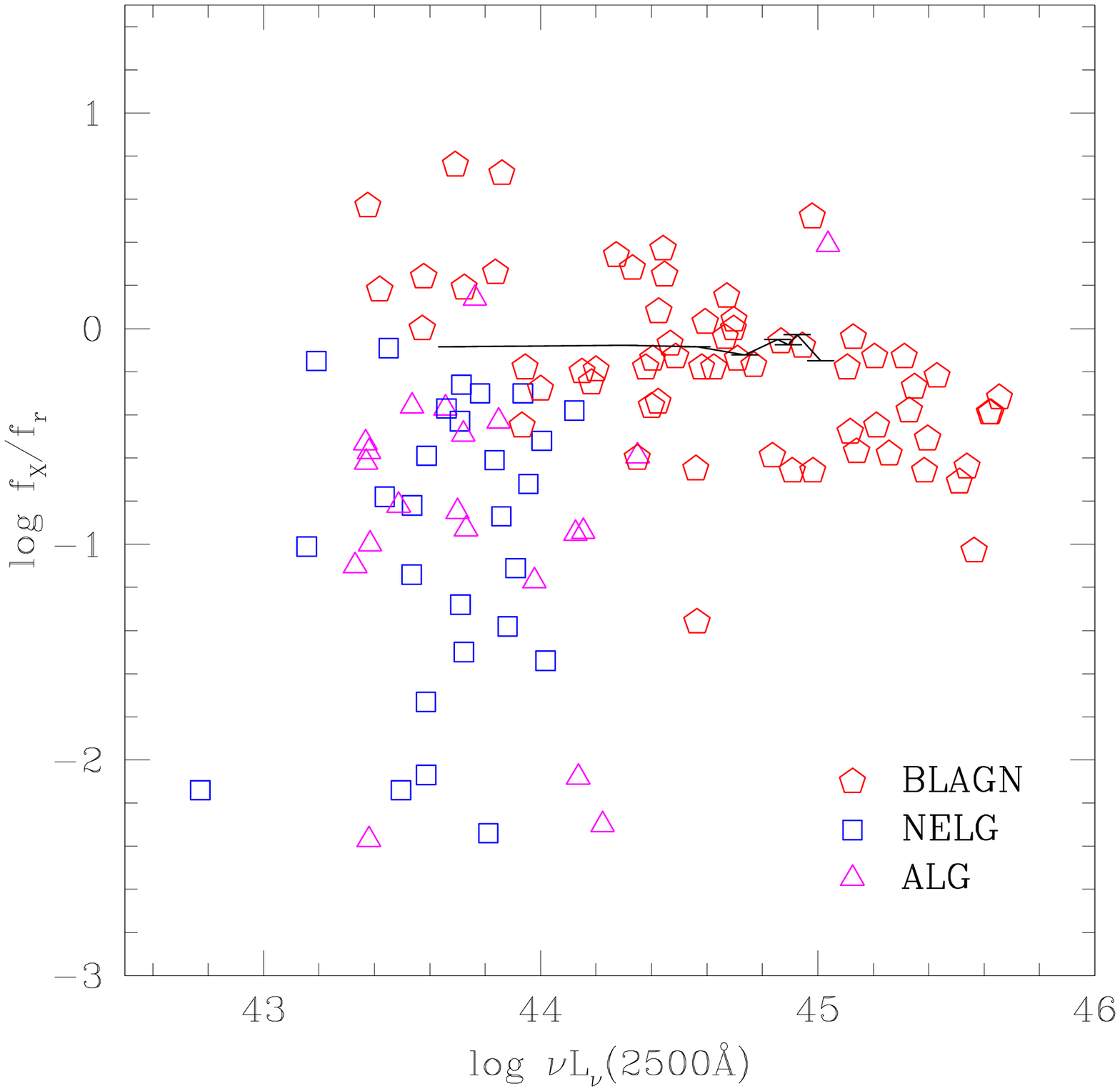}{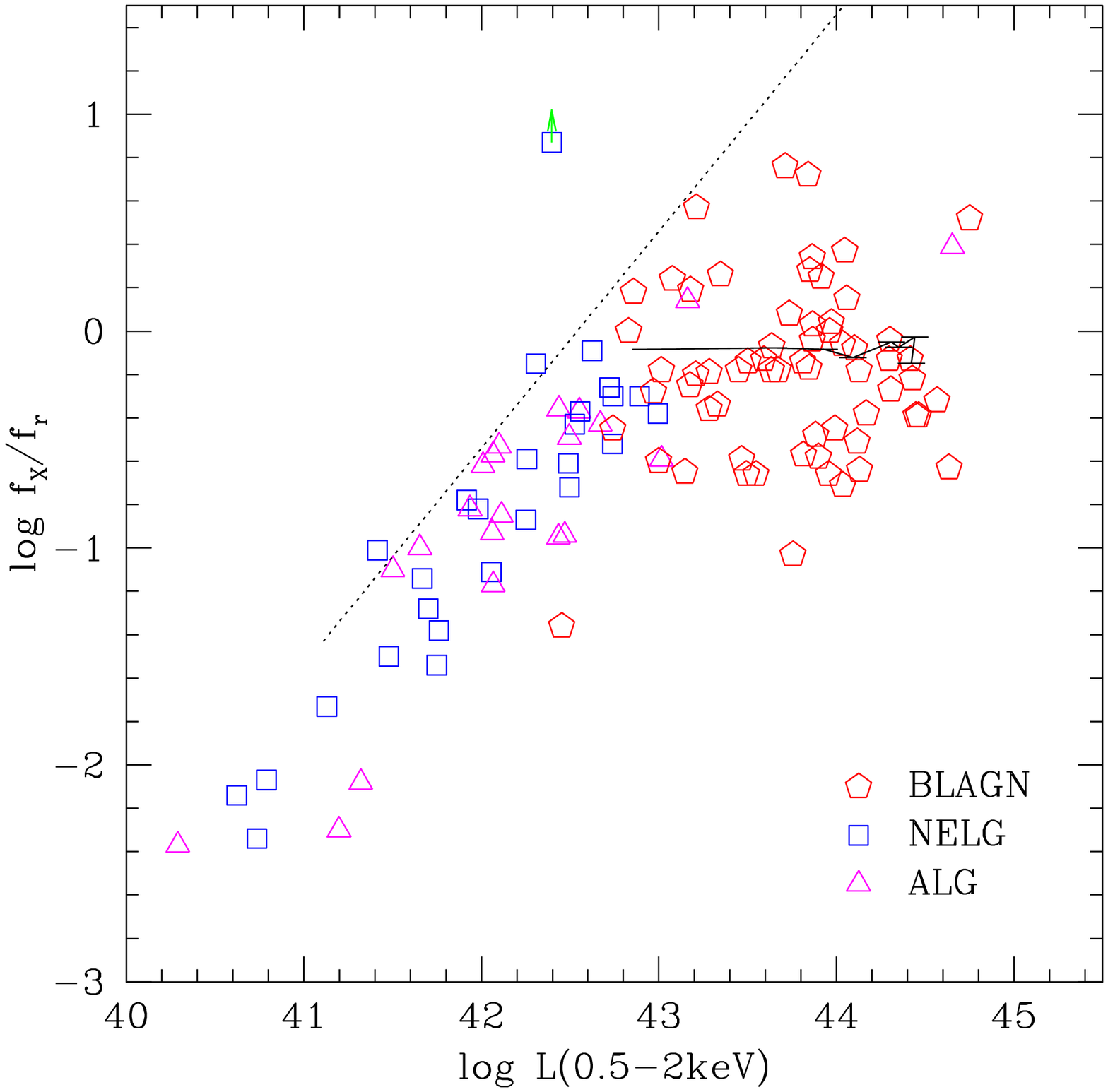}
\caption{\small 
LEFT: \logxr\ vs. \loglopt\, shows that galaxies span a range
of X-ray to optical flux ratio at least as wide as do BLAGN.  The
curve traces the mean expected quasar track for $z=0.5$ 
redshift bins for the simulation described in \S~\ref{results} and
imposed limits of $\logfx =-15$ and 
$\rs<22$.  All input simulated quasars have constant $\logxr= -0.3$
(corresponding to $\aox \sim 1.4$). The high luminosity point has
$\overline{z}=0.37$, while the faint end has $\overline{z}=4.2.$   
RIGHT: \logxr\ vs. \loglx\, correlates strongly for galaxies but not
BLAGN.  The dashed line corresponding to $\rs=22$ and $z=0.5$ is shown
for  a range of \logfx\, from --16 to --12 to illustrate the effect
of our spectroscopic limit.  The wide range of \logxr\ for galaxies may
reflect the unlinking of nuclear X-ray luminosity from the
optical luminosity, because the latter is strongly affected by  
host galaxy contributions.
\label{ebl_logxr}}
\end{figure}

We detect no discernible dependence of \aox\ on either redshift
\citep{MSHIZX02} or optical luminosity.  Figure~\ref{ebl_logxr} shows
no correlation between \logxr\ and optical luminosity (which we
calculate as $\nu L_{\nu}$ at 2500\AA, assuming optical spectral slope
$\alpha= -0.5$).  Galaxies span a lower but equally wide range of
\logxr\ as do BLAGN.  The curve in Figure~\ref{ebl_logxr} traces the
simulated mean quasar track with $\Delta z=0.5$ redshift bins and
imposed limits of $\logfx =-15$ and $\rs<22$.  All input simulated
quasars have constant $\logxr= -0.12$ ($\aox \sim 1.4$).  The high
luminosity point has $\overline{z}=0.37$, while the faint end has
$\overline{z}=4.2.$  No trend is seen in BLAGN for either data or
simulation, because the accretion dominates the spectral energy
distribution.  By contrast,  Figure~\ref{ebl_logxr} shows a strong
correlation between \logxr\ and X-ray luminosity for galaxies.  The
trend is likely a combintation of 2 effects, one intrinsic,
and one extrinsic: (1) The wide range of \logxr\ for galaxies may
reflect the unlinking of nuclear X-ray luminosity from the
optical luminosity, where the latter is strongly affected by  
host galaxy contributions.  (2) The current optical spectroscopic
magnitude limit excludes sources that are very weak in optical
relative to X-ray emission (have large \fxfr) and have small \fx.    

\subsection{Colors}
\label{colors}

The optical colors of ChaMP sources depend on object type,
redshift, and reddening.  The SDSS provides excellent statistics
on the colors of the objects that fall into their spectroscopic
sample. The ability to compare the photometric and color properties
of X-S sources with SDSS sources is an important ChaMP
feature which will be treated in upcoming papers. 

\begin{figure}[ht]
\epsscale{0.8}
\centering
\vspace*{-0.5in}
\plotone{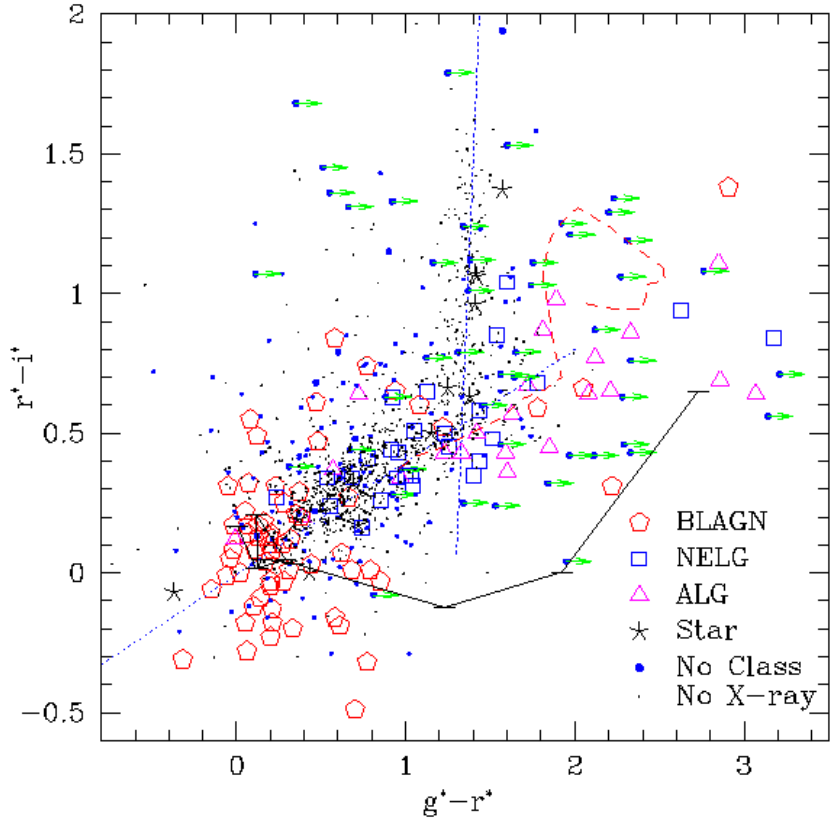}
\caption{\small 
SDSS \gmrs\, vs. \rmis\
color-color diagram for objects in 6 ChaMP fields. 
Optical objects with no X-ray detection are shown as small black
points.  The typical mean locus of SDSS objects that we use for a
sanity check of our photometric calibrations is plotted as 2
short-dashed straight lines.  The reddest BLAGN in both colors is at
$z=4.93$  \citep{SJGP02}. The solid curve traces the mean
expected quasar track for $z=0.5$ redshift bins for the simulation
described in \S~\ref{results}, with imposed flux limits of $\logfx=
-15$ and $\rs=25$.  The first point on the track that is away from the
stellar locus is at $z=3.2$, while the reddest \gmrs\ point on the
track corresponds to $\overline{z}=4.6$.  Without X-ray information,
normal quasars are virtually indistinguishable from stars in this
color plane until $z>3.5$.  The mean track of an E/S0 galaxy is shown
as a red long-dashed line, using the HyperZ code of \citet{BMP00}. The 
\gmrs\ colors of E galaxies redden to $\gmrs\sim2$ as the break moves
redward from the \gp\ to the \rp\ band at $z\sim 0.4$.   The loop back
blueward in \gmrs\ occurs at $z\sim 1$, where an $L^*$ E/S0 has
$\rs=23.5$ \citep{blanton01}.  
\label{gmr_rmi}}
\end{figure}

Figure~\ref{gmr_rmi} shows the optical color distribution of ChaMP
sources.  The mean stellar locus in these SDSS colors is shown
as two light short-dashed lines.  The range of BLAGN is similar to O-S
quasars, where the expected span of $z<3$ quasars in both \gmrs\ and
\rmis\ is from about --0.5 to 1 \citep{RGT01b}.  We plot the curve of
the simulated mean  quasar track for $z=0.5$ redshift 
bins for our simulation (with imposed flux limits of $\logfx= -15$ 
and $\rs=25$).  The first point on the track to leave the stellar
locus is at $\overline{z}=3.2$.  Without X-ray information, normal
quasars are virtually indistinguishable from stars in this color plane
until higher redshifts, where quasars redden significantly in \gmrs, 
with less change in \rmis.  The reddest \gmrs\ point on the track 
corresponds to $\overline{z}=4.6$.   Galaxies are generally redder
than BLAGN.  For ALGs, the 4000\AA\, break is the strongest color
determinant.  The mean track of an E/S0 galaxy is shown
as a red long-dashed line, using the HyperZ code of \citet{BMP00}, assuming
a Bruzual-Charlot E/S0 galaxy spectrum and typical {\em E(B-V)}$=0.05$
for Galactic dereddening. The \gmrs\ colors of E galaxies redden to
$\gmrs\sim2$ as the break moves redward from the \gp\ to the \rp\ band
at $z\sim 0.4$, and the track achieves its reddest \gmrs\, at about
$z=1$ before looping back blueward to end at $z=1.2$.
Several ALGs with $0.4<z<0.6$ are shown with $\gmrs$
close to 3; with $\loglx\sim42$, these are likely to be dust-reddened,
low luminosity AGN.  

\begin{figure}[ht]
\epsscale{1}
\plotone{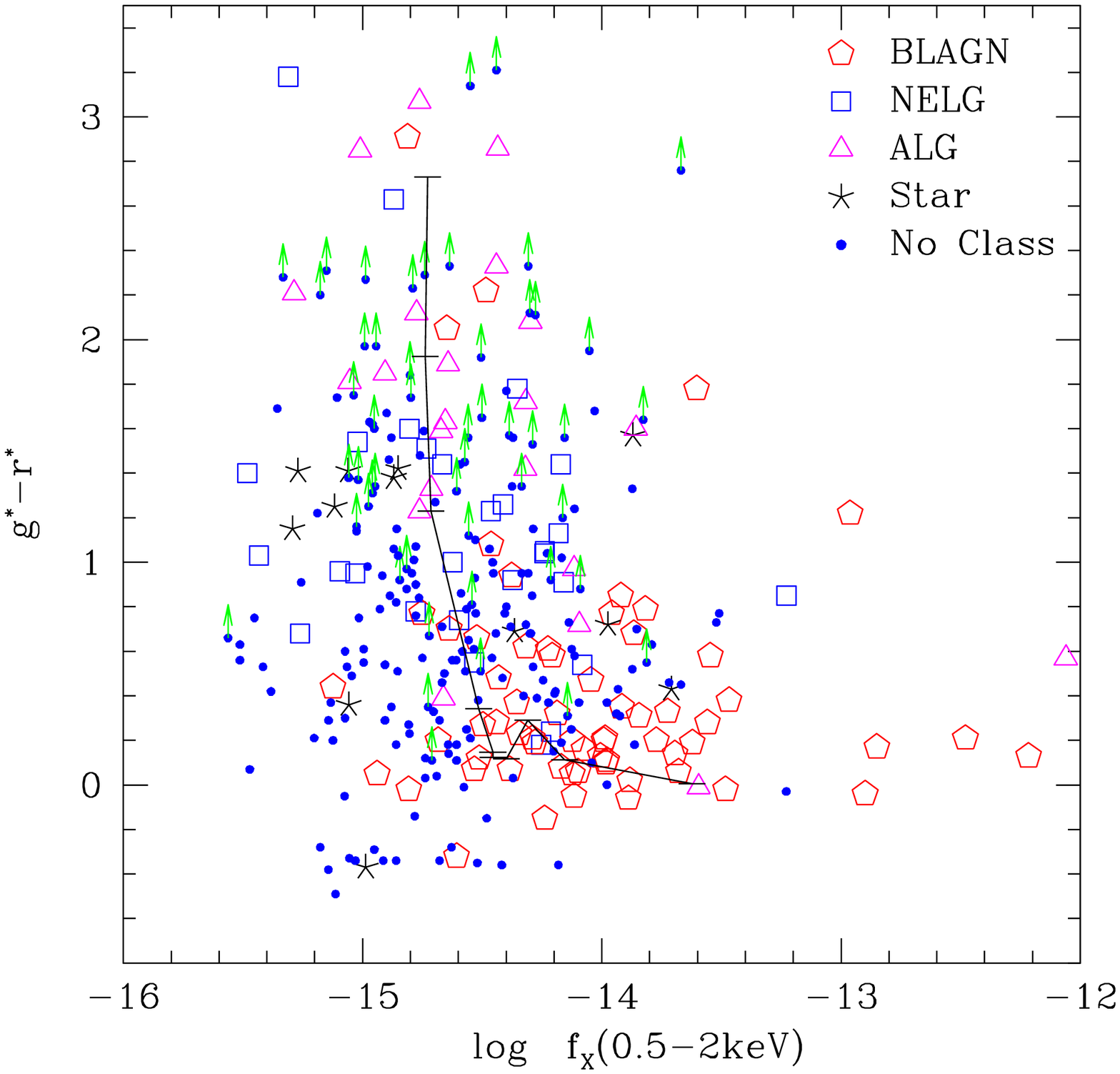}
\caption{\small 
SDSS \gmr\, colors derived from our NOAO Mosaic imaging
vs. \Chandra\ \logfx\, calculated in the 0.5--2\,~keV band.  Colors
include limits (denoted by arrows) when the object is only detected in a
single optical filter.  The brightest X-ray sources are blue
in \gmrs.  The curve traces the mean expected quasar track for $z=0.5$ 
redshift bins for the simulation described in \S~\ref{results}, with
imposed flux limits of $\logfx= -15$ and $\rs=25$.  The bright end
point has $\overline{z}=0.37$, while the faint end has
$\overline{z}=4.6$   
\label{gmr_x}}
\end{figure}

The distribution of optical colors changes with X-ray flux as shown in
Figure~\ref{gmr_x}.  In our spectroscopic sample to date, BLAGN inhabit
the full range of X-ray fluxes, but a narrow range of \gmrs. The
brightest X-ray sources 
are all blue in \gmrs.  The curve in Figure~\ref{gmr_x} traces the
simulated quasar track with imposed flux limits of $\logfx= -15$ and
$\rs=25$.  The X-ray bright end point has $\overline{z}=0.37$, while
the faint end  has $\overline{z}=4.6$  Many of the faint red objects
are likely to be high redshift and/or highly absorbed AGN.  X-ray
faint blue objects may be the wings of the low-redshift quasar \fxfr\ 
distribution. 

\begin{figure}[ht]
\plotone{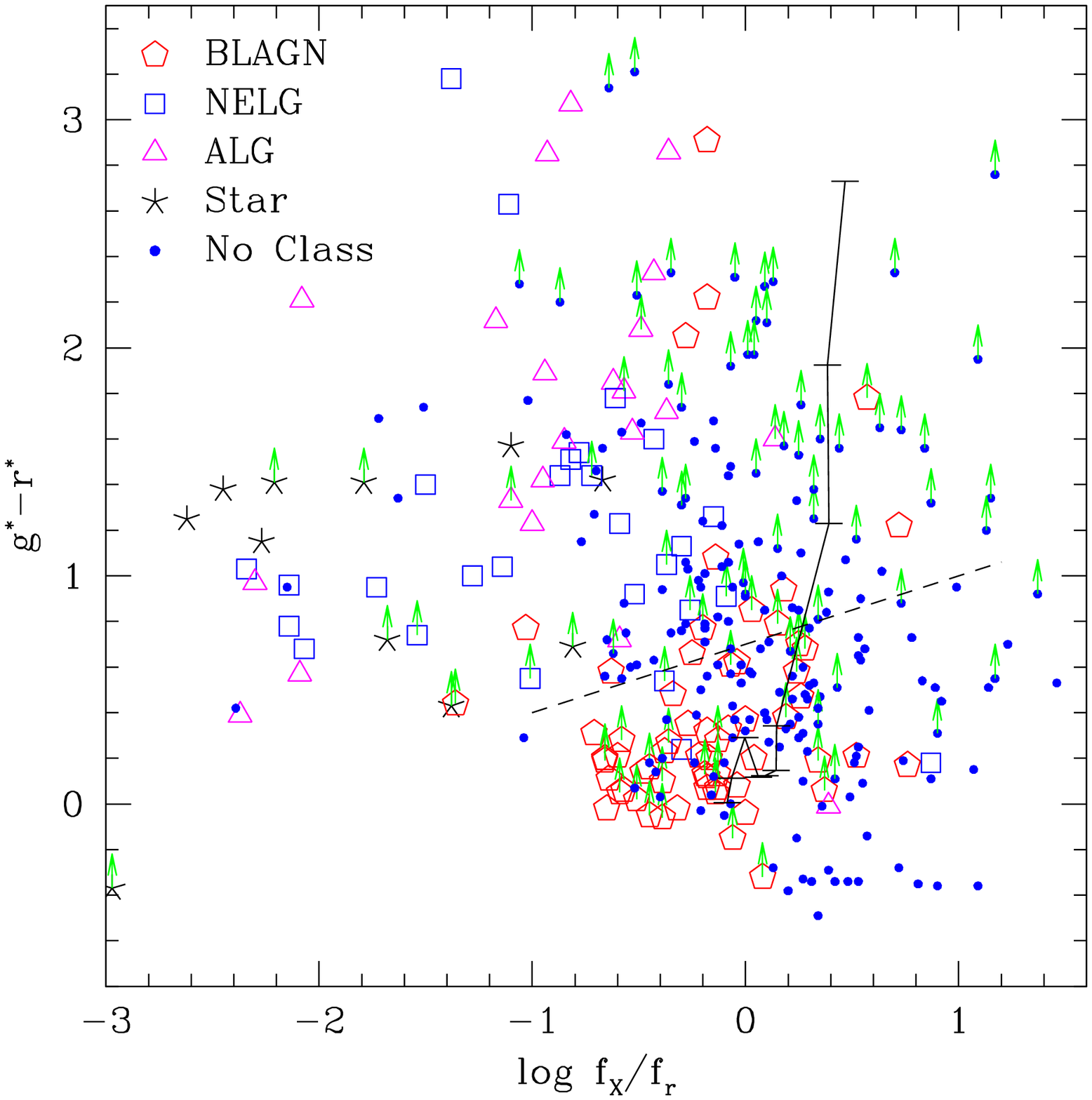}
\caption{\small 
SDSS \gmr\, colors derived for \Chandra\ sources from our Mosaic imaging
plotted against \logxr\, in the 0.5--2\,~keV band.  
The solid curve traces the mean expected quasar track for $z=0.5$
redshift bins for the simulation described in \S~\ref{results}, with
imposed flux limits of $\logfx= -15$ and $\rs=25$.  The
bright end point has $\overline{z}=0.37$, while the faint end
has $\overline{z}=4.6.$  All input simulated quasars have constant
$\logxr = -0.3$, so the shift towards larger \logxr\ arises
only from the imposed flux limits.  Objects in the lower right
part of the plot are largely consistent with O-S quasars and can be
classified as such with high confidence even without spectroscopy; a
possible demarcation is shown as the dotted line $\gmrs<0.3\,
\logxr\, + 0.7$.   Clustered with stars at low \logxr\, are 
galaxies that likely have little X-ray contribution from an active
nucleus.   
\label{gmr_logxr}}
\end{figure}

Figure~\ref{gmr_logxr} shows SDSS \gmr\, colors plotted against
\logxr\, in the 0.5--2\,~keV band.  Stars are all at $\logxr <-1$ and
$\gmrs<1.6$.  Bright BLAGN cluster strongly at high \logxr\ and blue
\gmr\ colors.  Objects in this part of the plot are largely consistent
with O-S quasars and can be classified as such with high confidence
even without spectroscopy; a suggested demarcation is shown as the
line $\gmrs<0.3\,\logxr\, + 0.7$ in Figure~\ref{gmr_logxr}.  
Redder objects with $\logxr>-1$ but $\gmrs > 1$ (many undetected
in our \gs-band imaging here) could be one of 2 types.  The simulated
quasar track shows that some are high redshift ($z>4$) quasars,
but many will be obscured AGN, as shown by the high fraction of objects
without broad lines in this region.

\subsection{X-ray Bright Galaxies with no Emission Lines}

A small number of X-ray bright, optical normal galaxies
\citep{comastri02} have also been found in XMM and \Chandra\, surveys.
These objects show at best weak optical emission lines.  We propose
three possible explanations for X-ray luminous 
objects showing no broad optical emission lines.  (1) A 'buried AGN' - a
nucleus that is intrinsically similar to that of a  quasar but whose
optical continuum and broad emission line region are shrouded by a
large obscuring column.  (2) A low luminosity
AGN (LLAGN) where the host galaxy light dominates light in the
spectral aperture.  One such type of LLAGN may host a nucleus with a
massive black hole accreting at very low rates, i.e., an advection
dominated accretion flow (ADAF). 
(3) A BL~Lac object. 

\subsubsection{Buried AGN}
\label{buried}

The unified model explains the Seyfert classes 1 \& 2 as different
views of the same phenomenon \citep{AR93}.  Quasars  are thought to be
luminous versions of Seyfert galaxies, so this unified (orientation +
obscuration) scheme should hold for quasars, i.e., they should have a
similar nucleus and dusty torus system in their center.  Seyfert~2
galaxies outnumber Seyfert~1 galaxies 4:1 in the local universe, which
is consistent with the covering factor of the obscuring tori
\citep{MR95}, and if 
that ratio can be extrapolated to higher luminosities, then we should
expect many Type~2 quasars to be detectable in the high-redshift
universe.  Only a few Type~2 quasars have been convincingly detected to date
(Akiyama et al. 2002, Norman et al. 2002, Stern et al. 2001, Schmidt
et al. 2002), implying a strong, perhaps luminosity-dependent evolution 
\citep{FXRB02}.  However, infrared surveys are currently finding
{\em bona fide} Type~2 quasars.  Several of the IRAS Hyperluminous
Galaxies (Beichman et al. 1986; Low et al. 1988, 1989) have been shown
to be Type~2 quasars (e.g., Wills et al. 1992; Hines \& Wills 1995; Hines 
et al. 1995).  More recently, samples selected from the 2MASS 
include a large number of Type~2 objects with quasar-like near-IR
luminosities.  Fully 1/3 of the spectroscopically confirmed AGN in the
2MASS are Type~2 objects \citep{CR01}.  Extrapolating to the entire
sky, 2MASS should detect roughly 5,000 such objects with $K < 15$,
many of which will have IR luminosities in the quasar range
\citep{SP2MASSQSO02}.  Near-IR surveys are less biased against
absorption and less sensitive to orientation than optical surveys,
but will preferentially select samples with high star formation rates
and copious dust.  

The X-ray characteristics of 2MASS quasars suggest strong absorption
($\nh\sim10^{22}\cmsq$; Wilkes et al. 2002) with contributions
from direct and unabsorbed, scattered, and/or extended emission. The 
possibility remains that they are intrinsically X-ray-weak.  The
sample of 2MASS quasars studied with \Chandra\ by \citet{WBSG02} are
all at $z<0.3$ (due to the 2MASS $K\sim 14.5$ limit).  At higher
redshifts, the reduced effective column in the X-ray bandpass should
be effective in discovering Type~2 quasars, and help resolve the
question of their intrinsic X-ray luminosities.  
How can and should a Type~2 quasar be recognized in an X-ray survey?
Not simply on the basis of X-ray hardness.  High column density
X-ray-absorbing gas (e.g., a warm absorber) may be interior to the
putative obscuring torus, so that even normal optical Type~1 (broad
line) quasars may show low luminosity and/or high column absorption in
the X-rays (however in these cases absorption in the restframe UV is
usually apparent; \citealt{GCXOBals01,BLW00}).  
Luminosity criteria may be subtle, since the claim for high-$L$ could
be based on either $L_{opt}$ or $L_X$, perhaps including large
absorption-corrections.  We find objects in our sample that show signs
of X-ray or optical absorption, or both, but with no strong evidence
that these properties are coupled.  

If an X-S AGN is found showing no broad emission lines, oftentimes the
search for broad lines may be very limited, including  only what is
seen on a discovery optical spectrum.  A search for broad 
H$\alpha$ is  sometimes possible from the ground, and may reveal
instead a luminous Type~1.9 quasar \citep{AU02}, but the classification
is by nature somewhat arbitrary.  Spectropolarimetry can reveal
broad line flux, but since those same photons must also be in the total
flux spectrum, broad line detection becomes a question of adequate
\snr. Indeed, many IR-selected Type~2 quasars show lower fractional
polarization than Type~1 quasars, presumably because of strong stellar
light contributions \citep{SP2MASSQSO02}. Furthermore, there are few objects
for which we can convincingly claim an optical Type~2 classification
without examining line ratios in detail in spectra of higher \snr.
Instead we define a ``buried AGN'' here as an object that has either
no or only narrow emission lines in optical spectra, strong evidence
for log$\nh>22$ in the rest-frame, and $\loglx>43$ without
absorption-correction.  If the signature of X-ray absorption is
clearly detected, then an absorption-corrected $\loglx>44$
qualifies as a Type~2 quasar candidate, with confirmation pending high
$\snr$ followup spectroscopy. We find no such objects in these 6
fields.  

\subsubsection{Low-Luminosity AGN}
\label{llagn}

Many of these X-ray selected galaxies are so distant that their
angular diameters are 
comparable to the slit (or fiber) widths used in ground-based
spectroscopic observations.  Using integrated spectra of a sample of
nearby, well-studied Seyfert~2  galaxies, \citet{moran02} demonstrate
that the defining spectral signatures of Seyfert~2s can be hidden by
light from their host galaxies. At $z\sim0.6$, a 1.5\arcsec\, slit
encompasses a region about 10\,kpc, which includes a substantial
fraction of the host starlight emission.  Some 60\% of the observed
objects would not be classified as Seyfert~2s on the basis of their
integrated spectra, which is comparable to the fraction of \Chandra\,
sources identified as  ``normal galaxies'' in deep surveys
(\citealt{HA01, BAJ02}).   

These 'buried AGN' may host low rate (advection dominated) accretion
flows (ADAFs; \citealt{narayan95}). The X-ray spectra produced by ADAFs
are relatively hard, so ADAF sources  may contribute a fair share of
the hard ($>$2\,keV) background \citep{yi98}.  Half of the 2-10\,keV
CXRB could consist of low-luminosity ($L_X < 10^{41}$ ergs/s) sources
if the comoving   number density is $\sim 3\times
10^{-3}$\,Mpc$^{-3}$, comparable to the density of $L^*$ galaxies
(e.g., Peebles 1993).   ADAF sources should be  characterized by
inverted radio spectra $I_{\nu}\propto \nu^{0.4}$



\begin{figure}[ht]
\centering
\vspace*{0.0in}
\epsscale{1.1}
\plottwo{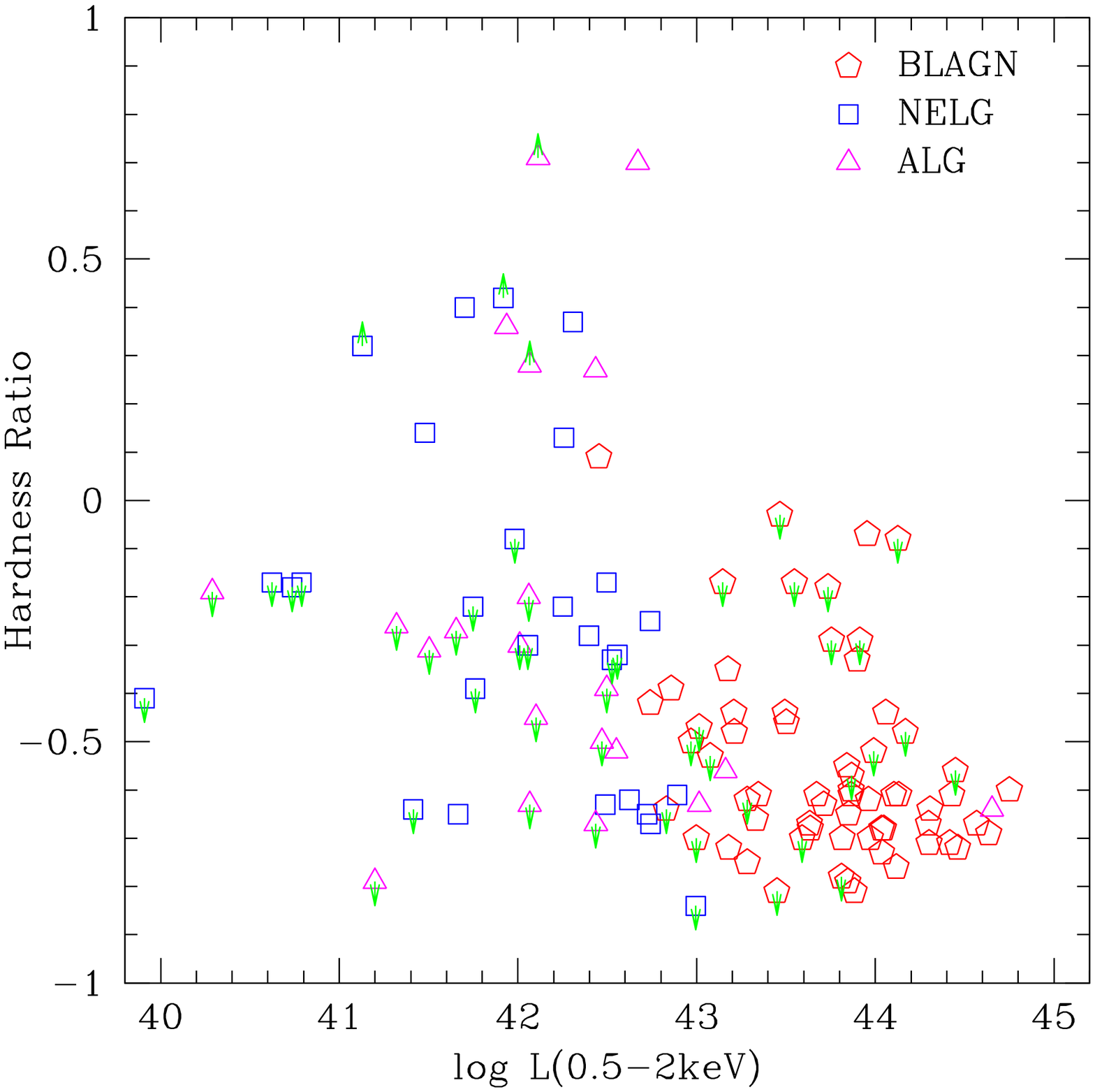}{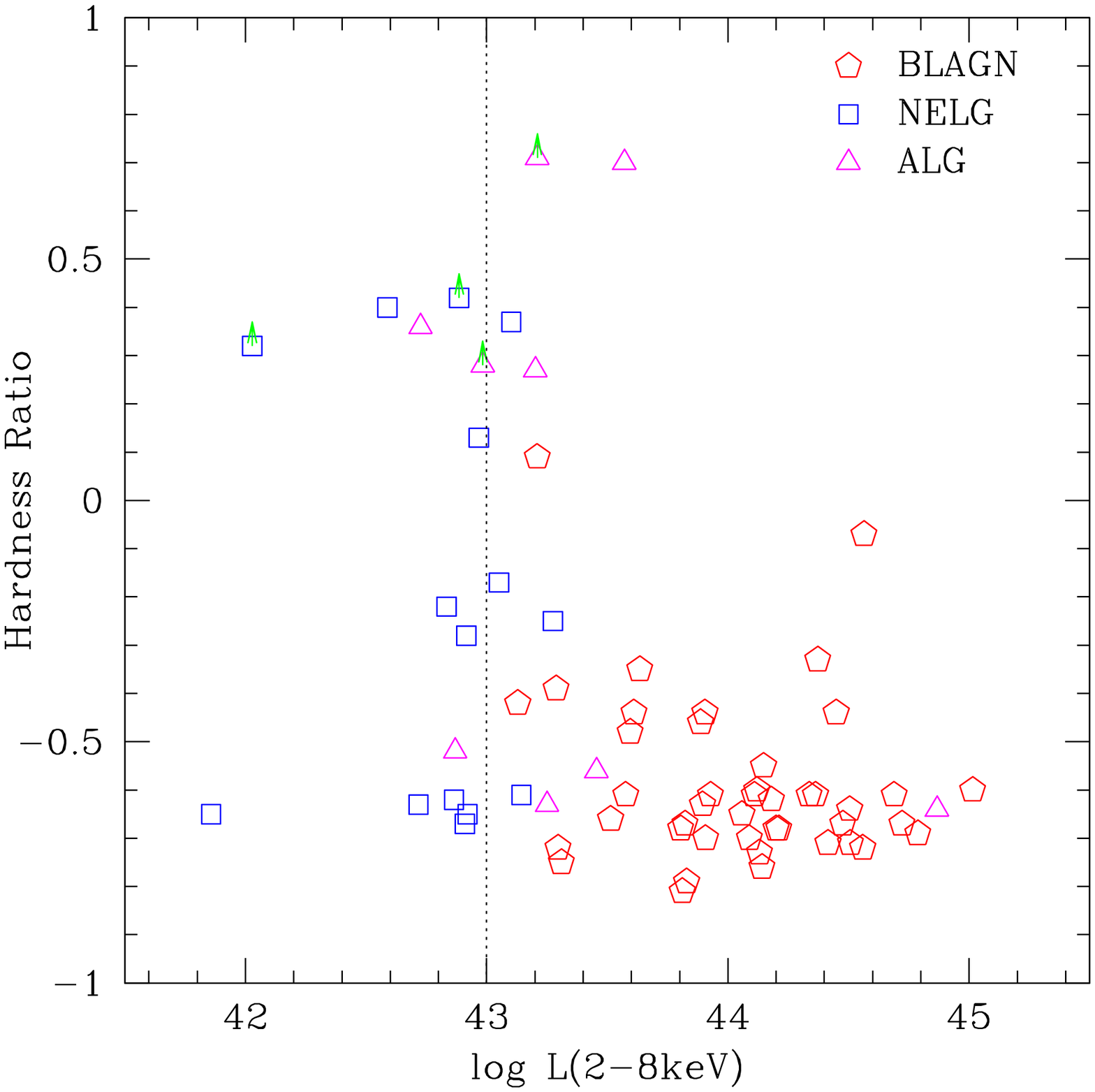}
\caption{\small 
Hardness ratio vs. X-ray luminosity in 2 bands. Most of the luminous
objects (quasars) are soft. This may arise both because they are at high
redshift (where absorption causes smaller changes in hardness) and
because they are intrinsically less obscured.  RIGHT: Plotting against a
harder $L_X$(2-8~keV) does not significantly change the plot.  A dashed
line shows our criterion for objects with no broad lines to be
classified as 'buried AGN'; log$L_X$(2-8\,~keV)$=43$.  
\label{xebl_hr}}
\end{figure}

\begin{figure*}[ht!]
\centering
\epsscale{1.2}
\plotone{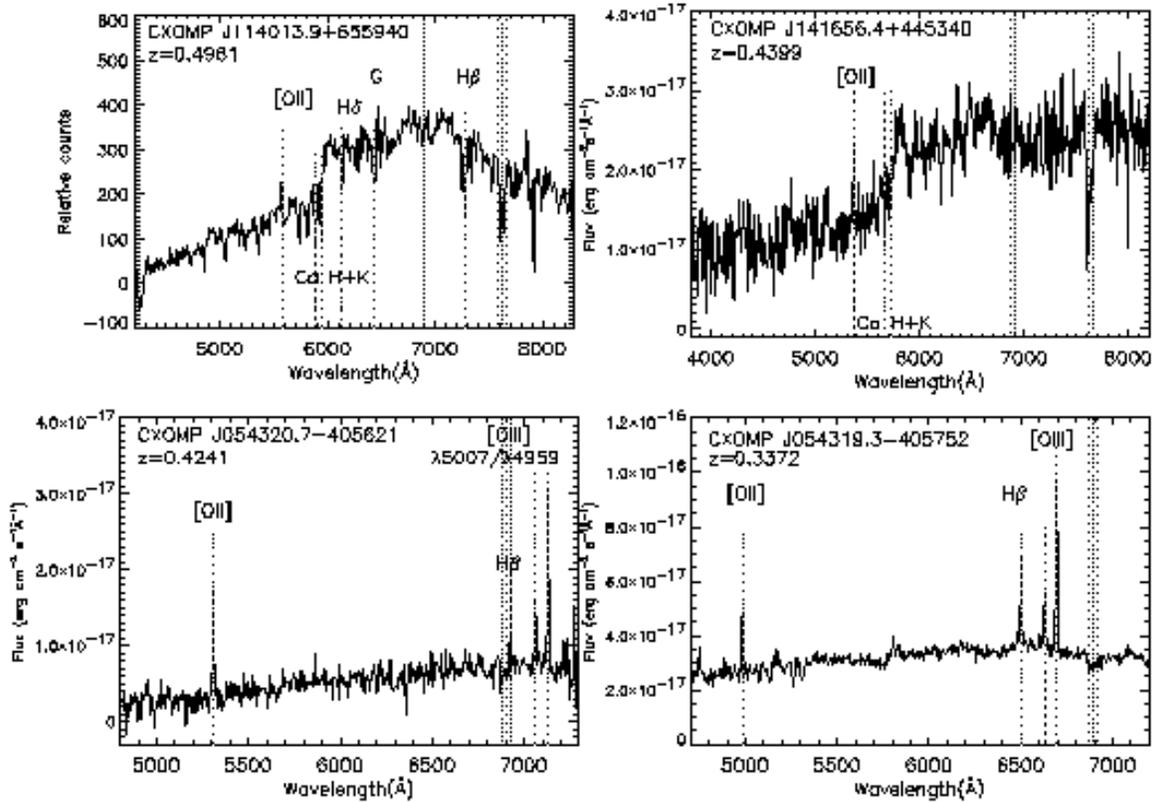}
\vspace*{-5cm}
\caption{Spectra of ChaMP candidate absorbed AGN.  These optical
spectra correspond to 4 X-ray sources with luminosities
(log$L_X$(2-8~keV)$>43$ erg s$^{-1}$), hard X-ray spectra, and no
signs of broad emission lines characteristic of optically-selected
quasars.  Grey regions mark the telluric absorption bands. 
 }
\label{type2}
\end{figure*}

There are 8 optical counterparts to X-ray sources in the current sample
that have hard band\footnote{While our $H$ band is nominally 2.5-8~keV,
we use 2-8~keV luminosities here to facilitate comparison to other
studies.} luminosities log$L$(2--8~keV) $>43$ and no signs of broad
optical emission lines in their spectra. 
The luminosity in this band is plotted against hardness ratio
in the right panel of Figure~\ref{xebl_hr}.  (A similar $L_X$
criterion in the 0.5--2~keV band would be log$L$(0.5-2~keV) $>42.4$)
Half of these Type~2 candidates are hard, with $HR_0>0$.  Optical
spectra of this sample are shown in Figure~\ref{type2}.  Several of
these objects have spectra of absorption line galaxies with only a
weak narrow [OII] emission line. Several counterparts are also
classified as narrow emission line galaxies with strong lines of
[OII], [OIII] and H$\beta$.  All these objects are at redshifts less
than unity.

Newer CXRB models attempt to include the observation that most hard
faint sources from \Chandra\,and XMM Newton surveys are objects at
$z<1$, which requires a very different evolution rate for Type~1 and
Type~2 objects.  This must at least modify the meaning of Unification;
all Type~2 AGN are not necessarily Type~1 from a different angle.
The covering fraction of the putative obscuring torus may also
change, and probably evolves statistically with cosmic time. The
relative dearth of high redshift Type~2 quasars could be explained   
by selection effects.  As discussed in \S~\ref{rxz}, these objects may
simply fall below our optical detection limits, or at least our
$\rs=22$ spectroscopic limit.  Since these objects have low surface
densities, as the ChaMP area grows, a volume-limited subsample will
help define their relative space densities. 

\subsubsection{BL~Lacs?}
\label{bllacs}

The first known BL~Lacs provided the three observational characteristics
of the class (e.g., \citealt{stein76,angel80}), repeated here as
presented by \citet{jan93}: "1.) They are intrinsically luminous with
strong and `rapid' variability. (BL~Lac objects have been observed to
vary at X-ray, optical, infrared, and radio wavelengths.)   
 2.) They have `featureless' optical spectra. (How to define
"featureless" has been a question for debate [...]).
 3.) The electro-magnetic radiation from BL~Lac objects is observed to
be linearly polarized."  
In general the polarization requirement has been for the observed
degree of polarization to be larger than that possible from differential
extinction (at most wavelengths a percentage polarization of greater
than 3\%) or for the degree and/or position angle of the polarization
to be variable.  Together, these observationally based criteria
reflect what we now know is the distinguishing physical characteristic
of all BL~Lacs, the presence of a strong source of synchrotron
radiation in these sources.  

We are interested in the identification of BL~Lacs because it is
possible that with only single epoch photometry and spectroscopy (and
no polarimetry data), objects which we have currently identified as
buried AGN might be BL~Lac objects. Lacking strong emission lines, the
optical and near-IR spectra of many BL~Lacs (particularly X-ray
selected objects; see EMSS survey results of \citealt{morris91}
and \citealt{jan94}) can be dominated by host galaxy starlight and
look very similar to what we call absorption line galaxies. 

While more data on these objects would be advantageous, we have 
checked whether BL~Lacs might be hidden in our absorption
line galaxy sample.  First, we checked for known radio emission
from our X-ray sample.  Since all known BL~Lac objects are strong radio
sources (e.g. \citealt{jan90,stocke90}),
the presence of strong emission (from a compact synchrotron core)
might indicate the presence of a BL~Lac. Unfortunately, no data are
available of sufficient sensitivity for our ALGs.  Our cross-correlation
of our X-ray source list to the FIRST \citep{white97} and NVSS
\citep{condon98} radio surveys with a 5\arcsec\, search radius  
did yield 2 matches to other sources.  One is a
bright nearby radio-quiet Sy~1 and the other is likely to be a distant
radio-loud quasar (see Appendix~\ref{indiv}).  Second, we checked the strength
of the 4000\AA\, break in our spectra of all sources we classed as
ALGs or BL~Lacs (all those with the strongest emission line equivalent 
width $W_{\lambda}<5$\AA).  As discussed by the EMSS team, a weak
break might indicate the presence of non-thermal (featureless)
emission contributing to the total emission -- again a possible sign
of a BL~Lac being present in the source.  We measured the 4000\AA\,
break contrast using the using the definition of \citet{dressler87} 
$${\rm Br}_{4000\AA} = \frac{f^+ - f^-}{f^+}$$ where $f^+$ is the
average flux in the rest wavelength range 4050--4250\AA\ and $f^-$ the
average flux between 3750--3950\AA.  As long as the \snr\ in this
wavelength region exceeds 5 per pixel, we count break strengths
$<0.25$ as BL~Lac candidates pending confirmation from any of the 3
criteria listed above. 

We found just one optical spectroscopic BL~Lac candidate; 
CXOMP~J05421.5-410206, with $\rs=19.6$ at $z=0.637$.  
While in the same range of luminosity as known
BL~Lac objects, it has a relatively low $\logxr=-0.6$.
We note that for optical magnitudes $V<20$, the optical luminosity
function for X-ray-selected BL~Lacs yields a surface density of
about 160~steradian$^{-1}$ or about 0.05~deg$^{-2}$ \citep{wolter94}. 
The optical QSO surface density is about 600 times as high at similar
magnitudes \citep{meyer01}, so few BL~Lacs are expected overall in the
ChaMP.

\subsection{Signs of Absorption?}
\label{abs}

\begin{figure*}[ht!]
\epsscale{1}
\plotone{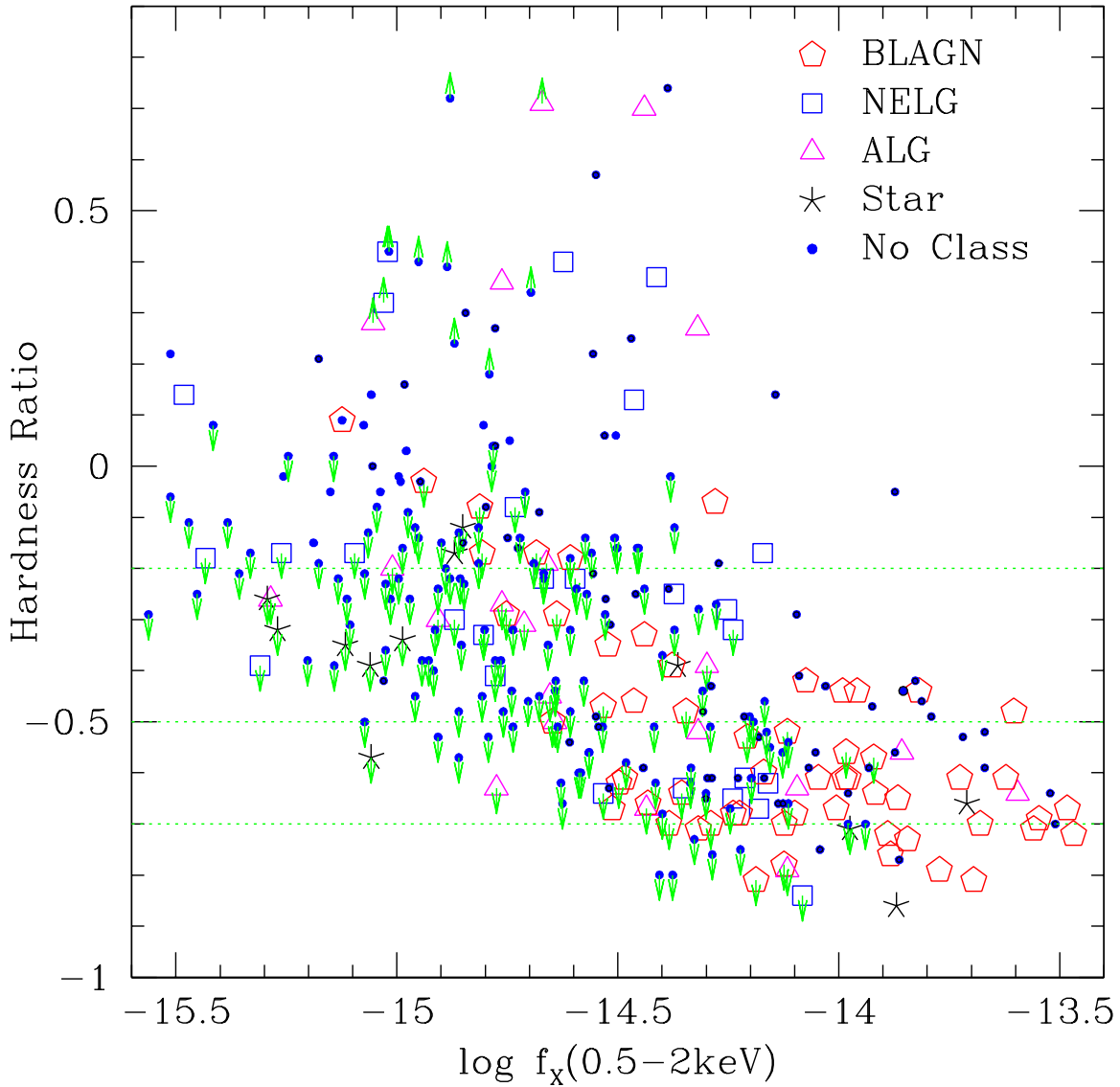}
\caption{\small
Corrected hardness ratios $HR=(H-S)/(H+S)$ as a function of X-ray
flux.  $HR$ values here are corrected to represent a source on-axis on
a front-illuminated (FI) chip, and to include upper or lower limits to
$HR_0$ (arrows), based on $2\sigma$ upper limits to $H$ or $S$ counts,
respectively.  Horizontal dashed lines show the $HR_0$ corresponding
to a typical unabsorbed AGN (bottom; $\Gamma=1.9/\nh=0$ \cmsq), the
average hardness ratio of the total CXRB spectrum (middle;
$\Gamma=1.4/\nh=0$ or $\Gamma=1.9/\lognh\ge 21.5$), and a highly
absorbed AGN (top; $\Gamma=1.9/\lognh\ge 22$).  Most BLAGN likely
suffer some moderate absorption, whereas a large fraction of NELG and
ALG show evidence for $\lognh\ge 22$.  The brightest 8 sources between
$-13.4 < \logfx < -12$ (typically soft) are left outside the plot
range for clarity.
\label{logfx_hr}}
\end{figure*}

Figure~\ref{logfx_hr} shows $HR$ values corrected as described
in \S~\ref{xanalysis}. Many previous studies have assumed that when 
total source counts exceed 30 or 50, that the $HR$ values are reliable
without accounting for errors. In fact, the combination of source
spectrum and telescope effective area may require a surprisingly large
number of counts just to achieve detection in all the bands composing an
$HR$ calculation. The fraction of limits (mostly upper limits, since
\Chandra's $H$ band sensitivity is significantly lower) decreases
below 10\% only above about 120  counts (for these observations
typically at about $\logfx=-14$ in Figure~\ref{logfx_hr}).  About half
the sources have $HR_0$ upper limits even for sources with 30 counts or
more.  Analysis of only those points detected in both bands would lead
to an apparent hardening at low fluxes.  Correctly incorporating the
upper limits via survival analysis (ASURV Rev 1.1 Isobe \& Feigelson
1990; LaValley, Isobe \& Feigelson 1992), we find no significant
evidence for hardening of the spectra toward lower fluxes.  Excluding
sources with fewer than 30 (0.3-8~keV) counts, the median (mean) of
$HR_0$ for sources with $<120$ counts is  --0.66 ($-0.58\pm0.02$),
while for sources with more counts  we find --0.62 ($-0.56\pm0.03$).
This is not inconsistent with the conclusion in \citet{kim03b} that spectral
hardening at faint fluxes is significant only in the softest band (S1;
0.3-0.9 keV) and thus most likely due to absorption.  
While spectral hardening towards fainter fluxes may be real, careful
accounting for errors in hardness ratios is important: spectral
stacking (e.g., Brandt et al. 2001; Tozzi et al. 2001) is a powerful
alternative for these analyses.  

Assuming that $\Gamma\sim 1.9$ for most BLAGN, the hardness ratios
indicate that most BLAGN likely suffer some absorption, whereas strong
absorption is rare.  A much larger fraction of NELG and ALG show
evidence for $\lognh\ge 22$.  


X-S quasar samples may be less biased against absorbers
(both intrinsic and line-of-sight) than are O-S 
samples.  This advantage is expected to be especially important at high
redshifts for two reasons.  First, X-rays from high-$z$ objects can be
detected through a larger intrinsic absorbing column of gas and dust
because the observed-frame X-ray bandpass corresponds to higher
energy, more penetrating X-rays at the source; the
observed-frame, effective absorbing column is $N_{\rm H}^{\rm eff}\sim
N_{\rm H}/(1+z)^{2.6}$ \citep{WRFA99}.  Second, optical color
criteria for high-$z$ candidates (e.g., blue filter dropouts) tend to
select objects with strong Ly$\alpha$ forest decrements.  

\begin{figure}[ht]
\plotone{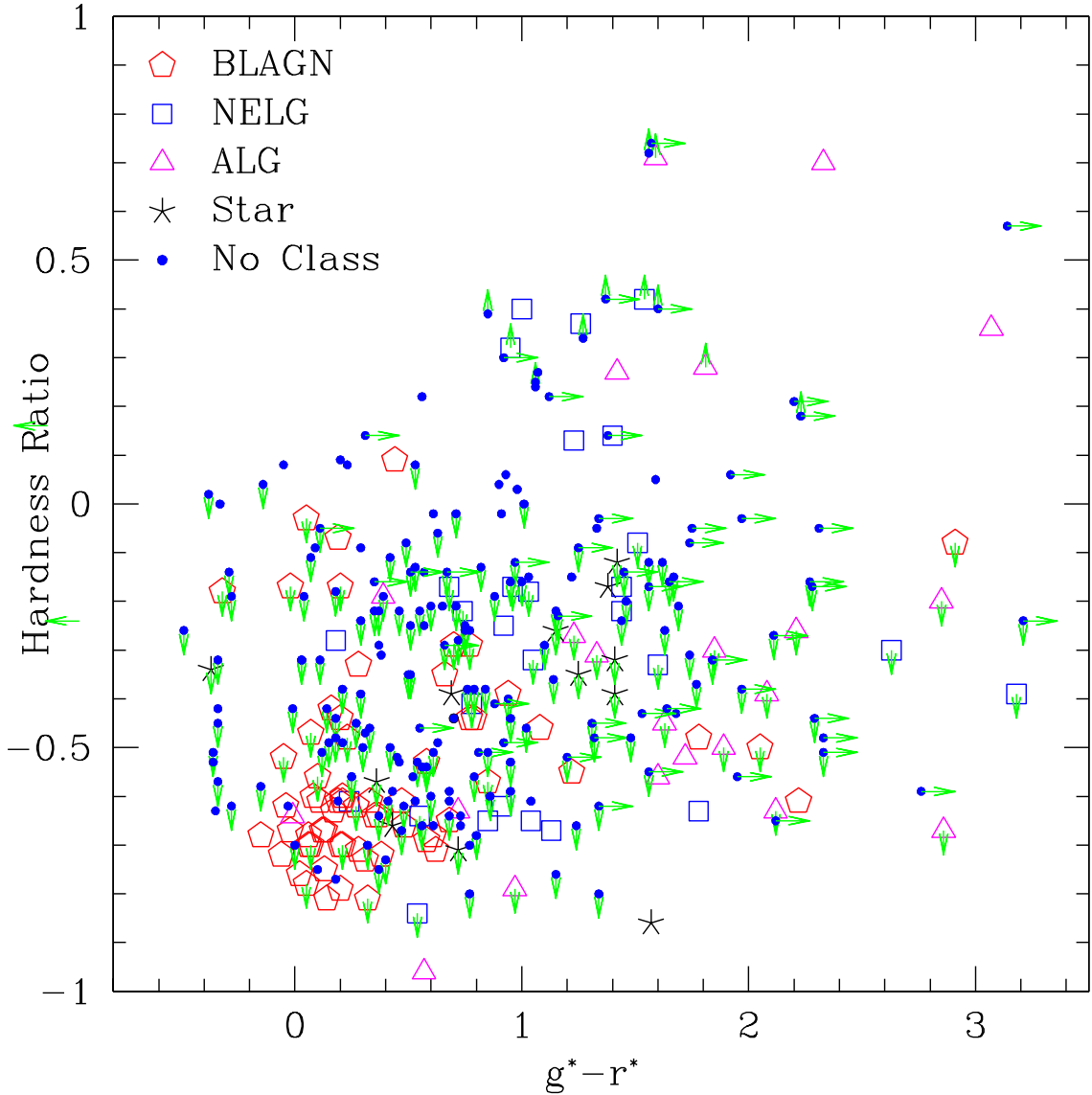}
\vspace*{1.0in}
\caption{\small 
SDSS \gmr\, colors derived from our 4m Mosaic imaging vs. corrected
\Chandra\ hardness ratio $HR_0$.  Since a range of redshifts
and optical object types may pertain, \gmrs\ is at best a crude measure
of optical reddening - as is $HR_0$ for X-ray absorbing column.
Nevertheless, above $HR_0\sim 0$ a weak correlation  
exists between these indicators ($P=10\%$; see \S~\ref{abs}). 
This suggests that at high columns, the effects of intrinsic dusty
absorbers may dominate the dispersion induced by factors such as
redshift, orientation, and/or reflection.  
\label{gmr_hr}}
\end{figure}

Do red optical colors reliably predict hard X-ray spectra or
vice-versa?  Figure~\ref{gmr_hr} shows two quantities, $HR_0$ and
\gmrs, which may serve as crude proxies for absorption measurements.
While \gmrs\ is affected by dust via optical reddening, it is also
strongly dependent on object type, host galaxy fraction, and
redshift.  While the hardness ratio $HR_0$ is 
affected by the total X-ray absorbing column, it is also dependent on
redshift since the effective column decreases as noted above.  We find
no overall correlation between these two absorption 
indicators, as is evident from Figure~\ref{gmr_hr}.  X-ray and
optical absorption are effectively decoupled for several 
physical reasons:  1) The gas-to-dust ratio varies, depending 
on densities, abundances, and ionization level \citep{Mdust01}.  2) 
Physically distinct absorbing regions \citep{WMgrains02}.  3)
Variations in geometry, reflection, and self-shielding
\citep{EModel00}.   

We note that for the 35 X-ray hard objects ($HR_0>0$) suggestive of
strong absorption, there is some weak evidence for a trend in
Figure~\ref{gmr_hr}.  Using ASURV \citep{LMIT92}, we include the
limits in both axes,\footnote{We employ a generalization of Kendall's
nonparametric rank correlation test developed by Brown, Hollander \&
Korwar (1974, hereafter BHK test).  This method assumes no
distribution for censored data and allows either type of censoring
simultaneously in both axes.} and find no significant correlation
(probability of no correlation is $9.8\%$).  At high columns, other 
scatter-inducing factors such as redshift, orientation, and/or
reflection may be overcome by the effects of intrinsic dusty
absorbers.  This suggests that the correlation should be tested with a
larger sample, as will become rapidly available from the ChaMP.  

Hard red objects in the
upper right hand corner of the plot (e.g., $HR_0>0$ and $\gmrs>1.5$)
likely contain a high fraction of obscured AGN; none with spectral
classifications to date are BLAGN.   Soft red objects in
the lower right (e.g., $HR_0<0$ and $\gmrs>1.5$) could be high
redshift AGN (where unabsorbed X-rays are shifted into the \Chandra\
passband) or soft galaxies above $z\sim0.3$.  Spectra of the
hard blue BLAGN (e.g., $HR_0>-0.2$ and $\gmrs<0.5$) show mostly
weak narrow emission lines, and several appear to have significant
blueshifted CIV absorption.  These objects are similar to 
BALQSOs, which are only slightly reddened \citep{reichard03a}
but known to be heavily absorbed in X-rays \citep{GCXOBals01}.


\subsection{Clusters of Galaxies}

Clusters are important probes of cosmology, especially at high mass and
redshift.  In general, a surface density of $>0.003\,$deg$^{-2}$ for
clusters more massive than Coma above $z>1$ violates $\Lambda$CDM
cosmologies at 95\% confidence \citep{Ehizclust02}.  Evolution of and
correlations between X-ray temperature, luminosity, and gas mass in
clusters \citep{Vclustevol02} enables the strongest available
constraints on hierarchical models. \Chandra's high sensitivity, low
background and good spatial resolution facilitate the detection of
spatially extended emission which at high galactic latitudes
corresponds to emission from groups or clusters of galaxies.

Our prototype X-ray extended source detection algorithm works by
performing photometry in annuli around sources detected by {\tt
wavdetect}, and compares the best-fit gaussian and beta models to the
expected PSF size.  All of the targeted clusters are detected.  Some
clusters also have point-like emission from a cD galaxy or quasar, which
can confuse the automated algorithm.  These are generally noticed
during our visual inspection, which also provides an opportunity to
flag any X-ray source surrounded by a concentration of optical
galaxies.  Using this approach, we have found 3 serendipitous clusters
in these 6 fields, each of which described further below.   The ChaMP
makes feasible simultaneous optical and 
X-ray searches for clusters.  Further refinement and testing of
cluster selection is underway \citep{DAP04} that combines automated
and by-eye X-ray source extent determinations, Voronoi Tesselation and
Percolation analysis on color-filtered optical images, and red
sequence identification in optical color-magnitude plots. 


\subsubsection{CXOMP\,J033912.2-352614}

\begin{figure}[ht]
\centering
\plotone{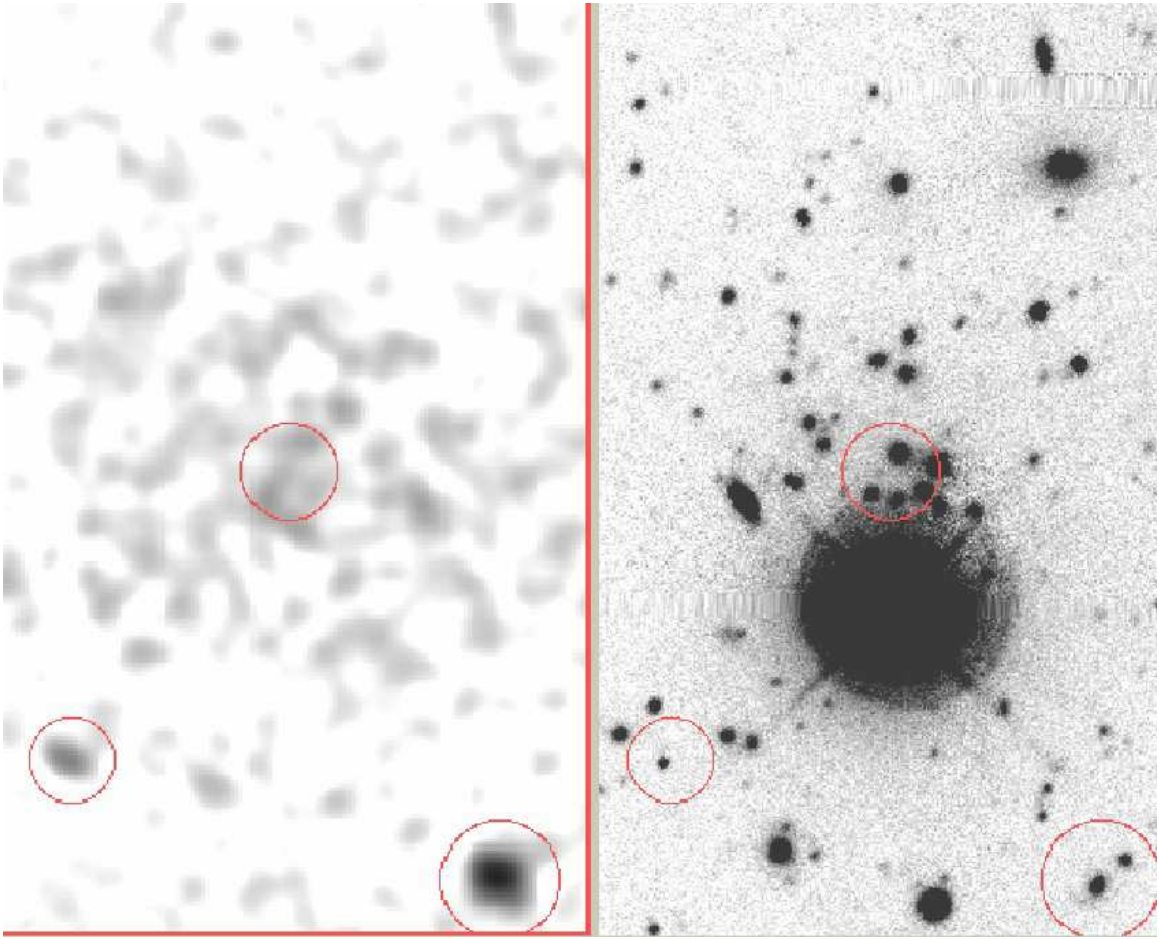}
\caption{\small 
CXOMP\,J033912.2-352614. 
LEFT: We note extended X-ray emission centered near $\alpha =
03~39~12.2,~\delta = -35~26~15$
in this ACIS-S image from ObsID 624.  The field is approximately
2\arcmin\ NS, and 1.3\arcmin\ EW.  The orientation (in this and all
images herein) is N up and E to the left. Circles show the positions 
of detected X-ray sources, with circle size corresponding to the
\Chandra\ 95\% EE at this off-axis angle.
RIGHT: Our CTIO4m/Mosaic imaging indeed shows a cluster of galaxies just N of
a $V=12$ star. The bright star is not the X-ray source since it is
19.5\arcsec\ South of the X-ray centroid, and stars do not generally
produce diffuse X-ray emission.   
\label{XS00624B7_045}}
\end{figure}

During our standard visual inspection of each X-ray source,
we noted extended X-ray emission centered near $03^h 39^m 12.2^s\,
-35^{\circ} 26\arcmin\ 15.0\arcsec$.
Figure~\ref{XS00624B7_045} shows the X-ray and optical \rp\ band
image of the region. A $\beta$ model to fit the cluster profile is
unconstrained, but fixing to an average cluster value of $\beta=0.6$
\citep{JC84}, we derive a core-radius $r_c=47\pm12\arcsec$ and 2821
counts (0.7-2~keV).\footnote{The cluster is only detectable to about
3$r_c$, within which there are 1630 counts.}  The (0.5-2~keV) flux
is therefore $3\times10^{-13}$\fcgs.  
This estimate may be affected (at the $\sim20\%$ level) by emission
from the extended halo of the central bright (10th mag) galaxy  of the
Fornax cluster, NGC~1399, which lies $9\arcmin$ to the west outside
the \Chandra\ field of view (Paolillo et al. 2002).  

In our optical imaging, the source corresponds to what appears to be a
cluster of galaxies just N of a bright ($V=12$ mag) star.
Figure~\ref{XS00624B7_045} shows the optical \rp\ band image of the
cluster next to the (2-pixel gaussian) smoothed X-ray image.  
We are confident that the X-ray emission does not originate from the
star.  Stars do not produce significantly extended X-ray emission.
Even if they did, the star could not reasonably be associated with the
diffuse emission situated as it is at 1.7 core radii (20\arcsec) from
the emission centroid.  As explored by \citet{DAP04},
the $(g^{\ast}-r^{\ast})\sim2.5$ colors of galaxies within 18\arcsec\,
of the X-ray centroid define a red sequence corresponding to cluster
elliptical galaxies at $z\sim0.4 - 0.5.$

\begin{figure}[ht]
\centering
\plotone{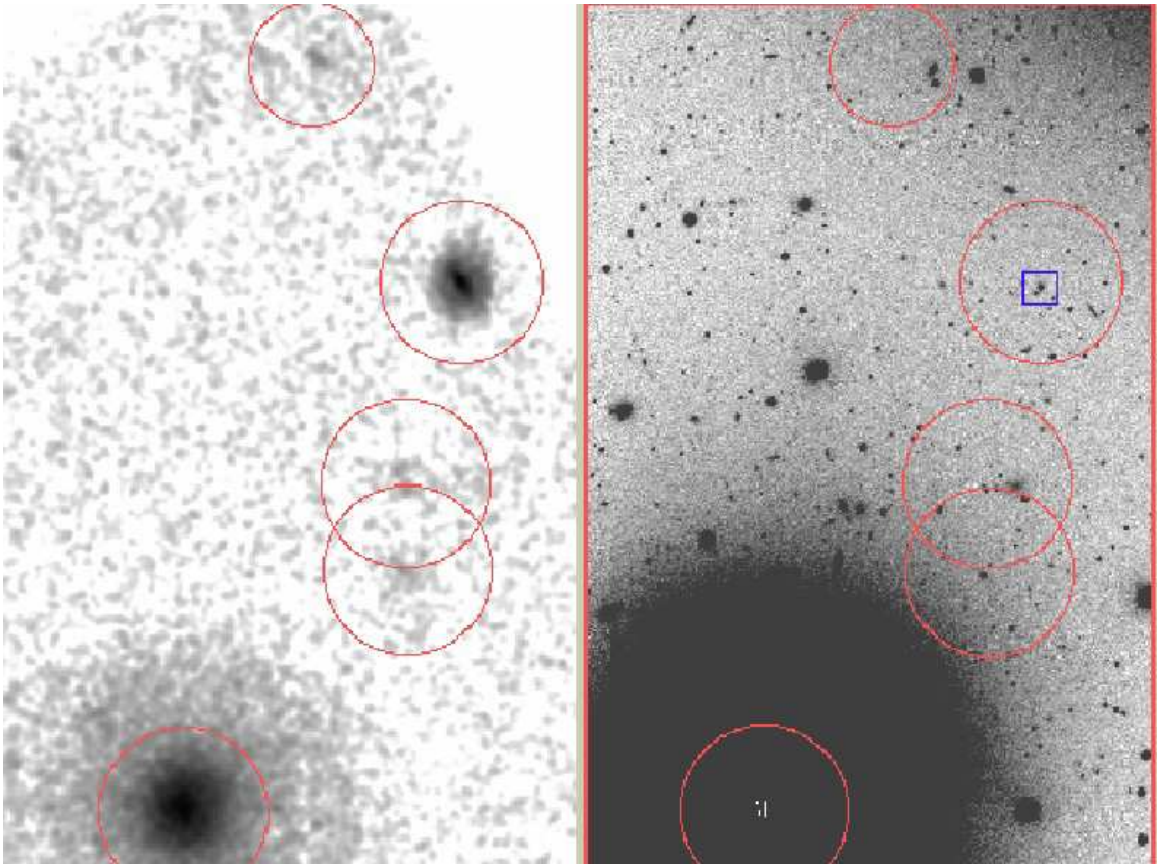}
\caption{\small 
CXOMP\,J033841.4-353133. Smoothed ACIS-I (LEFT) and optical \rp\ image
(RIGHT) of the field in ObsID 624 containing a bright background
quasar (see \S~\ref{lp944}). Each image is oriented N
up, E to the left, covering 6.7\arcmin N-S and 4.3\arcmin 
E-W. The BLAGN at $z=0.3597$ is marked by a box in the optical
image. Circles represent the \Chandra\ 95\% EE radius.  The large
(37\arcsec) EE radius and apparent non-point morphology of the X-ray
source reflect the PSF at this large (12.6\arcmin) off-axis angle.
\label{XS00624B3_002fig}}
\end{figure}

\subsubsection{CXOMP\,J054240.9-405515}
\label{XS00914B0_001}

\begin{figure}[ht]
\centering
\plotone{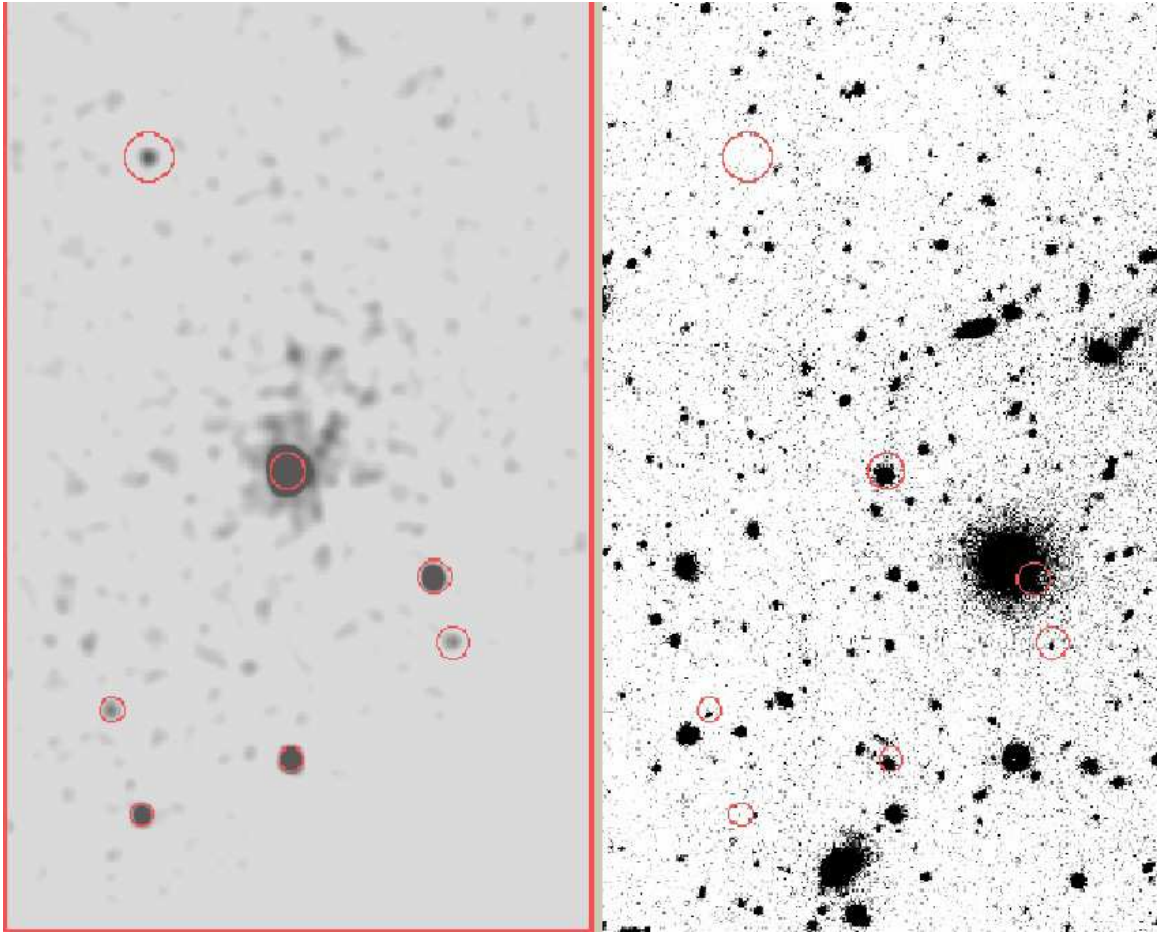}
\caption{\small CXOMP\,J054240.9-405515.
This source has well over 5000 counts in this ACIS-I image (LEFT), and
shows a sharp peak with significant ($\sim40\arcsec$) 
extent (see also the top panel of Figure~\ref{VIexamp} for a
close-up).  Our optical spectrum of the corresponding $\rs=18.3$ object  
(RIGHT) centered in each of these images (230\arcsec\ N-S, 140\arcsec\
E-W) reveals a $z=0.724$ BLAGN  (see \S~\ref{XS00914B0_001}).  The extended
emission quite likely arises in a cluster of galaxies whose members
are several mag fainter 
than the quasar.  
\label{XS00914B0_001fig}}
\end{figure}

The bright EXOSAT and ROSAT All-Sky Survey X-ray source EXO~0541.0-4056 
was studied in the ASCA Medium Sensitivity Survey as
1AXG~J054242-4054. Our spectrum reveals a BLAGN at $z=0.7243$, but the
X-ray image clearly shows extended emission out to $\geq40\arcsec$ from the
quasar (Figure~\ref{XS00914B0_001fig}).  A different stretch and zoom
of the same region is shown at the top of Figure~\ref{VIexamp}.
To better examine the cluster emission, we first created an image in
the 0.7-2~keV band only, and masked out all point sources and the quasar
with a circle of 4.5\arcsec radius.  After smoothing the masked image
with a 6\arcsec\ Gaussian, we determined the peak of the diffuse
emission to be at $05^h 42^m 40.4^s\, -40^{\circ} 55\arcmin\
7.0\arcsec$, 
the NE of the quasar, corresponding to a projected distance of
$\sim80$kpc, assuming the quasar and cluster are at the same redshift.
Since the cluster emission centroid is offset by
nearly $r_c$ from the brightest cluster galaxy (the quasar), this cluster
is not relaxed, and is most likely still sloshing from a merger.  The
fraction of such offsets as a function of redshift is potentially a
powerful probe of cluster evolution \citep{Cclustmerg01}. 

Extracting a profile around the cluster centroid, we 
find the cluster is well-fit with a $\beta$-model of core radius
$r_c=12\pm2.5\arcsec$, with $\beta=0.6\pm0.07$ and total flux of
$340\pm40$ counts\footnote{The errorbar includes the uncertainty due
to the value of the $\beta$.}, fitting down to the background
intensity of 0.03 counts/pixel.  The corresponding cluster flux of
$3.6\times 10^{-14}$\fcgs corresponds to a cluster luminosity
of $10^{44}$erg~s\mone.  

Deep imaging at fine spatial resolution also reveals evidence
in the optical and near-infrared for clusters of galaxies around
(radio loud) quasars \citep{yee93} at similar redshifts
\citep{sanchez02}.  These quasars often do not inhabit the cluster
cores, which may support the interaction/merger hypotheses for their
fueling, since lower velocities at large radii enhance merging
efficiency \citep{bekki94}. 

\subsubsection{CXOMP\,J234815.7+005350}
\label{XS00861B7_041}

\begin{figure}[ht]
\centering
\plotone{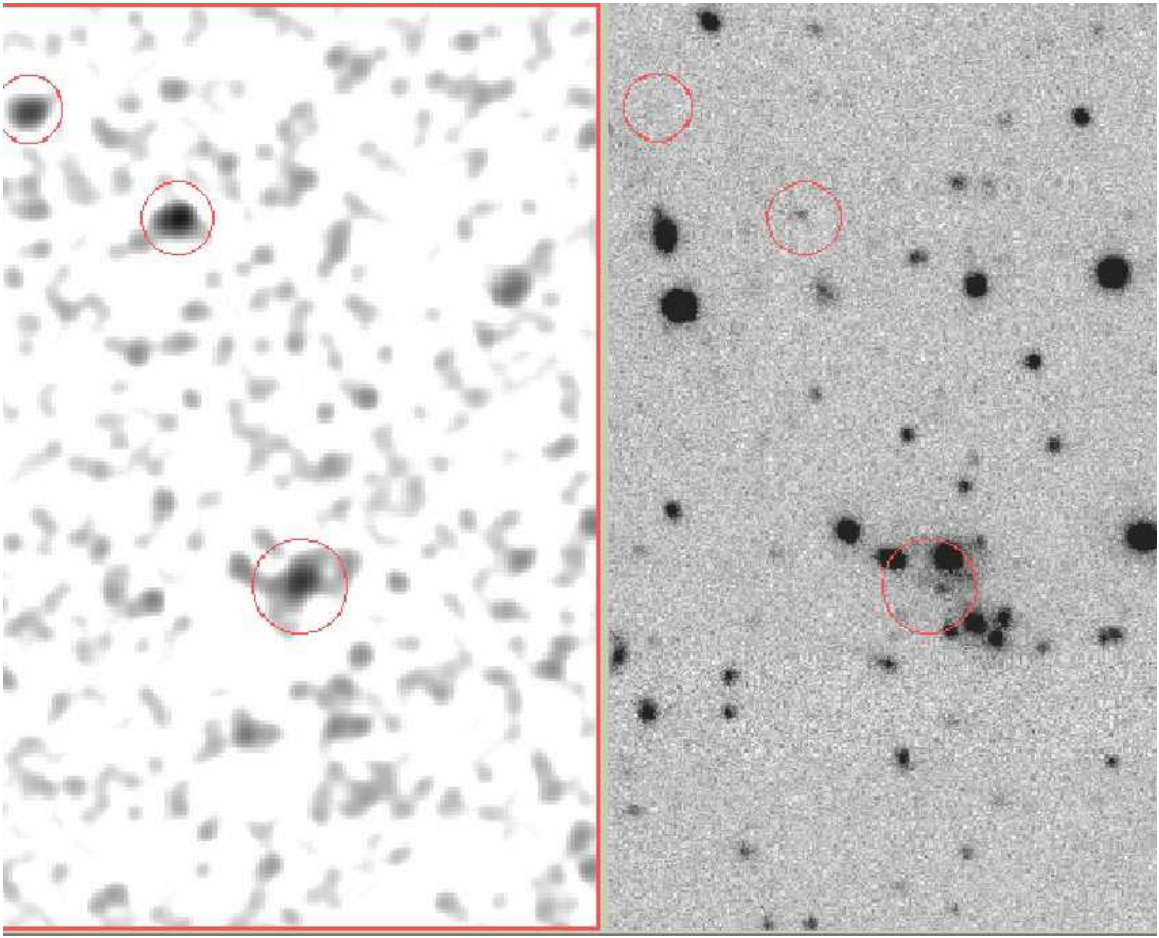}
\caption{\small 
CXOMP\,J234815.7+005350.  
LEFT: The X-ray source has just $31\pm8$ counts in this ACIS-S image
and morphology consistent with a point source.  Circles show the 95\%
EE PSF size, which has radius 6\arcsec\ at this off-axis angle
(4.8\arcmin). 
RIGHT: The \rp\ band image shows a group of galaxies coincident
near the X-ray centroid.  From our optical spectrum (Magellan 6.5m
with  LDSS2 03 Dec 2002), we tentatively identify the brightest
$\rs=19.7$ galaxy just NW of the X-ray centroid as a $z=0.4$
absorption line galaxy  (see \S~\ref{XS00861B7_041}).   Both images are
240\arcsec\ N-S, 90\arcsec\ E-W in size.  
\label{XS00861B7_041fig}}
\end{figure}

The X-ray emission from this object is consistent with a point source,
but the optical image looks convincingly like an intermediate redshift
group or cluster of galaxies (Figure~\ref{XS00861B7_041fig}).  We
tentatively assign a redshift of $z=0.41$ to the cluster, based on 
a spectrum (Magellan 6.5m with LDSS2 03 Dec 2002) of the brightest
$\rs=19.7$ galaxy.  The source flux of $\fx = 1.0\times10^{-15}$\fcgs\,
at this redshift corresponds to $\loglx=41.84$.  If there is no
significant contribution from cluster emission, the evidence suggests
that an X-ray bright elliptical galaxy dominates the group. The object
is not likely to be a BL~Lac because the CaII break strength is normal
for an elliptical galaxy ($0.57\pm0.1$).
Some diffuse optical emission may be from the bright extended disk
of the host galaxy ($\sim 15$\,kpc in radius), in which case a knot
of optical emission near the X-ray centroid suggests the possibility
of an ultraluminous X-ray source (ULX) in the host.  However, this
would be 30 times brighter than the brightest known ULX in the Antenna
galaxies \citep{zezas02} or M82 \citep{kaaret01}.  Also, we would
expect to see strong optical emission lines from the associate star
formation regions in the disk.  

\section{Multiple Sources}
\label{pairs}

 Close X-ray sources ($<10\arcsec$) with optical counterparts are good lens
candidates, especially if there is also extended emission signifying a
massive potential.  They are also interesting for study of the
effects of interactions/mergers on activity \citep{GQ234502}.  Sources
with  larger separations where one lacks an optical counterpart are good
candidates for X-ray jets, whose emission barely dims at all with
distance \citep{SDA02}.  We have so far identified 2 close pairs
of X-ray sources in these 6 fields.

\begin{figure}[ht]
\centering
\plotone{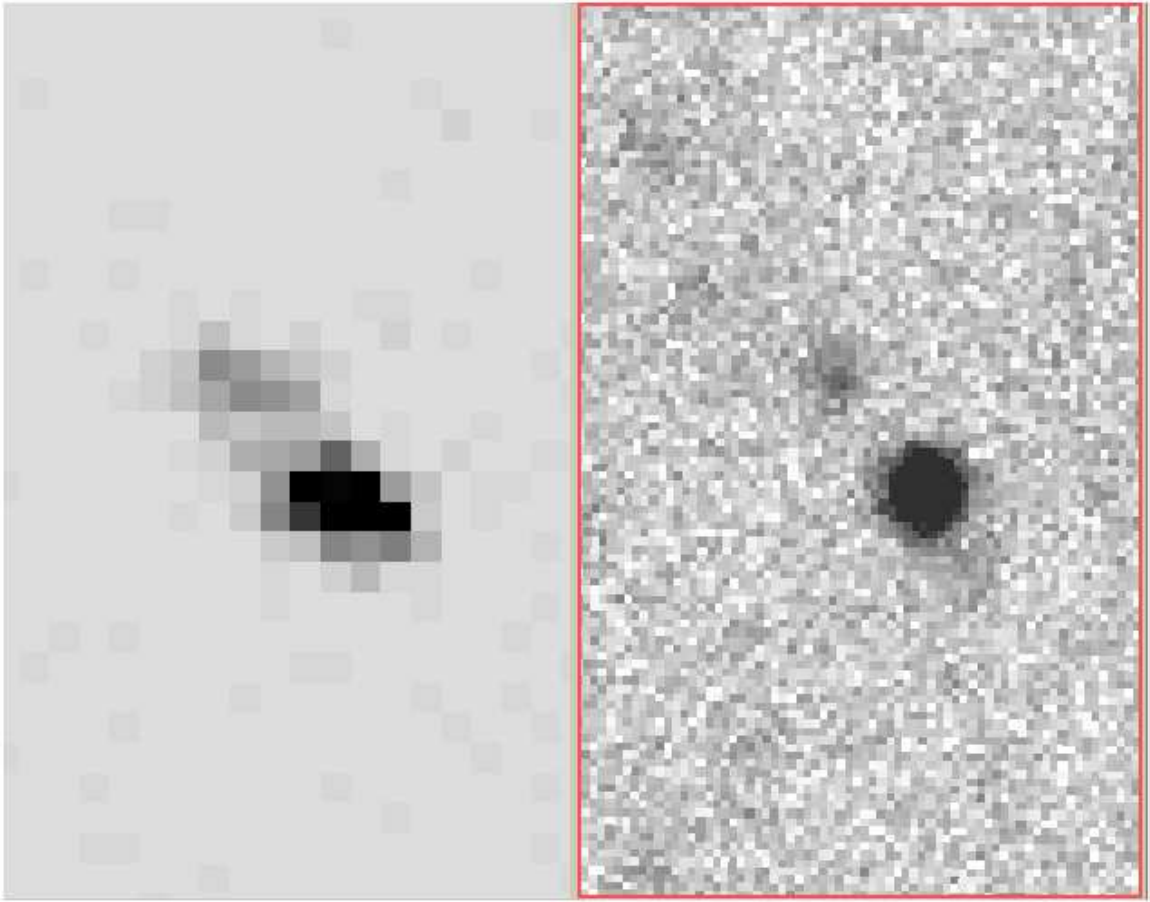}
\caption{\small 
The pair of X-ray sources shown here - CXMOP\,J234835.3+005832 and
CXOMP\,J234835.5+005836 - are separated by 4.9\arcsec\ in the
ACIS-S image (LEFT), and both have reliable optical counterparts in
our Mosaic \rp\-band image (RIGHT).  The field shown is approximately
30\arcsec\ NS, and 18\arcsec\ EW.  Our spectrum of the
brighter more southern object shows it to be a BLAGN at $z=0.9469$.
Spectroscopic follow-up of the fainter object $\is=23.3$
should help determine whether the system is lensed.  
\label{XS00861B7_014}}
\end{figure}

We find a new close pair of X-ray sources with optical counterparts in
the field of ObsID 861, as shown in Figure~\ref{XS00861B7_014}.  
A Magellan spectrum from  14 July 2002 reveals that the brighter 
object CXOMP\,J234835.3+005832 is a $z=0.95$ BLAGN. Since the 90\%
encircled energy (EE) PSFs of these sources overlap, we estimated their
individual counts as follows.  We first measure the counts for each
source in a core radius of 3pixels. We then measure the total counts
in the union of 2 aperture of radius of 10.4 pixels (5\arcsec)
encompassing both sources, and apportion the total counts to each
according their core counts ratio.   
The fainter source CXOMP\,J234835.5+015836 is beyond our spectroscopic
survey limit, but should be followed up as a lens candidate.
The 2 objects' \fxfr\ values differ significantly, 
(with $\logxr=0.72$ and 0.34 for the fainter and brighter object,
respectively) so we believe that they are unlikely to
be lensed.  However, not all lensed objects are achromatic across a
wide range in frequency;  microlensing, time variability and
differential absorption are possible reasons (e.g., Green et al. 2002;
Morgan et al. 2001).  

\begin{figure}[ht]
\centering
\plotone{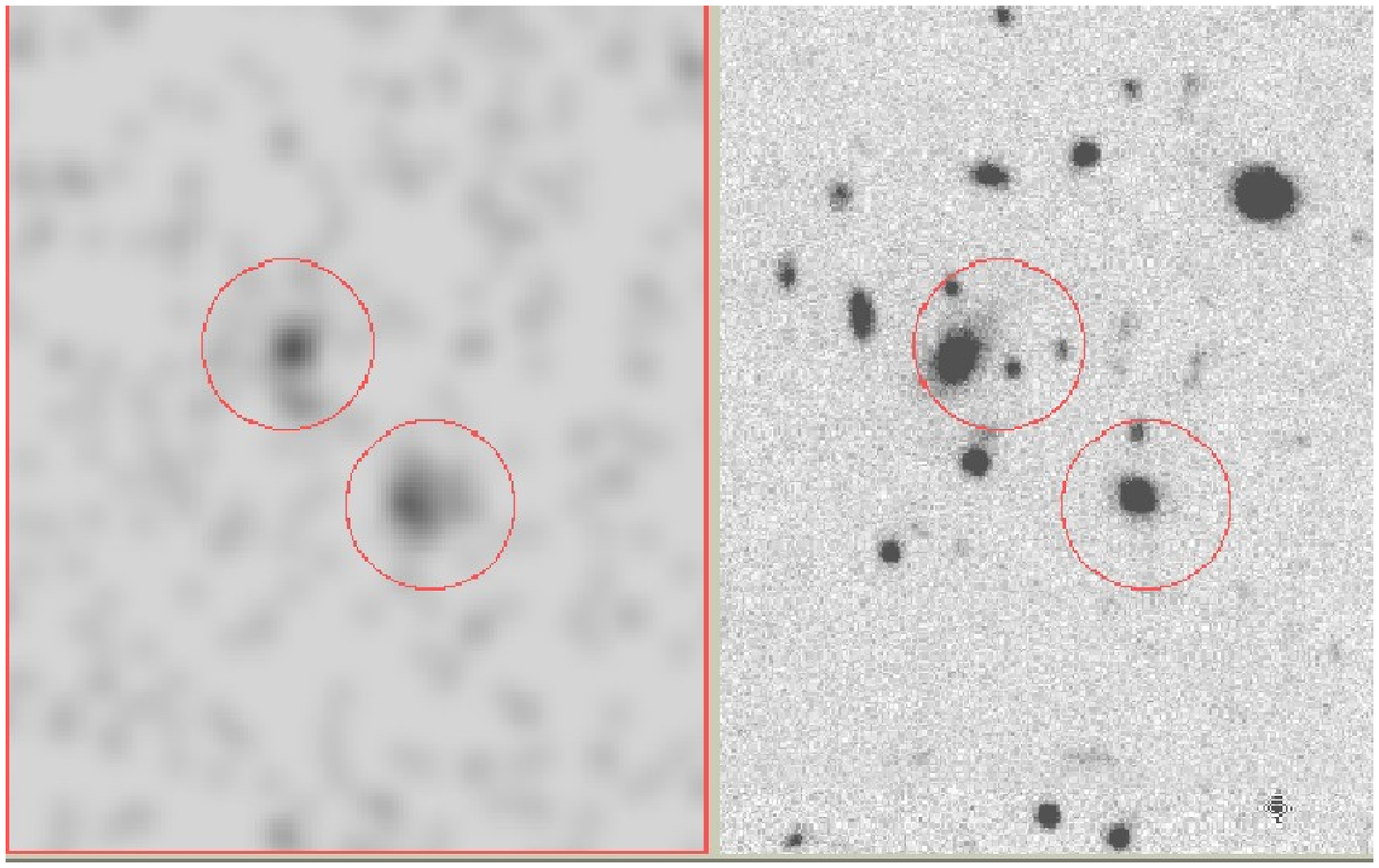}
\caption{\small 
X-ray (LEFT) and optical \rp -band (RIGHT) images of
CXOMP\,J234758.9+010343 and CXOMP\,J234758.0+010329. Circles show the PSF
size (50\% EE) at this (7\arcmin) off-axis angle.  Each
image is 90\arcsec NS and 70\arcsec\ EW.  The apparent X-ray extent of
the individual sources is consistent with the \Chandra\ PSF.  The
southern object CXOMP\,J234758.0+010329 is a $\rs=20.1$ ALG at
$z=0.2481$.  The optical counterpart centroids are separated by
19\arcsec.  While these X-ray emitting objects may belong to a group or
cluster, we detect no significant extended emission.  
\label{XS00861B6_028tile}}
\end{figure}

CXOMP\,J234758.0+010329 is an $\rs=20.1$ ALG at $z=0.2481$.
As shown in Figure~\ref{XS00861B6_028tile},  CXOMP\,J234758.9+010343 and
CXOMP\,J234758.0+010329 are separated by only 10\arcsec, and both appear
to belong to a group of galaxies.  We detected no extended emission
from the group.

\section{Plans and Prospects}
\label{prospects}

Significant ($>10\%$) cosmic variance occurs between the \lognlogs\,
in imaged X-ray fields \citep{RPTP02} at {\em brighter} fluxes
($\logfx>-14.3$ cgs; 0.5-2~keV), where cumulative source densities are
just 130deg\mtwo\, ($\sim$10 sources per ACIS-I field). 
How strong is the X-ray cosmic variance, and is it associated with
cluster-mass lenses?  Using \Chandra\ data, an X-ray point-source
excess over the expectations of blank fields in the fields of several
clusters has been claimed \citep{CMMP01,MPKD02}. In the field of A2104,
\citet{MPKD02} found an unexpectedly high fraction of these sources to
be red cluster member galaxies both with and without optical
spectroscopic signatures of AGN.  The ChaMP has a large
fraction\footnote{Of the full field list, 25\% have cluster targets.
Of the fields with 20 ksec exposures or greater, half have cluster
targets.} of target clusters in the field list, and so is well-suited
for testing for any such effect. \citet{kim03b} examine ChaMP
cluster and non-cluster field subsamples and compare the co-added
\lognlogs, finding no significant difference in point source counts.

In these 6 fields we have discovered 3 new clusters of galaxies.
At least two are above $z=0.4$, while the third still lacks a
redshift.
Two show X-ray extent, of which one hosts a bright quasar.  Further 
spectroscopy is needed to verify membership of faint 
galaxies seen optically. Our optical imaging program will allow a new
generation of studies of the clustering of X-S AGN and of
AGN-galaxy correlations.  With their low surface density and high
luminosities, do AGN trace high peaks of the density field, or do they
randomly sample the galaxy distribution? Is there a change in form of
the AGN correlation function at small separations, implying
interactions, or triggering \citep{EGRM88, KFM99}?  What is the
lifetime of activity in galactic nuclei?  With the assumption that
luminous quasars reside in the most massive host halos, weaker
clustering implies shorter quasar lifetimes \citep{HH01, MW01}.

Multi-wavelength data, beyond the optical and X-ray, is also being
planned in the longer term.  A subset of ChaMP
sources are too faint optically to be identified in our imaging
program, but we have initiated ground-based near-IR imaging.  This
reveal many sources which are optically faint due to obscuration likely
including a significant subset of the AGN population.   We
have also begun a program to obtain high resolution radio followup
observations of
the bright radio sources (NVSS and/or FIRST identifications) in the
ChaMP sample using the VLA A array at 3.5 cm. This not only allows us
to study the radio properties of the X-S sample, but also
assists in source classification. Most of the sources observed to date
are unresolved in the radio, as expected for typical broad-lined
AGN. However a significant subset have complex double/triple structure
and another subset prove to be mis-identifications due to the low
resolution of the radio survey data.  Once these observations are
complete, we anticipate both extending the wavelength coverage for
these bright radio sources to 20 cm and planning deeper observations
to study the radio-quieter part of the X-ray sample.  

Beyond the mechanics of source identification and classification, a
primary science goal of the ChaMP is to obtain an AGN sample which is more
representative of the AGN population as a whole than previous samples,
due to selection in a broad, hard X-ray band.  With this sample we seek
to understand the intrinsic population, study the range of spectral
energy distributions and fractions of various source types in order to
better constrain models for the geometry and physics of the inner
regions.  To some extent such studies will be facilitated by the
multi-wavelength data described above, but in the longer term we also
expect to follow-up one/more well-defined subsets in more detail to
better understand the emission mechanisms involved in the various
wavebands.  

\subsection{Public Data}
\label{public}

Quantities discussed and plotted here are available in catalog
form along with image data for the 6 fields in this paper 
through our web site {\url http://hea-www.harvard.edu/CHAMP/}.
For scientists using ChaMP catalog or image data, we appreciate
an acknowledgment: ``This work made use of images and/or data products
provided by the Chandra Multiwavelength Project (ChaMP; Green et
al. 2002), supported by NASA.  Optical data for the ChaMP 
are obtained in part through the National Optical Astronomy
Observatory (NOAO), operated by the Association 
of Universities for Research in Astronomy, Inc. (AURA) under
cooperative agreement with the National Science Foundation.''

\acknowledgments

We gratefully acknowledge support for this project under
NASA under CXC archival research grant AR2-3009X.
RAC, AD, PJG, DK, DM, and BW also acknowledge support through NASA
Contract NASA contract NAS8-39073 (CXC). 

We are indebted to the staffs at Kitt Peak, CTIO, Las Campanas,
Keck, and MMT for assistance with optical observations.
Thanks to Arne Henden for accurate coordinates for standard
stars, and to Warren Brown, Perry Berlind, Michael Calkins, and Susan 
Tokarz for providing or helping with spectroscopy on Mt Hopkins. 
Thanks to the referee for a thoughtful and thorough review.

Hearty thanks to the whole Chandra team for an exquisite observatory
and a quality archive.  We are particularly grateful to the late Leon
VanSpeybroeck for allowing us early access to his proprietary
\Chandra\, data, and for the astonishing design quality of the
\Chandra, mirrors.  Leon uniquely combined brilliance, integrity
and humility. That inspiration and his legacy, \Chandra, live on.



{}

\clearpage






\clearpage

\appendix

\section{Individual Objects in 6 ChaMP Fields}
\label{indiv}

Below we note idiosyncracies of particular fields
and objects therein, if they were not discussed previously.
We also mention previously known sources, based on a
NED search within 12\arcsec\ of all X-ray sources.  We do not tabulate
or reference sources that have been detected in the growing number of
all-sky or serendipitous surveys (e.g., ROSAT, SDSS, 2MASS, APM)
unless a spectroscopic classification has been published.

\subsection{MS1137.5+6625 }	

At $z=0.7824$, the target cluster MS1137.5+6625 (Leon van Speybroek,
PI) is the second most distant of the EMSS clusters \citep{DonMeg99},
and is part of a GTO program to combine X-ray and radio measurements
to constrain $H_0$ and $q_0$ through the  Sunyaev Zel'dovich
effect. The \Chandra\ image of the target cluster has been analyzed by
\citet{GMS113702}.  

CXOMP\,J114148.0+660603, with 125 $H$ and 19 $S$ band counts, this is
the hardest amongst all the sources in this study ($HR=0.74\pm0.08$).
It is relatively bright ($\rs=21.9,~\is=20.5$), but not detected
($\gs>25.3$) in the \gp\ band.  


CXOMP\,J114013.9+655940: 
The very red color ($\gmr>2.9$) of this $\rs=21$ absorption line
galaxy is unexpected for a $z=0.5$ elliptical
galaxy such as this \citep{EGALCOLORS01}, and even more so given that
the object has a very {\em soft} X-ray spectrum ($HR_0 < -0.67$).
With log\,$L_X =42.56$, this galaxy certainly contains an AGN,
but the optical light may be dominated by a reddened host galaxy.

\subsection{V1416+4446}      

V1416+4446 is a $z=0.40$ cluster included in the same GTO program 
(Leon van Speybroek, PI) as MS1137+6625.

Included within the ACIS-I field is the bright $z=0.114$ Seyfert~1
galaxy PG\,1415+451.  Here there are 5067$\pm 72$ ACIS counts,
corresponding to $9\times 10^{-13}$\fcgs\, (0.5-2keV).  This
source is also detected in the Faint Images of 
the Radio Sky at Twenty cm (FIRST) survey \citep{white97}, with
$1\pm0.14$ mJy, but is nevertheless radio-quiet \citep{xu99}. 
%
Using a 5\arcsec\, search radius, the only other FIRST source we find
in these 6 fields is CXOMP\,J141655.7+445452.  With $\rp=20.94$, an
(0.5-2keV) flux $2\times 10^{-14}$\fcgs\, and $1.37\pm0.34$ mJy in the
FIRST survey, it is likely to be a radio-loud quasar.



\subsection{LP944-20}       
\label{lp944}

While the target of this observation 
(PI Gibor Basri; see Rutledge et al. 2000) is a brown dwarf star,
the field is near ($\sim 13\arcmin$ from) the center of the Fornax
cluster, which includes the two bright galaxies NGC~1399 and
NGC~1404.  The field has been studied in the optical in detail
recently by Hilker et al. (1999a,b) who discovered a background galaxy
cluster at z=0.11 behind the center of the Fornax cluster as part of
the 2dF Fornax Cluster Spectroscopic Survey (FCSS; Drinkwater et
al. 2000).   


%
FCSS~J033922.5-352530 (CXOMP\,J033922.4-352530) is a $B_J=17$,
$\rs=17.56$ galaxy at $z=0.04664.$ 


NGC~1404 at $v=1947$\kms\ is an 11th mag elliptical galaxy
detected in the field with 10710 counts.
ROSAT HRI, PSPC and \Chandra\ data on the field were
studied by Paolillo et al. (2002), in which they find that the
NGC 1404 halo is well represented by a single symmetric $\beta$-model
and follows the stellar light profile within the inner 8kpc. 
While it decreases the effective sky area available in this ObsID for
serendipitous sources, our analyses do not include this object.

CXOMP\,J033841.4-353133;
Our HYDRA-CTIO spectroscopy of this bright (1710 count, \rp=20,
$HR= -0.675$) X-ray source shown in Figure~\ref{XS00624B3_002fig} reveals
a BLAGN at $z=0.3597$. Since it is in the background of the Fornax
cluster, the X-ray spectrum of this quasar provides a direct measure
of all the gas both neutral and ionized, along one line of sight through the 
cluster, the halos of both of its giant member galaxies, and the
immediate environment of the quasar.   We use the ACISABS
model\footnote{http://www.astro.psu.edu/users/chartas/xcontdir/xcont.html}
within CIAO Sherpa to account for the time-dependent quantum efficiency
degradation of ACIS and fit an absorbed power-law model to the X-ray
spectrum of the quasar.  This yields a standard AGN slope of
$\Gamma=1.81\pm0.06$ (90\% confidence limits), with a marginal excess 
absorbing column of $1.6\pm 1.5$ over the local Galactic column (which
is 1.5, all columns in units of $10^{20}$ cm$^{-2}$). Our result is
consistent with those of \citet{WM02} who find a considerable deficit
of HI-rich galaxies in the centre of the cluster, but our fit further
suggests that little ionized gas may be present, which is worthy of
further investigation.



CXOMP\,J033942.9-352410
is a previously known $\rs=19.0$, $z=1.04295$
quasar FCSS~J033942.9-352410. From our HYDRA-CTIO spectra, we measure
$z=1.0417$ 

FCSS~J033943.1-352200 (CXOMP\,J033943.0-352159) is listed as an $r=17.3$
NELG at $z=0.06208$  

FCSS~J034015.4-352850 (CXOMP\,J034015.4-352849) is a $\rs=18.8$ quasar
listed at $z=0.85504$, which assumes the bright emission line at
5225\AA\ to be MgII.  Instead, we classify this as a $z=1.74$ BLAGN,
because on the blue side of this emission line we detect the
characteristic  shoulder of AlIII superposed  on CIII]$\lambda$1909.

%
%

\subsection{Q2345+007}       

The target of this \Chandra\ image (Green et al. 2002)
is a wide ($7\arcsec$.3) separation quasar pair Q2345+007A,B
($z=2.15$), whose components have strikingly similar optical spectra
(Steidel \& Sargent 1991).  A deep search for a lensing cluster
revealed nothing, and differences in the absorption X-ray hardness led
the authors to posit that the system is a binary and not a lens.
Evidence for or against the lens hypothesis has been sought in optical
photometric variability studies, allowing a search for variable
objects other than the quasars.

Three of the X-ray sources fainter than $\rp=20$
in this field are flagged as highly variable in Kochanski et al. and
as such are likely to be AGN.  
CXOMP\,J234811.6+005700: This $B_J=22$ highly variable object (G5158)  
has 269 counts in our ACIS image. 
CXOMP\,J234811.9+010301 (G7676) with just 14 ACIS counts is another such
source. 
CXOMP\,J234812.8+005750 (G5740) has 203 ACIS counts.

CXOMP\,J234813.8+005640: We classify this $\rs=20.46$ object as a $z=1.0416$
quasar\footnote{Confusingly, an object at these coordinates is listed
\citep{VCV00} as a $z=2.653$ quasar with the name 2345+007C, but these
coordinates are not near the target objects that comprise 2345+007,
which in any case has $z=2.15$.} based on a single broad emission line
at 5715\AA.  

CXOMP\,J234840.1+010753: 
A $z=0.71868$ quasar published in the SDSS Early Data Release
(2001) as SDSS~J234840.05+01075 quasar.  We include the SDSS redshift
in our tabulations.

\subsection{CLJ0542.8-4100} 	

The intended target of this field (Harald Ebeling, PI)
is a high redshift cluster.





CXOMP\,J054320.2-410156.1:
The extremely X-ray bright cataclysmic variable TX~Col  is imaged in
this field.  TX Col is a highly variable member of the intermediate
polar subclass of CVs.  A complete study of the Chandra observation of
TX Col is in preparation \citep{Salin03}.

CXOMP\,J054328.5-410119: Our imaging reveals an $\is=21$ object with
complicated morphology, possibly 3 objects.  A spectrum of the object
shows an NELG at $z=0.3987$.

\subsection{MS2137.3-2353}  	

This $z=0.313$ cluster was chosen as a target (PI Mike Wise)
for study of cooling flows.  None of the X-ray sources in this field
yielded matches in NED. 

CXOMP\,J213945.0-234655 is an X-ray selected $z=4.930$ QSO 
discovered by \citep{SJGP02} from the ChaMP.  

CXOMP J214018.0-234920 has an extremely blue optical spectrum, 
and only 1 broad line at 6735\AA, which we take to be rest-frame
Mg\,II$\lambda\,2800$ at $z=1.402\pm0.002$. 




\end{document}